\documentclass[twoside,english,authoryear]{elsarticle}
\usepackage[scaled=1.3]{beramono}
\usepackage[T1]{fontenc}
\usepackage[latin9]{inputenc}
\usepackage{geometry}
\usepackage{array}
\usepackage{amsmath}
\usepackage{amssymb}
\usepackage{graphicx}
\usepackage{esint}

\makeatletter

\providecommand{\tabularnewline}{\\}

\makeatother

\usepackage{babel}

\usepackage{lineno}

\journal{Journal of Mathematical Psychology}

\usepackage{babel}

\usepackage{amsthm}

\newtheorem{theorem}{Theorem}
\newtheorem{corollary}{Corollary}

\makeatother

\usepackage{babel}
\begin{document}

\title{Phase-Oscillator Computations as Neural Models of Stimulus-Response
Conditioning and Response Selection\tnoteref{1}}

\author[CSLI]{P.~Suppes}

\ead{psuppes@stanford.edu}

\author[LS]{J.~Acacio~de~Barros}

\ead{barros@sfsu.edu}

\author[CSLI]{G.~Oas}

\ead{oas@stanford.edu}

\address[CSLI]{CSLI, Ventura Hall, 220 Panama Street, Stanford University, Stanford,
CA 94305-4101}

\address[LS]{Liberal Studies Program, San Francisco State University, 1600 Holloway
Ave., San Francisco, CA 94132}
\begin{abstract}
The activity of collections of synchronizing neurons can be represented
by weakly coupled nonlinear phase oscillators satisfying Kuramoto's
equations. In this article, we build such neural-oscillator models,
partly based on neurophysiological evidence, to represent approximately
the learning behavior predicted and confirmed in three experiments
by well-known stochastic learning models of behavioral stimulus-response
theory. We use three Kuramoto oscillators to model a continuum of
responses, and we provide detailed numerical simulations and analysis
of the three-oscillator Kuramoto problem, including an analysis of
the stability points for different coupling conditions. We show that
the oscillator simulation data are well-matched to the behavioral
data of the three experiments. \end{abstract}
\maketitle

\tnotetext[t1]{This article is dedicated to the memory of William K. Estes, who was one of the most important scientific psychologists in the second half of the twentieth century.  His path-breaking 1950 Psychological Review article, "Toward a statistical theory of learning," (vol.57, pp. 94-107) is the "grandfather" of the present work, because of his mentorship of the first author (PS) of this article in the 1950s.  It was through Estes' postdoctoral tutoring at the Center for Advanced Study in the Behavioral Sciences in 1955-1956 that he learned stimulus-response theory.}

\section{Introduction}

\tnotetext[1]{This document is a collaborative effort.}

With the advent of modern experimental and computational techniques,
substantial progress has been made in understanding the brain's mechanisms
for learning. For example, extensive computational packages based
on detailed experimental data on ion-channel dynamics make it now
possible to simulate the behavior of networks of neurons starting
from the flow of ions through neurons' membranes \citep{BowerBeeman2003}.
Yet, there is much to be determined at a system level about brain
processes. This can be contrasted to the large literature in psychology
on learning, psychophysics, and perception. Among the most developed
behavioral mathematical models of learning are those of stimulus-response
(SR) theory \citep{estes1950towarda2,EstesSuppes1959,suppes1959alinear,suppes1960stimulus,SuppesAtkinson1960,SuppesGinsberg1963}.
(From the large SR literature, we reference here only the results
we use in some detail.) In this article, we propose weakly coupled
phase oscillators as neural models to do the brain computations needed
for stimulus-response conditioning and response conditioning. We then
use such oscillators to analyze two experiments on noncontingent probabilistic
reinforcement, the first with a continuum of responses and a second
with just two. A third experiment is on paired-associate learning.

The oscillator assumptions we make are broadly based on neurophysiological
evidence that neural oscillators, made up of collections of synchronized
neurons, are apparently ubiquitous in the brain. Their oscillations
are macroscopically observable in electroencephalograms \citep{Freeman1979,GerstnerKistler2002,WrightLiley1995}.
Detailed theoretical analyses have shown that weakly interacting neurons
close to a bifurcation exhibit such oscillations \citep{GerstnerKistler2002,HoppensteadtIzhikevich1996a,HoppensteadtIzhikevich1996b,Izhikevich2007}.
Moreover, many experiments not only provide evidence of the presence
of oscillators in the brain \citep{EckhornEtAl1988,FriedrichEtAl2004,KazantsevEtAl2004,LutzEtAl2002,MurthyFetz1992,ReesEtAl2002,RodriguezEtAl1999,SompolinskyEtAl1990,SteinmetzEtAl2000,Tallon-BaudryEtAl2001,Wang1995},
but also show that their synchronization is related to perceptual
processing \citep{FriedrichEtAl2004,KazantsevEtAl2004,LeznikEtAl2002,MurthyFetz1992,SompolinskyEtAl1990}.
They may play a role in solving the binding problem \citep{EckhornEtAl1988}.
More generally, neural oscillators have already been used to model
a wide range of brain functions, such as pyramidal cells \citep{LyttonSejnowski1991},
effects of electric fields in epilepsy \citep{ParkEtAl2003}, activities
in the cat visual cortex \citep{SompolinskyEtAl1990}, learning of
songs by birds \citep{TrevisanEtAl2005}, learning \citep{vassilieva2011learning},
and coordinated finger tapping \citep{YamanishiEtAl1980}. \citet{SuppesBrain2000}
showed that a small number of frequencies can be used to recognize
a verbal stimulus from EEG data, consistent with the brain representation
of language being neural oscillators. From a very different perspective,
in \citet{billock2005sensory} synchronizing neural oscillators were
used to rescale sensory information, and \citet{billock2011tohonor}
used synchronizing coupled neural oscillators to model Steven's law,
thus suggesting that neural synchronization is relevant to cognitive
processing in the brain. 

Our working hypothesis is that main cognitive computations needed
by the brain for the SR experiments we consider can be modeled by
weakly coupled phase oscillators that together satisfy Kuramoto's
nonlinear equations \citep{AcebronEtAl2005,HoppensteadtIzhikevich1996a,HoppensteadtIzhikevich1996b,Kuramoto1984,Strogatz2000,Winfree2002}.
When a stimulus is sampled, its corresponding neural oscillator usually
phase locks to the response-computation oscillators. When a reinforcement
occurs, the coupling strengths will often change, driven by the reinforcement
oscillator. Thus, SR conditioning is represented by phase locking,
driven by pairwise oscillator couplings, whose changes reflect learning.

Despite the large number of models of processing in the brain, we
know of no detailed systematic efforts to fit neural oscillator models
to behavioral learning data. We choose to begin with SR theory for
several reasons. This theory has a solid mathematical foundation \citep{SuppesRISS2002};
it has been used to predict many non trivial quantitative features
of learning, especially in the form of experimentally observed conditional
probabilities \citep{Bower1961,SuppesAtkinson1960,SuppesGinsberg1963,SuppesEtAl1964}.
It can also be used to represent computational structures, such as
finite automata \citep{suppes1969stimulusresponse,SuppesRISS2002}.
Furthermore, because neural oscillators can interfere (see, for example,
our treatment of a continuum of responses below), they may provide
a basis for quantum-like behavior in the brain \citep{deBarrosSuppes2009a,SuppesdeBarros2007},
an area of considerable research in recent years \citep{BruzaBusemeyerGabora2009}.

In Section \ref{sec:Stimulus-Response-Theory} we start with a brief
review of SR theory, followed by a detailed stochastic process version
of an SR model for a continuum of responses. In Section \ref{sec:Oscillator-Model-SR}
we present in some detail the neural oscillator computation for the
SR continuum model. Extension to other SR models and experiments follows,
as already remarked. In Section \ref{sec:Comparison-with-Experiment}
we compare the computations of the oscillator simulations to empirical
data from three behavioral experiments designed to test the predictions
of stimulus-response theory.

\section{Stimulus-Response Theory\label{sec:Stimulus-Response-Theory}}

The first aim of the present investigation is to use the extension
of stimulus-response theory to situations involving a continuum of
possible responses, because the continuum case most closely corresponds
to the physically natural continuous nature of oscillators. The SR
theory for a finite number of responses stems from the basic paper
by \citep{estes1950towarda2}; the present formulation resembles most
closely that given for the finite case in \citet[Chapter 1]{SuppesAtkinson1960},
and in the continuum case \citep{suppes1960stimulus}.

The general experimental situation consists of a sequence of trials.
On each trial the participant (in the experiment) makes a response
from a continuum or finite set of possible responses; his response
is followed by a reinforcing event indicating the correct response
for that trial. In situations of simple learning, which are characterized
by a constant stimulating situation, responses and reinforcements
constitute the only observable data, but stimulus-response theory
postulates a considerably more complicated process which involves
the conditioning and sampling of stimuli, which are best interpreted
as patterns of stimuli, with a single pattern being sampled on each
trial. In the finite case the usual assumption is that on any trial
each stimulus is conditioned to exactly one response. Such a highly
discontinuous assumption seems inappropriate for a continuum or a
large finite set of responses, and it is replaced with the general
postulate that the conditioning of each stimulus is \textit{smeared}
over a set of responses, possibly the whole continuum. In these terms,
the conditioning of any stimulus may be represented uniquely by a
\textit{smearing probability distribution}, which we also call the
\textit{conditioning distribution} of the stimulus. 

The theoretically assumed sequence of events on any trial may then
be described as follows: \\
\medskip{}

\begin{tabular}{ccccccccc}
\multicolumn{1}{c}{{\footnotesize }%
\begin{tabular}{c}
{\footnotesize trial begins }\tabularnewline
{\footnotesize with each }\tabularnewline
{\footnotesize stimulus in a}\tabularnewline
{\footnotesize a certain}\tabularnewline
{\footnotesize state of }\tabularnewline
{\footnotesize conditioning}\tabularnewline
\end{tabular}} & {\footnotesize $\rightarrow$} & {\footnotesize }%
\begin{tabular}{c}
{\footnotesize a}\tabularnewline
{\footnotesize stimulus }\tabularnewline
{\footnotesize (pattern) is}\tabularnewline
{\footnotesize sampled}\tabularnewline
\end{tabular} & {\footnotesize $\rightarrow$} & {\footnotesize }%
\begin{tabular}{c}
{\footnotesize response occurs,}\tabularnewline
{\footnotesize being drawn from}\tabularnewline
{\footnotesize the conditioning }\tabularnewline
{\footnotesize distribution }\tabularnewline
{\footnotesize of the sampled}\tabularnewline
{\footnotesize stimulus}\tabularnewline
\end{tabular} & {\footnotesize $\rightarrow$} & {\footnotesize }%
\begin{tabular}{c}
{\footnotesize reinforcement}\tabularnewline
{\footnotesize occurs}\tabularnewline
\end{tabular} & {\footnotesize $\to$} & {\footnotesize }%
\begin{tabular}{c}
{\footnotesize possible}\tabularnewline
{\footnotesize change in}\tabularnewline
{\footnotesize conditioning}\tabularnewline
{\footnotesize occurs.}\tabularnewline
\end{tabular}{\footnotesize{} }\tabularnewline
\end{tabular}\\
\medskip{}

The sequence of events just described is, in broad terms, postulated
to be the same for finite and infinite sets of possible responses.
Differences of detail will become clear. The main point of the axioms
is to formulate verbally the general theory. As has already been more
or less indicated, three kinds of axioms are needed: conditioning,
sampling, and response axioms. After this development, more technically
formulated probabilistic models for the different kind of experiments
are given.

\subsection{\noindent General Axioms}

\noindent The axioms are formulated verbally but with some effort
to convey a sense of formal precision.\\

\noindent \textbf{Conditioning Axioms.} \\

\noindent C1. For each stimulus $s$ there is on every trial a unique
conditioning distribution, called the smearing distribution, which
is a probability distribution on the set of possible responses.\\

\noindent C2. If a stimulus is sampled on a trial, the mode of its
smearing distribution becomes, with probability $\theta$, the point
(if any) which is reinforced on that trial; with probability $1-\theta$
the mode remains unchanged.\\

\noindent C3. If no reinforcement occurs on a trial, there is no change
in the smearing distribution of the sampled stimulus.\\

\noindent C4. Stimuli which are not sampled on a given trial do not
change their smearing distributions on that trial. \\

\noindent C5. The probability $\theta$ that the mode of the smearing
distribution of a sampled stimulus will become the point of the reinforced
response is independent of the trial number and the preceding pattern
of occurrence of events. \\

\noindent \textbf{Sampling Axioms.} \\

\noindent S1. Exactly one stimulus is sampled on each trial.\\

\noindent S2. Given the set of stimuli available for sampling on a
given trial, the probability of sampling a given element is independent
of the trial number and the preceding pattern of occurrence of events.
\\

\noindent \textbf{Response Axioms.} \\

\noindent R1. The probability of the response on a trial is solely
determined by the smearing distribution of the sampled stimulus.\\

These axioms are meant to make explicit the conceptual framework of
the sequence of events postulated to occur on each trial, as already
described.

\subsection{Stochastic Model for a Continuum of Responses}

Four random variables characterize the stochastic model of the continuum-of-responses
experiment: $\Omega$ is the probability space, $P$ is its probability
measure, $S$ is the set of stimuli (really patterns of stimuli),
$R$ is the set of possible responses, $E$ is the set of reinforcements,
and \textbf{$\mathbf{S}$}, \textbf{$\mathbf{X}$}, and $\mathbf{Y}$,
are the corresponding random variables, printed in boldface. The other
random variable is the conditioning random variable $\mathbf{Z}$,
which carries the essential information in a single parameter $z_{n}$.
It is assumed that on each trial the conditioning of a stimulus $s$
in $S$ is a probability distribution $K_{s}\left(r|z\right),$ which
is a distribution on possible responses $x$ in $R$. So, in the experiments
analyzed here, the set of responses $R$ is just the set of real numbers
in the periodic interval $\left[0,\,2\pi\right]$. It is assumed that
this distribution $K_{s}\left(r|z\right)$ has a constant variance
on all trials and is defined by its mode $z$ and variance. The mode
changes from trial to trial, following the pattern of reinforcements.
This qualitative formulation is made precise in what follows. The
essential point is this. Since only the mode is changing, we can represent
the conditioning on each trial by the random variable $\mathbf{Z}$.
So $\mathbf{Z}_{s,n}=z_{s,n}$ says what the mode of the conditioning
distribution $K_{s}\left(r|z_{n}\right)$ for stimulus $s$ is on
trial $n$. We subscript the notation $z_{n}$ with $s$ as well,
i.e., $z_{s,\, n}$ when needed. Formally, on each trial $n$, $z_{n}$
is the vector of modes $\left(z_{1},\cdots,\, z_{m}\right)$ for the
$m$ stimuli in $S$. So $\mathbf{\Omega}=\left(\Omega,\, P,\, S,\, R,\, E\right)$
is the basic structure of the model.

Although experimental data will be described later in this paper,
it will perhaps help to give a schematic account of the apparatus
which has been used to test the SR theory extensively in the case
of a continuum of responses. A participant is seated facing a large
circular vertical disc. He is told that his task on each trial is
to predict by means of a pointer where a spot of light will appear
on the rim of the disc. The participant\textquoteright{}s pointer
predictions are his responses in the sense of the theory. At the end
of each trial the \textquotedblleft{}correct\textquotedblright{} position
of the spot is shown to the participant, which is the reinforcing
event for that trial. 

The most important variable controlled by the experimenter is the
choice of a particular reinforcement probability distribution. Here
this is the noncontingent reinforcement distribution $F\left(y\right)$,
with density $f\left(y\right),$ on the set $R$ of possible responses,
and ``noncontingent'' means that the distribution of reinforcements
is not contingent on the actual responses of the participant.

There are four basic assumptions defining the SR stochastic model
of this experiment.\\

\noindent A1. If the set $S$ of stimuli has $m$ elements, then $P\left(\mathbf{S}_{n}=s|s\,\epsilon\, S\right)=\dfrac{1}{m}$.
\\

\noindent A2. If $a_{2}-a_{1}\leq2\pi$ then $P\left(a_{1}\leq\mathbf{X}_{n}\leq a_{2}|\mathbf{S}_{n}=s,\,\mathbf{Z}_{s,\, n}=z\right)=\int_{a_{1}}^{a_{2}}k_{s}\left(x|y\right)dx.$
\\

\noindent A3. \textbf{\textit{(i)}} $P\left(\mathbf{Z}_{s,n+1}=y\,|\,\mathbf{S}_{n}=s,\,\mathbf{Y}_{n}=y\,\&\,\mathbf{Z}_{s,n}=z_{s,n}\right)=\theta$ 
\begin{description}
\item [{\textmd{and}}]~

\begin{description}
\item [{\textit{(ii)}}] $P\left(\mathbf{Z}_{s,n+1}=z_{s,n}\,|\,\mathbf{S}_{n}=s,\,\mathbf{Y}_{n}=y\,\&\,\mathbf{Z}_{s,n}=z_{s,n}\right)=\left(1-\theta\right).$
\end{description}
\end{description}
A4. The temporal sequence in a trial is: 
\begin{equation}
\mathbf{Z}_{n}\rightarrow\mathbf{S}_{n}\rightarrow\mathbf{X}_{n}\rightarrow\mathbf{Y}_{n}\rightarrow\mathbf{Z}_{n+1}.\label{eq:trial-1}
\end{equation}

Assumption A1 defines the sampling distribution, which is left open
in Axioms S1 and S2. Assumptions A2\textbf{ }and A3 complement Axioms
C1\textbf{ }and C2. The remaining conditioning axioms, C3, C4, and
C5, have the general form given earlier. The same is true of the two
sampling axioms. The two axioms C5 and S2 are just \textit{independence-of-path}
assumptions. These axioms are crucial in proving that for simple reinforcement
schedules the sequence of random variables which take as values the
modes of the smearing distributions of the sampled stimuli constitutes
a continuous-state discrete-trial Markov process. 

For example, using $J_{n}$ for the joint distribution of any finite
sequence of these random variables and $j_{n}$ for the corresponding
density, 

\[
P\left(a_{1}\leqq\mathbf{X}_{n}\leqq a_{2}|\mathbf{S}_{n}=s,\mathbf{\, Z}_{s,n}=z\right)=\int_{a_{1}}^{a_{2}}j_{n}\left(x|s,\, z\right)dx=K_{s}\left(a_{2};\, z\right)-K_{s}\left(a_{1};\, z\right).
\]
The following obvious relations for the response density $r_{n}$
of the distribution $R_{n}$ will also be helpful later. First, we
have that 

\[
r_{n}\left(x\right)=j_{n}\left(x\right),
\]
i.e., $r_{n}$ is just the marginal density obtained from the joint
distribution $j_{n}$. Second, we have \textquotedblleft{}expansions\textquotedblright{}
like 

\[
r_{n}\left(x\right)=\int_{0}^{2\pi}j_{n}\left(x,\, z_{s,n}\right)dz_{s,n},
\]

and

\[
r_{n}\left(x\right)=\int_{0}^{2\pi}\int_{0}^{2\pi}\int_{0}^{2\pi}j_{n}\left(x,\, z_{s,n},\, y_{n-1}x_{n-1}\right)dz_{s,n}dy_{n-1}dx_{n-1}.
\]

\subsection{Noncontigent Reinforcement}

Noncontingent reinforcement schedules are those for which the distribution
is independent of $n$, the responses, and the past. We first use
the response density recursion for some simple, useful results which
do not explicitly involve the smearing distribution of the single
stimulus. There is, however, one necessary preliminary concerning
derivation of the asymptotic response distribution in the stimulus-response
theory. 

\begin{theorem}In the noncontingent case, if the set of stimuli has
$m$ elements, 
\begin{equation}
r\left(x\right)=\lim_{n\rightarrow\infty}r_{n}\left(x\right)=\frac{1}{m}\sum_{s\epsilon S}\int_{0}^{2\pi}k_{s}\left(x;\, y\right)f\left(y\right)dy.\label{eq:theoremnoncontingentcase}
\end{equation}

\end{theorem}

We now use (\ref{eq:theoremnoncontingentcase}) to establish the following
recursions. In the statement of the theorem $E(\mathbf{X}_{n})$ is
the expectation of the response random variable $\mathbf{X}_{n}$,
$\mu_{r}\left(\mathbf{X}_{n}\right)$ is its $r$-th raw moment, $\sigma^{2}\left(\mathbf{X}_{n}\right)$
is its variance, and $\mathbf{\mathbf{X}}$ is the random variable
with response density $r$. 

\begin{theorem} 
\begin{equation}
r_{n+1}\left(x\right)=\left(1-\theta\right)r_{n}\left(x\right)+\theta r\left(x\right),\label{eq:theoremE(Xn)1}
\end{equation}
\begin{equation}
E\left(\mathbf{X}_{n+1}\right)=\left(1-\theta\right)E\left(\mathbf{X}_{n}\right)+\theta E\left(\mathbf{X}\right),\label{eq:theoremE(Xn)2}
\end{equation}
\begin{equation}
\mu_{r}\left(\mathbf{X}_{n+1}\right)=\left(1-\theta\right)\mu_{r}\left(\mathbf{X}_{n}\right)+\theta\mu_{r}\left(\mathbf{X}\right),\label{eq:theoremE(Xn)3}
\end{equation}
\begin{equation}
\sigma^{2}\left(\mathbf{X}_{n+1}\right)=\left(1-\theta\right)\sigma^{2}\left(\mathbf{X}_{n}\right)+\theta\sigma^{2}\left(\mathbf{X}\right)+\theta\left(1-\theta\right)\left[E\left(\mathbf{X}_{n}\right)-E\left(\mathbf{X}\right)\right]^{2}\label{eq:theoremE(Xn)4}
\end{equation}
\end{theorem}Because (\ref{eq:theoremE(Xn)1})-(\ref{eq:theoremE(Xn)3})
are first-order difference equations with constant coefficients we
have as an immediate consequence of the theorem: 

\begin{corollary}

\begin{equation}
r_{n}\left(x\right)=r\left(x\right)-\left[r\left(x\right)-r_{1}\left(x\right)\right]\left(1-\theta\right)^{n-1},\label{eq:corollary1}
\end{equation}
\begin{equation}
E\left(\mathbf{X}_{n}\right)=E\left(\mathbf{X}\right)-\left[E\left(\mathbf{X}\right)-E\left(\mathbf{X}_{1}\right)\right]\left(1-\theta\right)^{n-1},\label{eq:corollary2}
\end{equation}
\begin{equation}
\mu_{r}\left(\mathbf{X}_{n}\right)=\mu_{r}\left(\mathbf{X}\right)-\left[\mu_{r}\left(\mathbf{X}\right)-\mu_{r}\left(\mathbf{X}_{1}\right)\right]\left(1-\theta\right)^{n-1}\label{eq:corollary3}
\end{equation}

\end{corollary}

Although the one-stimulus model and the $N$-stimulus model both yield
(\ref{eq:theoremE(Xn)1})--(\ref{eq:corollary3}), predictions of
the two models are already different for one of the simplest sequential
statistics, namely, the probability of two successive responses in
the same or different subintervals. We have the following two theorems
for the one-stimulus model. The result generalizes directly to any
finite number of subintervals. 

\begin{theorem}For noncontingent reinforcement 

\begin{multline}
\lim_{n\rightarrow\infty}P\left(0\leqq\mathbf{X}_{n+1}\leqq c,\,0\leqq\mathbf{X}_{n}\leqq c\right)\\
=\theta R\left(c\right)^{2}+\left(1-\theta\right)\frac{1}{m^{2}}\sum_{s^{\prime}\epsilon S}\sum_{s\epsilon S}\int_{0}^{c}\int_{0}^{c}\int_{0}^{2\pi}k_{s}\left(x;\, z\right)k_{s^{\prime}}\left(x^{\prime};\, z\right)f\left(z\right)dx\, dx^{\prime}dz,
\end{multline}

and
\begin{multline}
\lim_{n\rightarrow\infty}P\left(0\leqq\mathbf{X}_{n+1}\leqq c,\, c\leqq\mathbf{X}_{n}\leqq2\pi\right)\\
=\theta R\left(c\right)\left[1-R\left(c\right)\right]+\left(1-\theta\right)\frac{1}{m^{2}}\sum_{s^{\prime}\epsilon S}\sum_{s\epsilon S}\int_{0}^{c}\int_{c}^{2\pi}\int_{0}^{2\pi}k_{s}\left(x;\, z\right)k_{s^{\prime}}\left(x^{\prime};\, z\right)f\left(z\right)dx\, dx^{\prime}dz,\label{eq:theoremnon-contingent2}
\end{multline}

where

\[
R\left(c\right)=\lim_{n\rightarrow\infty}R_{n}\left(c\right).
\]

\end{theorem}

We conclude the treatment of noncontingent reinforcement with two
expressions dealing with important sequential properties. The first
gives the probability of a response in the interval $\left[a_{1},\, a_{2}\right]$
given that on the previous trial the reinforcing event occurred in
the interval $\left[b_{1},\, b_{2}\right]$.

\noindent \begin{theorem}

\begin{multline}
P\left(a_{1}\leqq\mathbf{X}_{n+1}\leqq a_{2}|b_{1}\leqq Y_{n}\leqq b_{2},\, a_{3}\leqq\mathbf{X}_{n}\leqq a_{4}\right)\\
=\left(1-\theta\right)\left[R_{n}\left(a_{2}\right)-R_{n}\left(a_{1}\right)\right]+\dfrac{\theta}{F\left(b_{2}\right)-F\left(b_{1}\right)}\frac{1}{m}\sum_{s\epsilon S}\int_{a_{1}}^{a_{2}}\int_{b_{1}}^{b_{2}}k_{s}\left(x;\, y\right)f\left(y\right)dx\, dy.\label{eq:probabilityofaresponse}
\end{multline}
\end{theorem}

The expression to which we now turn gives the probability of a response
in the interval $\left[a_{1},\, a_{2}\right]$ given that on the previous
trial the reinforcing event occurred in the interval $\left[b_{1},\, b_{2}\right]$
and the response in the interval $\left[a_{3},\, a_{4}\right]$.

\noindent \begin{theorem}
\begin{multline}
P\left(a_{1}\leqq\mathbf{X}_{n+1}\leqq a_{2\,}|\, b_{1}\leqq Y_{n}\leqq b_{2},\, a_{3}\leqq\mathbf{X}_{n}\leqq a_{4}\right)\\
=\dfrac{\left(1-\theta\right)}{R_{n}\left(a_{4}\right)-R_{n}\left(a_{3}\right)}\frac{1}{m^{2}}\sum_{s^{\prime}\epsilon S}\sum_{s\epsilon S}\int_{0}^{2\pi}\int_{a_{1}}^{a_{2}}\int_{a_{3}}^{a_{4}}\int_{a}^{b}k_{s}\left(x;\, z\right)k_{s^{\prime}}\left(x^{\prime};\, z\right)g_{n}\left(z\right)dx\, dx^{\prime}dz.\\
+\dfrac{\theta}{F\left(b_{2}\right)-F\left(b_{1}\right)}\frac{1}{m}\sum_{s\epsilon S}\int_{a_{1}}^{a_{2}}\int_{b_{1}}^{b_{2}}k_{s}\left(x;\, z\right)f\left(y\right)dx\, dy.
\end{multline}
\end{theorem}

\subsection{More General Comments on SR theory\label{sub:General-Comments-on-SR}}

The first general comment concerns the sequence of events occurring
on each trial, represented earlier by equation (\ref{eq:trial-1})
in Assumption A4 of the stochastic model. We want to examine in a
preliminary way when brain computations, and therefore neural oscillators,
are required in this temporal sequence.
\begin{description}
\item [{\textmd{\textit{(i)}}}] $\mathbf{Z}_{n}$ sums up previous conditioning
and does not represent a computation on trial $n$; 
\item [{\textmd{\textit{(ii)}}}] $\mathbf{S}_{n}$, which represents the
experimentally unobserved sampling of stimuli, really patterns of
stimuli, uses an assumption about the number of stimuli, or patterns
of them, being sampled in an experiment; the uniform distribution
assumed for this sampling is artificially simple, but computationally
useful; in any case, no brain computations are modeled by oscillators
here, even though a complete theory would be required; 
\item [{\textmd{\textit{(iii)}}}] $\mathbf{X}_{n}$ represents the first
brain computation in the temporal sequence on a trial for the stochastic
model; this computation selects the actual response on the trial from
the conditioning distribution $k_{s}\left(x|z_{s,\, n}\right)$, where
$s$ is the sampled stimulus on this trial $\left(\mathbf{S}_{n}=s\right);$
this is one of the two key oscillator computations developed in the
next section;
\item [{\textmd{\textit{(iv)}}}] $\mathbf{Y}_{n}$ is the reinforcement
random variable whose distribution is part of the experimental design;
individual reinforcements are external events totally controlled by
the experimenter, and as such require no brain computation by the
participant; 
\item [{\textmd{\textit{(v)}}}] $\mathbf{Z}_{n+1}$ summarizes the assumed
brain computations that often change at the end of a trial the state
of conditioning of the stimulus $s$ sampled on trial $n$; in our
stochastic model, this change in conditioning is represented by a
change in the mode $z_{s,n}$ of the distribution $K_{s}\left(x|z_{s,\, n}\right)$.
From the assumptions A1-4 and the general independence-of-path axioms,
we can prove that the sequence of random variable $Z_{1},\ldots,\, Z_{n},\ldots$
is a first-order Markov process \citep{suppes1960stimulus}.
\end{description}
The second general topic is of a different sort. It concerns the important
concept of \textit{activation}, which is a crucial but implicit aspect
of SR theory. Intuitively, when a stimulus is sampled, an image of
that stimulus is activated in the brain, and when a stimulus is not
sampled on a trial, its conditioning does not change, which implies
a constant state of low or no activity for the brain image of that
unsampled stimulus.

Why is this concept of activation important for the oscillator or
other physical representations of how SR theory is realized in the
brain? Perhaps the most fundamental reason arises from our general
conception of neural networks, or, to put it more generally, ``purposive''
networks. For large networks, i.e., ones with many nodes, it is almost
always unrealistic to have all nodes active all of the time. For biological
systems such an inactive state with low consumption of energy is necessary
in many situations. The brain certainly seems to be a salient example.
Without this operational distinction, imagine a human brain in which
episodic memories of many past years were as salient and active as
incoming perception images. This critical distinction in large-scale
networks between active and inactive nodes is often ignored in the
large literature on artificial neural networks, but its biological
necessity has long been recognized, and theoretically emphasized in
psychology since the 1920s. In this paper, we accept the importance
of activation, but do not attempt an oscillator representation of
the physical mechanism of activating brain images arising from perceived
stimuli. This also applies to the closely related concept of \textit{spreading}
activation, which refers to brain images perceptually activated associatively
by other brain images, which are not themselves directly activated
by perception.

\section{Model overview}

The neural oscillator model we developed is significantly more complex,
both mathematically and conceptually, than SR theory. Furthermore,
it requires the use of some physical concepts that are probably unfamiliar
to some readers of this journal. So, before we present the mathematical
details of the model in Section \ref{sec:Oscillator-Model-SR}, we
introduce here the main ideas behind the oscillator model in a conceptual
way, and show each step of the computation and how it relates to SR
theory. Schematically, the oscillator model corresponds to the following
steps, described in more detail later in this section.
\begin{enumerate}
\item Trial $n$ starts with a series of stimulus and response neural oscillators
connected through excitatory and inhibitory couplings. 
\item A stimulus oscillator is activated, and by spreading activation the
response oscillators.
\item Activation leads to new initial conditions for the stimulus and response
oscillators; we assume such conditions are normally distributed.
\item The stimulus and response neural oscillators evolve according to non-linear
deterministic differential equations.
\item After a response time $\Delta t_{r}$, relative phase relations between
the stimulus and response oscillators determine the response made
on trial $n$. 
\item The reinforcement oscillator is then activated at time $t_{e}$ after
the response has occured. 
\item Activation of the reinforcement oscillator leads to other new ``initial''
conditions at $t_{e}$; such conditions we again assume are normally
distributed. 
\item The stimulus, response, and reinforcement neural oscillators, as well
as their couplings, evolve according to nonlinear deterministic differential
equations.
\item After a time interval $\Delta t_{e}$, reinforcement is completed,
and the excitatory and inhibitory couplings may have changed. 
\end{enumerate}
Step $1$ corresponds to the random variable $\mathbf{Z}_{n}$ in
equation (\ref{eq:trial-1}), Step $2$ to $\mathbf{S}_{n}$, Steps
3\textendash{}5 to $\mathbf{X}_{n}$, and Steps 6\textendash{}9 to
$\mathbf{Y}_{n}$ and \textbf{$\mathbf{Z}_{n+1}$}. Let us now look
at each process in more detail, from a neural point of view.

\paragraph{Sampling}

We start the description of our model by the sampling of a stimulus,
$\mathbf{S}_{n}$. For each element $s$ of the set of stimuli $S$,
we assume the existence of a corresponding neural phase oscillator,
$\varphi_{s}$. The sampling of a specific $s$ thus activates the
neural oscillator $\varphi_{s}$. As mentioned before, we do not present
a detailed theory of activation or spreading activation, but simply
assume that for each trial a $\varphi_{s}$ is activated according
to a uniform distribution, consistent with $\mathbf{S}_{n}$. Once
an oscillator $\varphi_{s}$ is activated, this activation spreads
to those oscillators coupled to it, including the response oscillators
$\varphi_{r_{1}}$ and $\varphi_{r_{2}}$.

\paragraph{Response}

In our model, we simplify the dynamics by considering only two response
oscillators, $\varphi_{r_{1}}$ and $\varphi_{r_{2}}$. We should
emphasize that, even though we are talking about two response oscillators,
we are not thinking of them as modeling two responses, but a continuum
of responses. Intuitively, we think of $\varphi_{r_{1}}$ as corresponding
to a certain extreme value in a range of responses, for example $1$,
and $\varphi_{r_{2}}$ to another value, say $-1$.%
\footnote{In reality, responses in our model have the same topological structure
as the unit circle. See Section \ref{sec:Oscillator-Model-SR} for
details. %
} 

Couplings between the stimulus and response oscillators can be of
two types: excitatory or inhibitory. Excitatory couplings have the
effect of promoting the synchronization of two oscillators in a way
that brings their phases together. Inhibitory couplings also promote
synchronization, but in such a way that the relative phases are off
by $\pi$. If oscillators are not at all coupled, then their dynamical
evolution is dictated by their natural frequencies, and no synchronization
appears. Figure 
\begin{figure}
\begin{centering}
(a)\includegraphics[width=7cm]{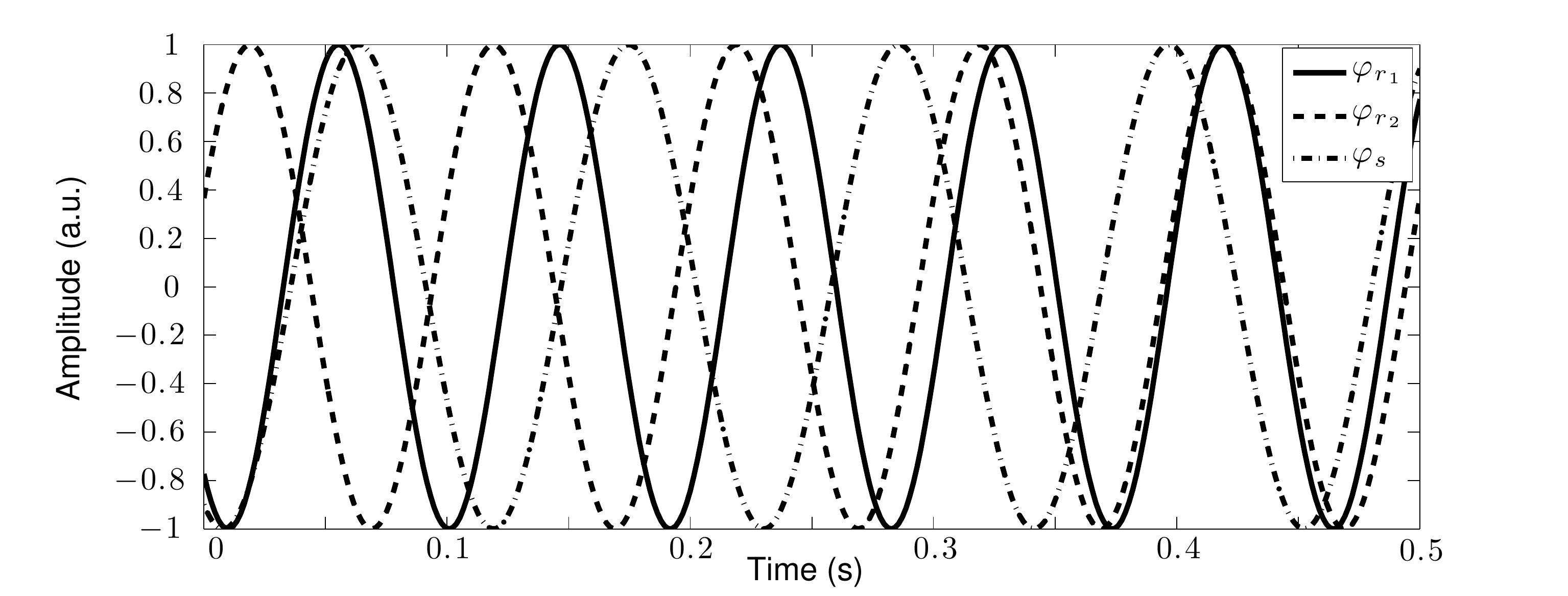}
\par\end{centering}

\begin{centering}
(b)\includegraphics[width=7cm]{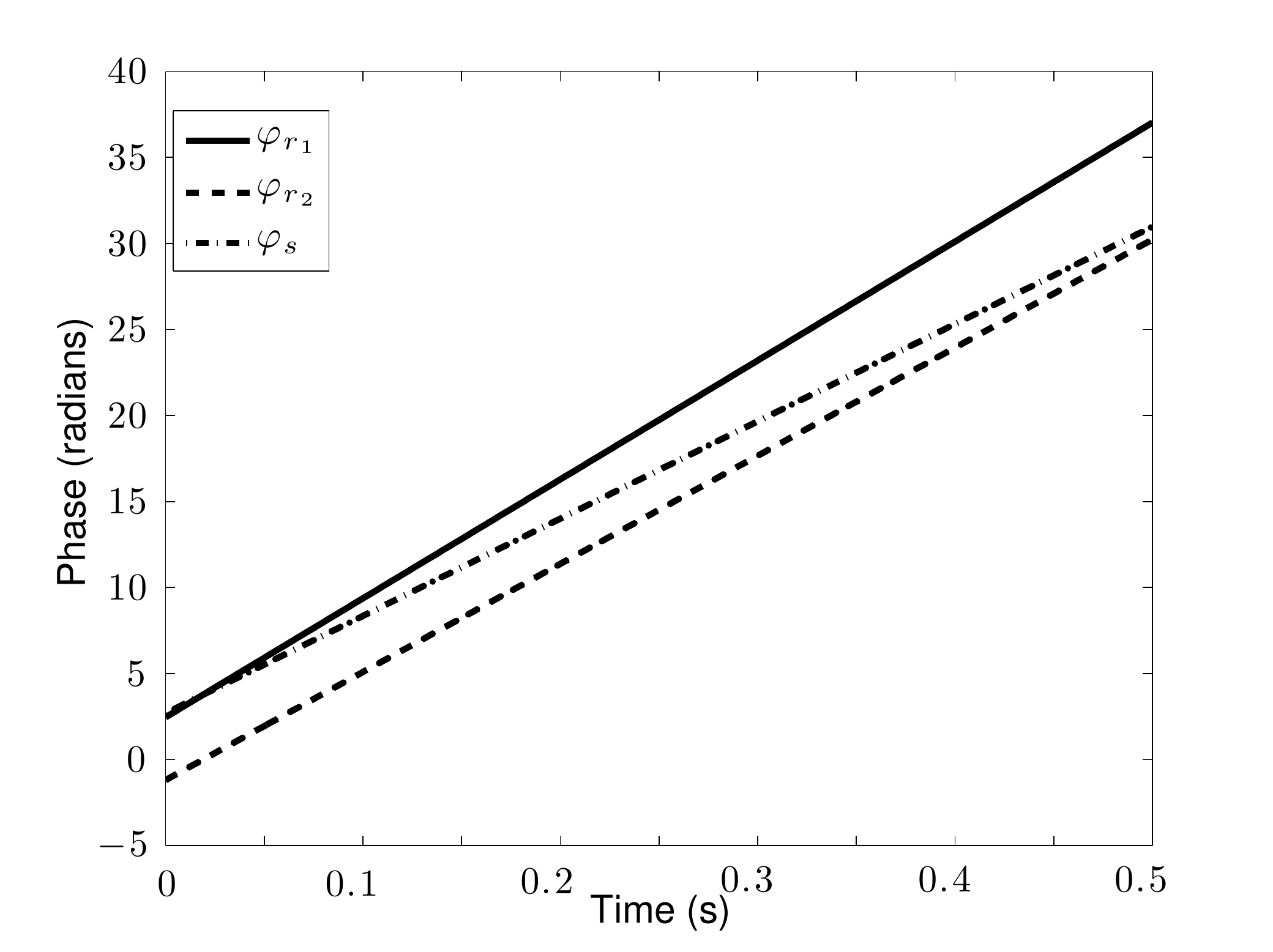}
\par\end{centering}

\caption{\label{fig:Time-evolution-uncoupled}Time evolution of three oscillators
represented by $A(t)=\cos\left(\varphi(t)\right)$, where $\varphi(t)$
is the phase. Graph (a) shows the oscillations for uncoupled oscillators
with frequencies $11$ Hz (solid line), $10$ Hz (dashed), and $9$
Hz (dash-dotted). Notice that though the oscillators start roughly
in phase, they slowly dephase as time progresses. Graph (b) shows
the phases $\varphi(t)$ of the same three oscillators in (a). }

\end{figure}
 \ref{fig:Time-evolution-uncoupled} shows the evolution for three
uncoupled oscillators. However, if a system starts with oscillators
randomly close to each other in phase, when oscillators are coupled,
some may evolve to be in-phase or out-of-phase. 

\begin{figure}
\begin{centering}
(a)\includegraphics[width=7cm]{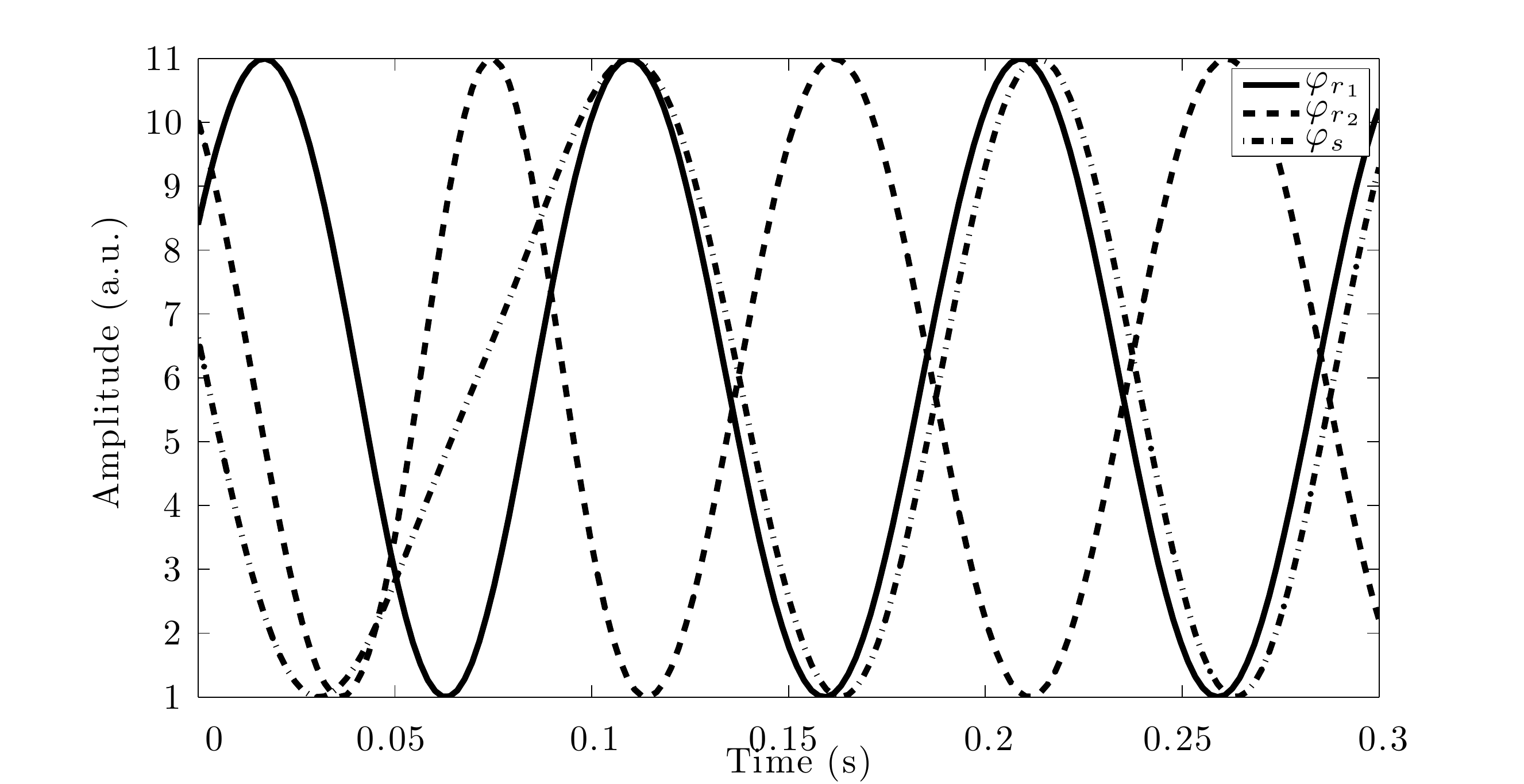}
\par\end{centering}

\begin{centering}
(b)\emph{\includegraphics[width=7cm]{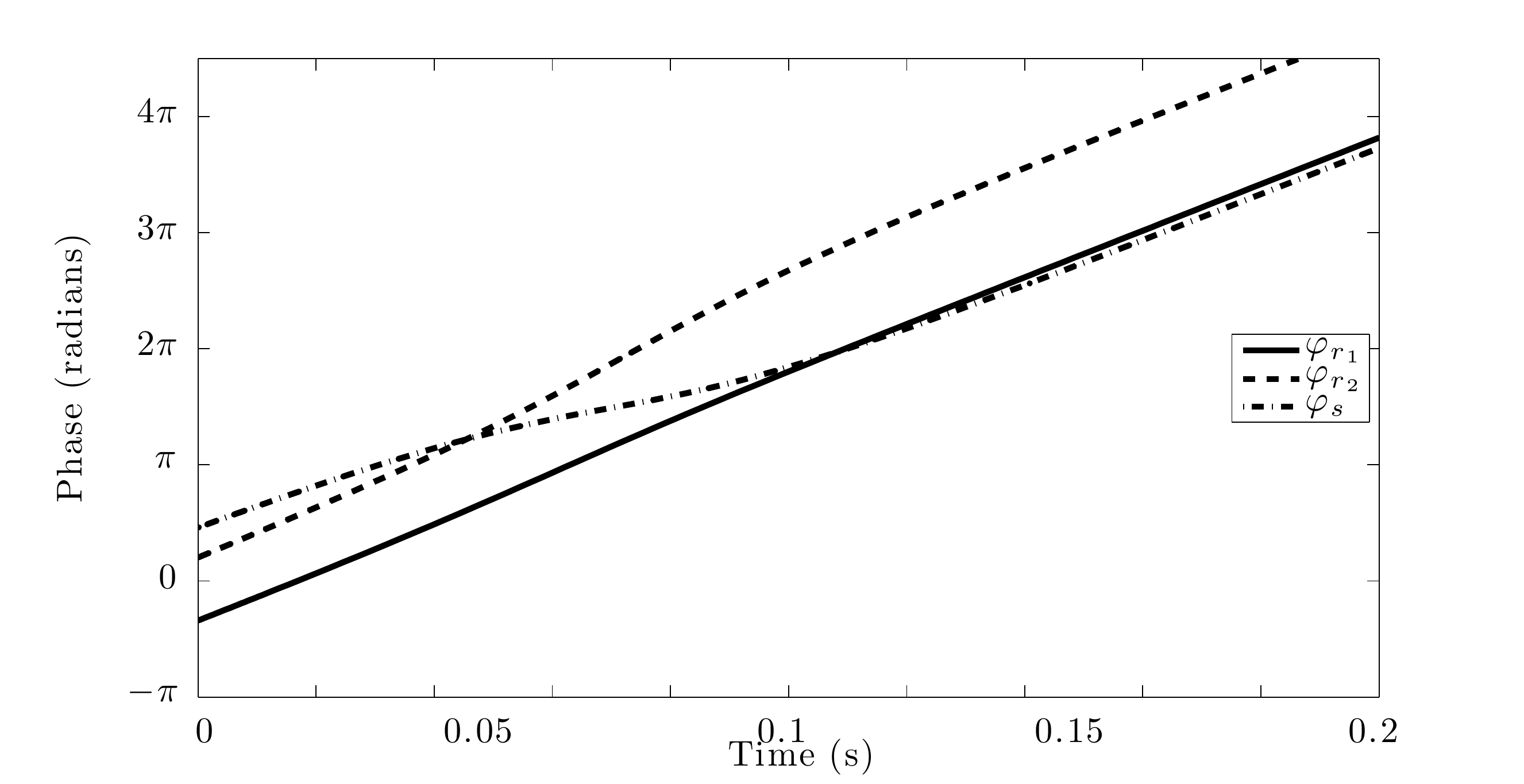}}
\par\end{centering}

\begin{centering}
(c)\includegraphics[width=7cm]{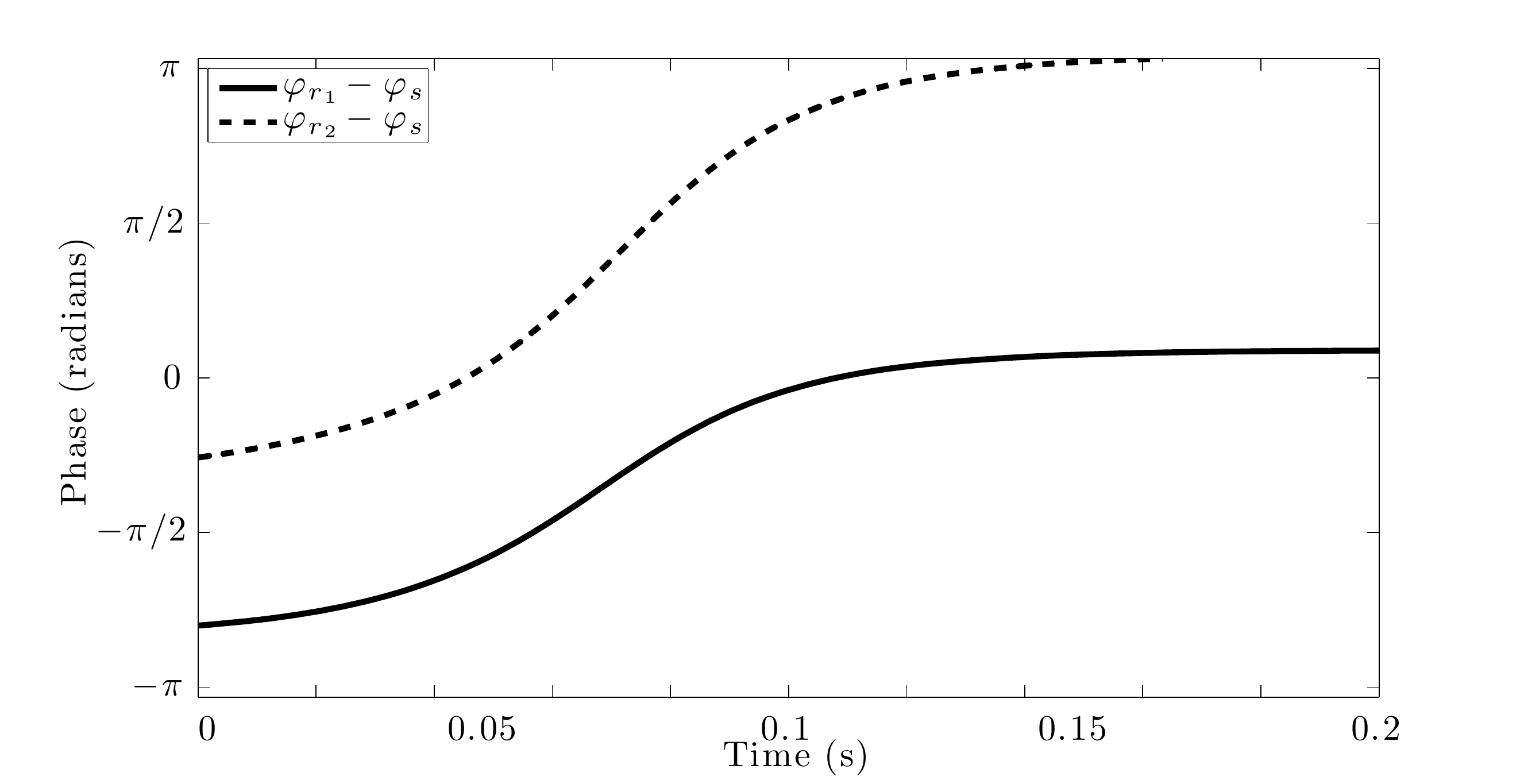}
\par\end{centering}

\begin{centering}
(d)\includegraphics[width=7cm]{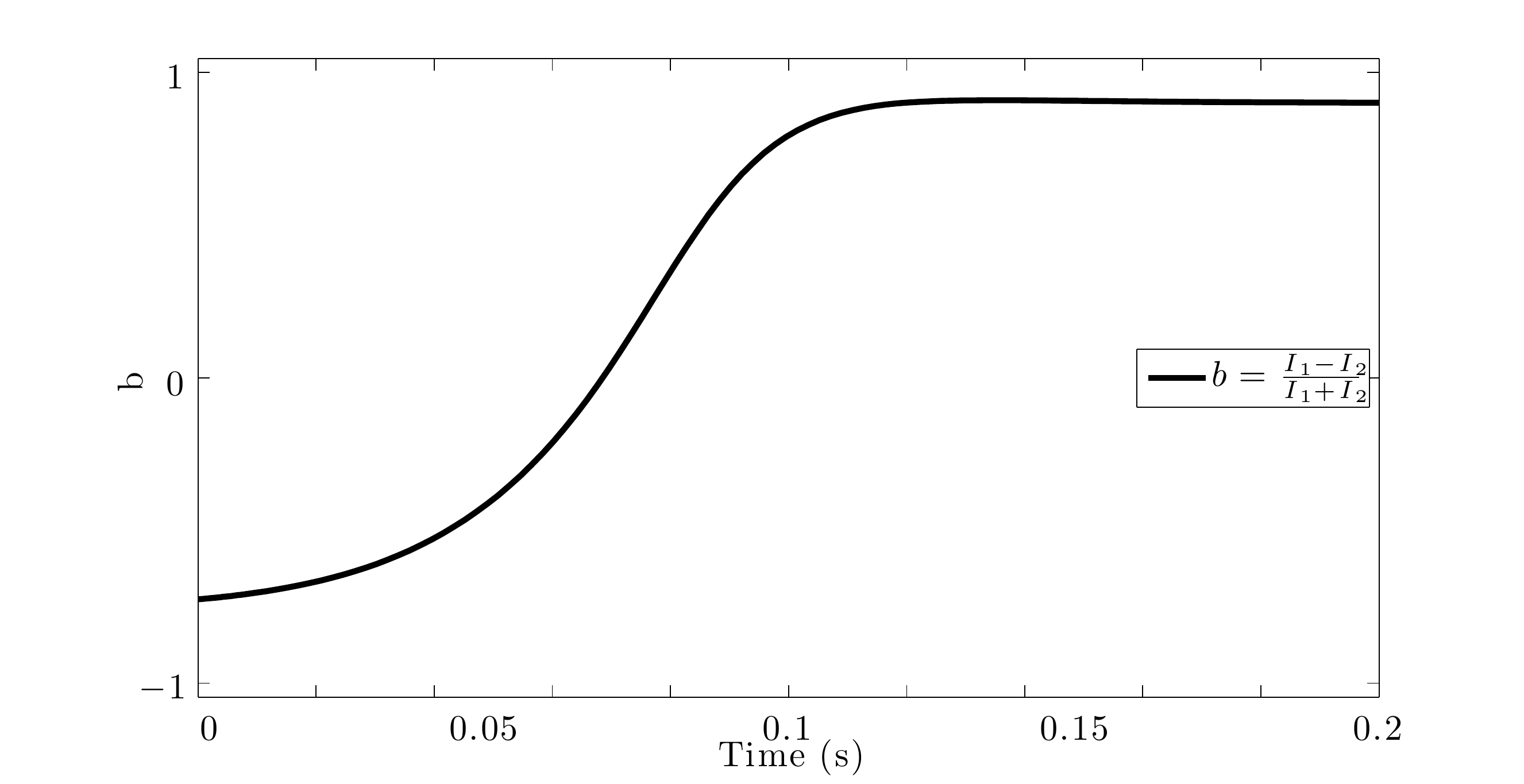}
\par\end{centering}

\caption{\label{fig:Osc-Evol-Response}Same three oscillators as in Figure
\ref{fig:Time-evolution-uncoupled}, but now coupled. We see from
(a) that even though the system starts with no synchrony, as oscillators
have different phases and natural frequencies, very quickly (after
$100$ ms) the stimulus oscillator $s$ (dash-dotted line) becomes
in synch and in phase with response $r_{1}$ (solid line), while at
the same time becoming in synch but off phase with $r_{2}$. This
is also represented in (b) by having one of the oscillators approach
the trajectory of the other. Graph (c) shows the phase differences
between the stimulus oscillators $s$ and $r_{1}$ (solid line) and
$s$ and $r_{2}$ (dashed line). Finally, (d) shows the relative intensity
of the superposition of oscillators, with $1$ corresponding to maximum
intensity at $r_{1}$ and minimum at $r_{2}$ and $-1$ corresponding
to a maximum in $r_{2}$ and minimum in $r_{1}$. As we can see, the
couplings lead to a response that converges to a value close to $1$. }

\end{figure}

To describe how a response is computed, we first discuss the interference
of neural oscillators. At the brain region associated to $\varphi_{r_{1}}$,
we have the oscillatory activities due not only to $\varphi_{r_{1}}$
but also to $\varphi_{s}$, since the stimulus oscillator is coupled
to $\varphi_{r_{1}}$. This sum of oscillations may result in either
constructive or destructive interference. Constructive interference
will lead to stronger oscillations at $\varphi_{r_{1}}$, whereas
destructive interference will lead to weaker oscillations. So, let
us assume that the couplings between $\varphi_{s}$ and the response
oscillators are such that the stimulus oscillator synchronizes in-phase
with $\varphi_{r_{1}}$ and out-of-phase with $\varphi_{r_{2}}$.
Synchronized in-phase oscillators correspond to higher intensity than
out-of-phase oscillators. The dynamics of the system is such that
its evolution leads to a high intensity for the superposition of the
stimulus $\varphi_{s}$ and response oscillator $\varphi_{r_{1}}$
(constructive interference), and a low intensity for the superposition
of $\varphi_{s}$ and $\varphi_{r_{2}}$ (destructive interference).
Thus, in such a case, we say that the response will be closer to the
one associated to $\varphi_{r_{1}}$, i.e., $1$. Figure \ref{fig:Osc-Evol-Response}
 shows the evolution of coupled oscillators, with couplings that lead
to the selection of response $1$. 

From the previous paragraph, it is clear how to obtain answers at
the extremes values of a scale ($-1$ and $1$ in our example). However,
the question remains on how to model a continuum of responses. To
examine this, let us call $I_{1}$ the intensity at $\varphi_{r_{1}}$
due to its superposition with $\varphi_{s}$, and let us call $I_{2}$
the intensity for $\varphi_{r_{2}}$ and $\varphi_{s}$. Response
$1$ is computed from the oscillator model when $I_{1}$ is maximum
and $I_{2}$ is minimum, and $-1$ vice versa. But those are not the
only possibilities, as the phase relations between $\varphi_{s}$
and the response oscillators do not need to be such that when $I_{1}$
is maximum $I_{2}$ is minimum. For example, it is possible to have
$I_{1}=I_{2}$. Since we are parametrizing $1$ ($-1$) to correspond
to a maximum for $I_{1}$ ($I_{2}$) and a minimum for $I_{2}$ ($I_{1}$),
it stands to reason that $I_{1}=I_{2}$ corresponds to $0$. More
generally, by considering the relative intensity between $I_{1}$
and $I_{2}$, defined by 
\begin{equation}
b=\frac{I_{1}-I_{2}}{I_{1}+I_{2}},\label{eq:b}
\end{equation}
which can be rewritten as 
\[
b=\frac{I_{1}/I_{2}-1}{I_{1}/I_{2}+1}\mbox{ if }I_{2}\neq0,
\]
any value between $-1$ and $1$ may be computed by the oscillator
model. For example, when $I_{2}=(1/3)I_{1}$, we see from equation
(\ref{eq:b}) that the computed response would be $1/2$, whereas
$I_{2}=3I_{1}$ would yield $-1/2$. We emphasize, though, that there
is nothing special about $-1$ and $1$, as those were values that
we selected to parametrize the response; we could have selected $\alpha$
and $\beta$ as the range for our continuum of responses by simply
redefining $b$ in Eq. (\ref{eq:b}). 

It can now be seen where the stochastic characteristics of the model
appear. First, Step 2 above corresponds to a random choice of stimulus
oscillator, external to the model. Once this oscillator is chosen,
the initial conditions for oscillators in Step 3 have a stochastic
nature. However, once the initial conditions are selected, the system
evolves following a deterministic (albeit nonlinear) set of differential
equations, as Step 4. Finally, from the deterministic evolution from
stochastic initial conditions, responses are computed in Step 5, and
we see in Figure 
\begin{figure}
\begin{centering}
\includegraphics[width=7cm]{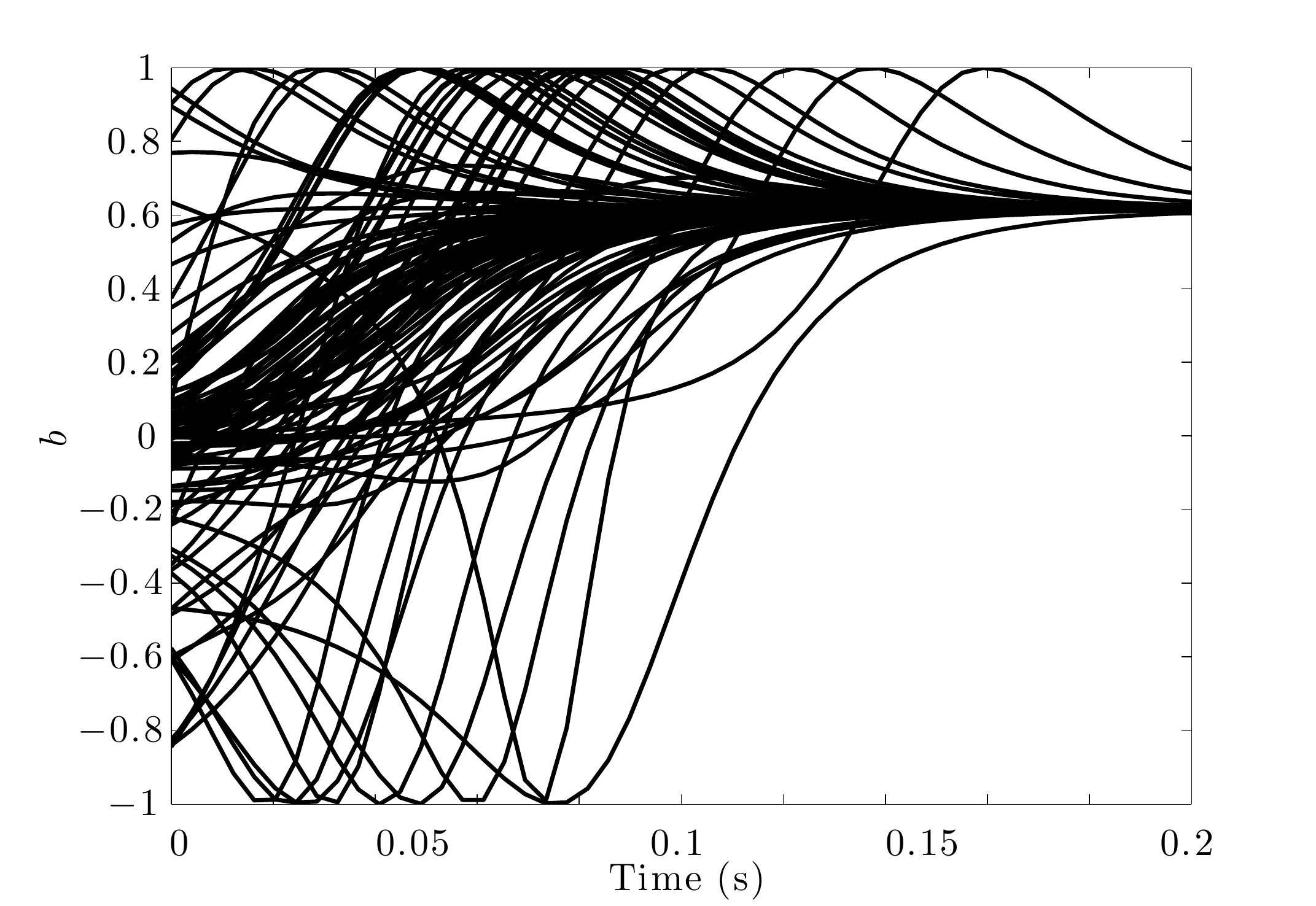}
\par\end{centering}

\caption{\label{fig:Response-computation-ensemble}Response computation trajectories
for 100 randomly-selected initial conditions for an oscillator set
conditioned to the response $1/2$.}

\end{figure}
 \ref{fig:Response-computation-ensemble} an ensemble of trajectories
for the evolution of the relative intensity of a set of oscillators.
It is clear from this figure that there is a smearing distribution
of computed responses at time $0.2$ s. This smearing is smaller than
the empirical value reported in Suppes et al. (1964). But this is
consistent with the fact that the production of a response by the
perceptual and motor system adds additional uncertainty to produce
a higher variance of the behavioral smearing distribution.

\paragraph{Reinforcement}

Reinforcement of $y$ is neurally described by an oscillator $\varphi_{e,y}$
that couples to the response oscillators with phase relations consistent
with the relative intensities of $\varphi_{r_{1}}$ and $\varphi_{r_{2}}$
producing the response $y$. The conditioning state of the oscillator
system is coded in the couplings between oscillators. Once the reinforcement
oscillator is activated, couplings between stimulus and response oscillators
change in a way consistent with the reinforced phase relations.\textbf{
}For example, if the intensity were greater at $\varphi_{r_{2}}$
but we reinforced $\varphi_{r_{1}}$, then the couplings would evolve
to push the system toward a closer synchronization with $\varphi_{r_{1}}$.
Figure \ref{fig:Time-evolution-after} 
\begin{figure}
\begin{centering}
\begin{tabular}{|c|c|}
\hline 
(a)\includegraphics[width=6cm]{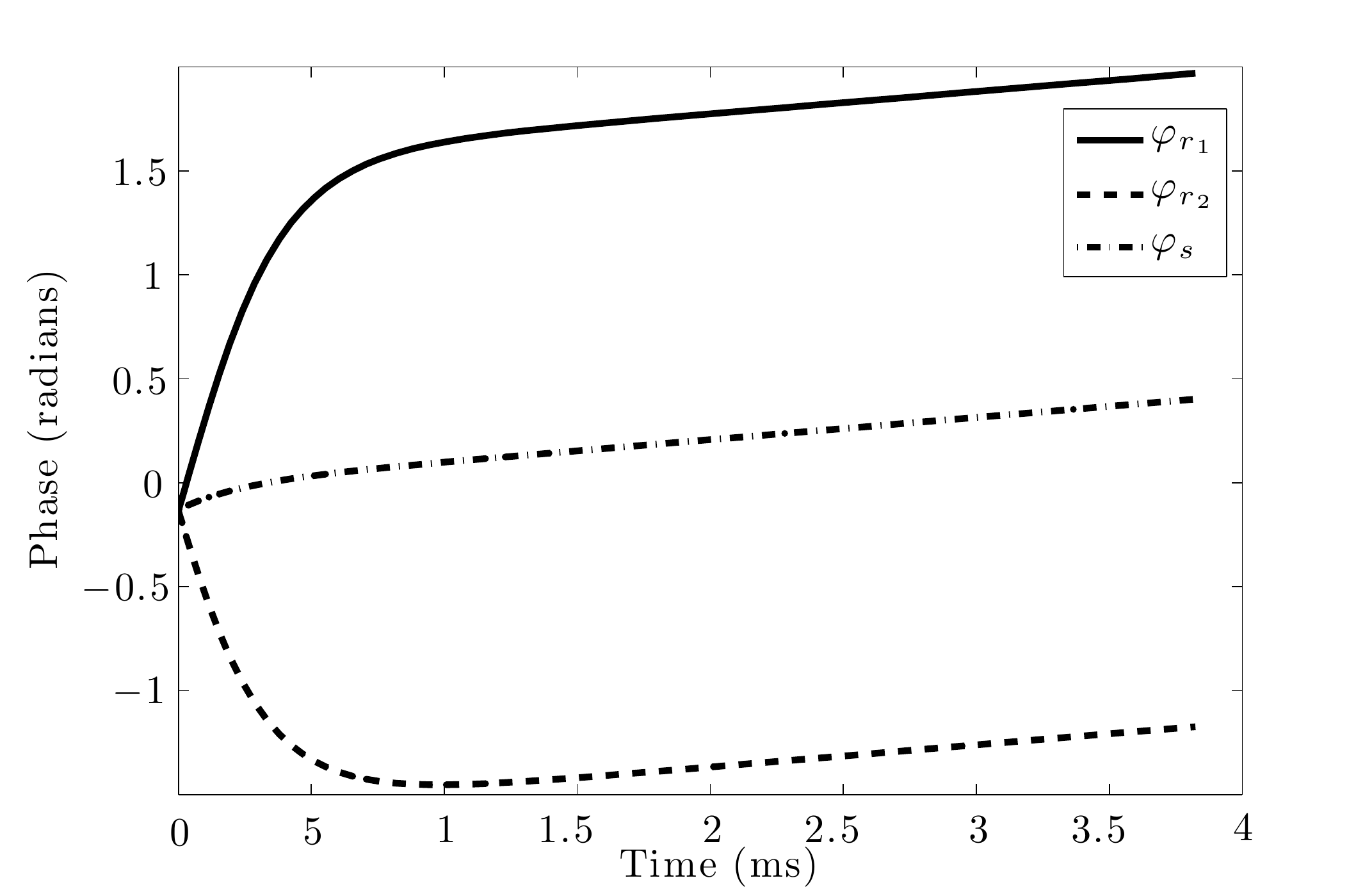} & (b)\includegraphics[width=6cm]{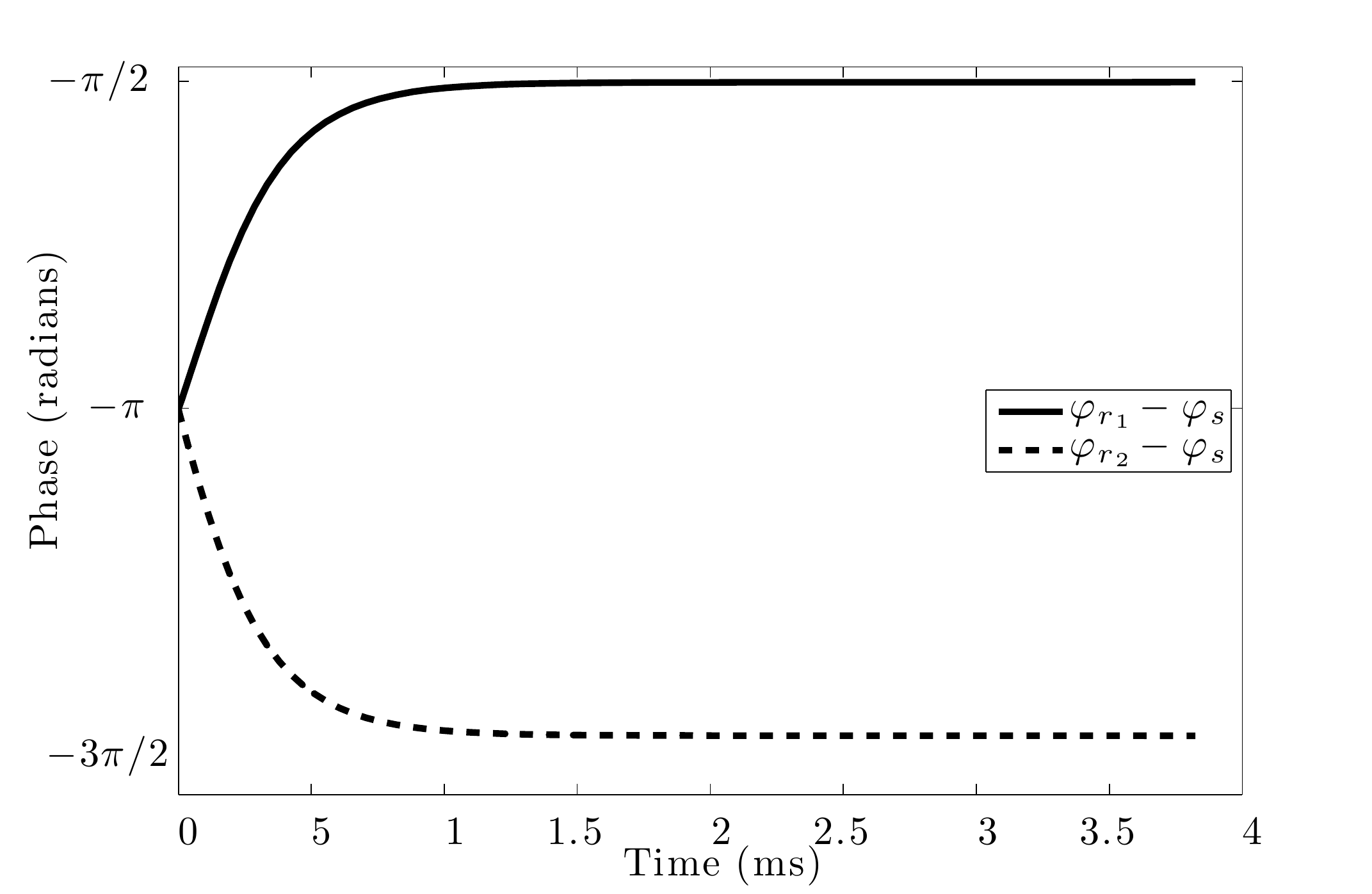}\tabularnewline
\hline 
\hline 
(c)\includegraphics[width=6cm]{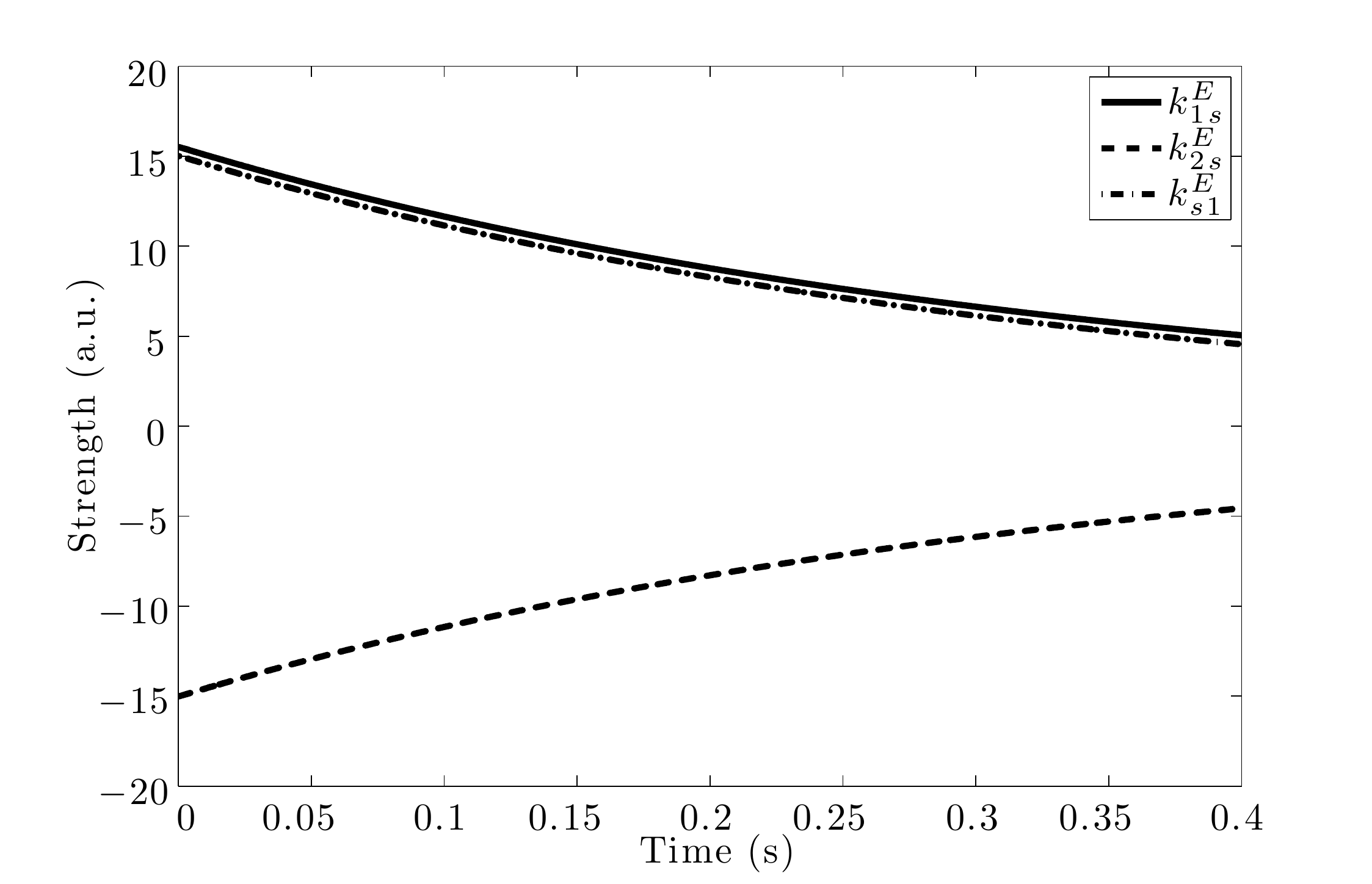} & (d)\includegraphics[width=6cm]{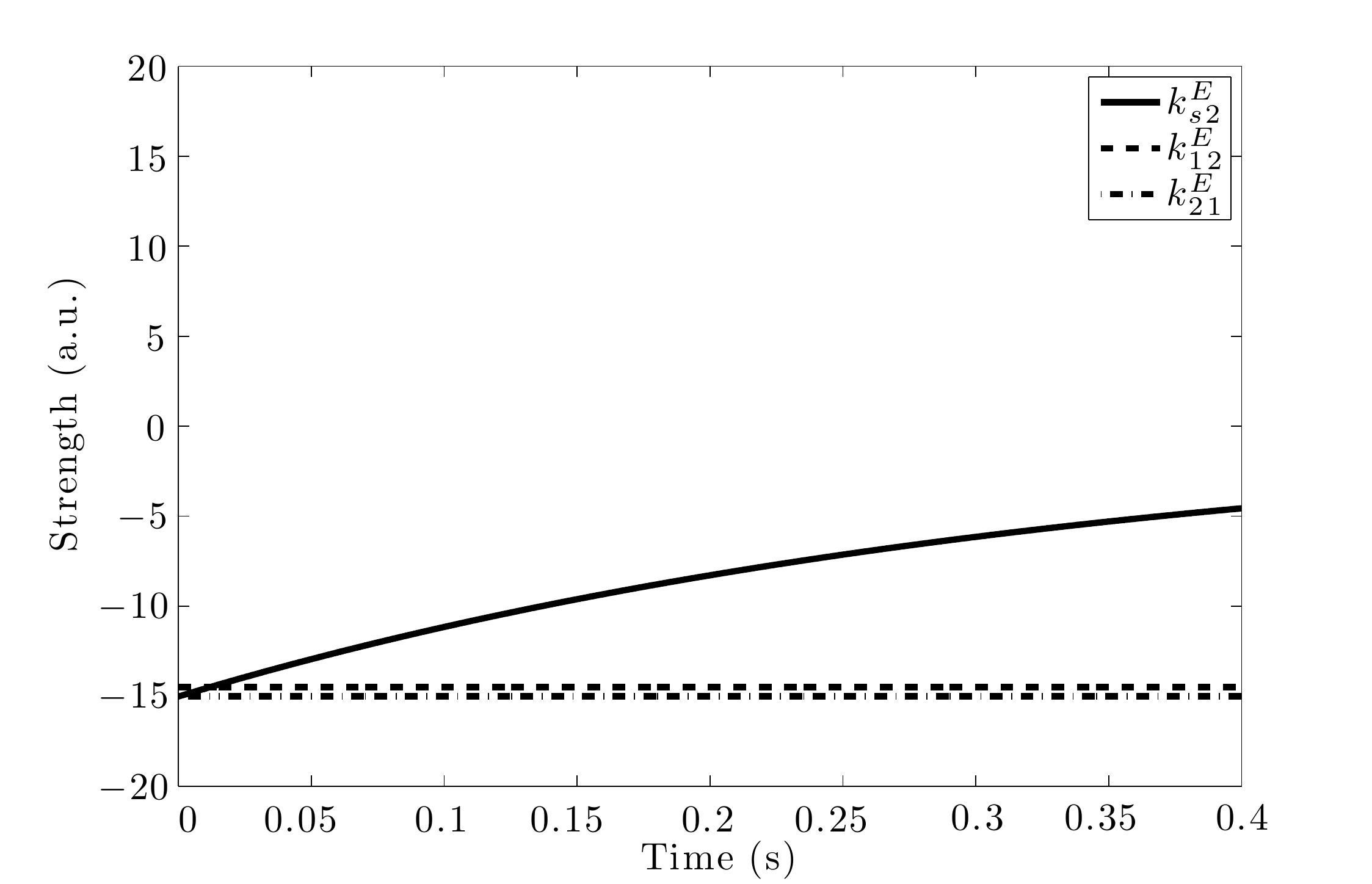}\tabularnewline
\hline 
(e)\includegraphics[width=6cm]{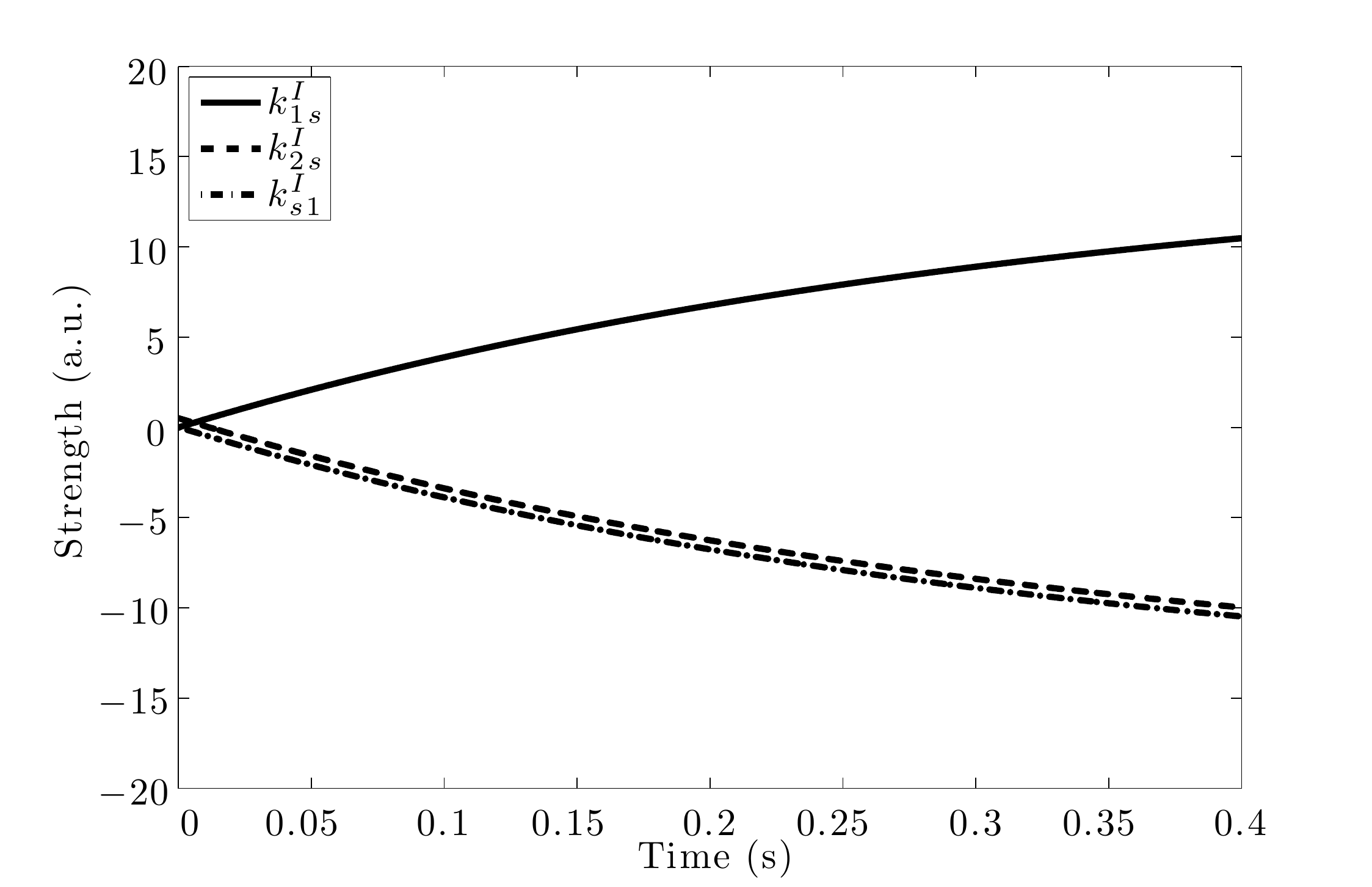} & (f)\includegraphics[width=6cm]{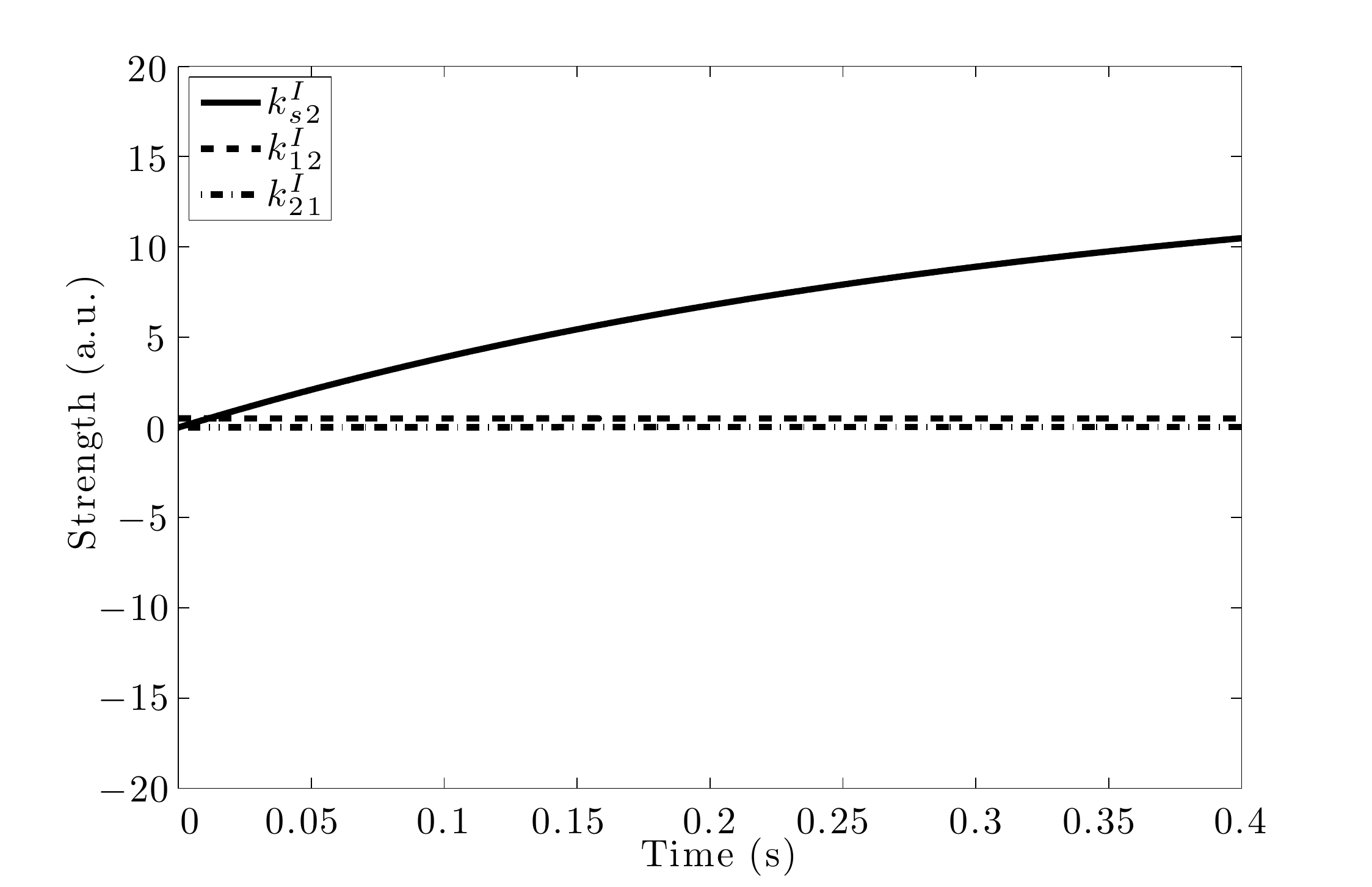}\tabularnewline
\hline 
\end{tabular}
\par\end{centering}

\caption{\label{fig:Reinforcement-Dynamics}(a) Dynamics of coupled oscillators
for the first $6\times10^{-3}$ seconds of reinforcement. Because
reinforcement couplings are strong, oscillators quickly converge to
an asymptotic state with fixed frequencies and phase relations. Graph
(b) shows the quick convergence to the phase relations $-\pi/2$ and
$\pi/2$ for $r_{1}$ and $r_{2}$ with respect to $s$. Lines in
(a) and (b) correspond to the same as Figures \ref{fig:Time-evolution-uncoupled}
and \ref{fig:Osc-Evol-Response}. Graphs (c), (d), (e) and (f) show
the evolution of the excitatory and inhibitory couplings. Because
the system is being reinforced with the phase differences shown in
(b), after reinforcement, if the same oscillators are sampled, we
should expect a response close to zero in the $-1$ and $1$ scale. }

\end{figure}
 shows the dynamics of oscillators and couplings during an effective
reinforcement, and Figure \ref{fig:Osc-Evol-Response}
\begin{figure}
\begin{centering}
(a)\includegraphics[width=7cm]{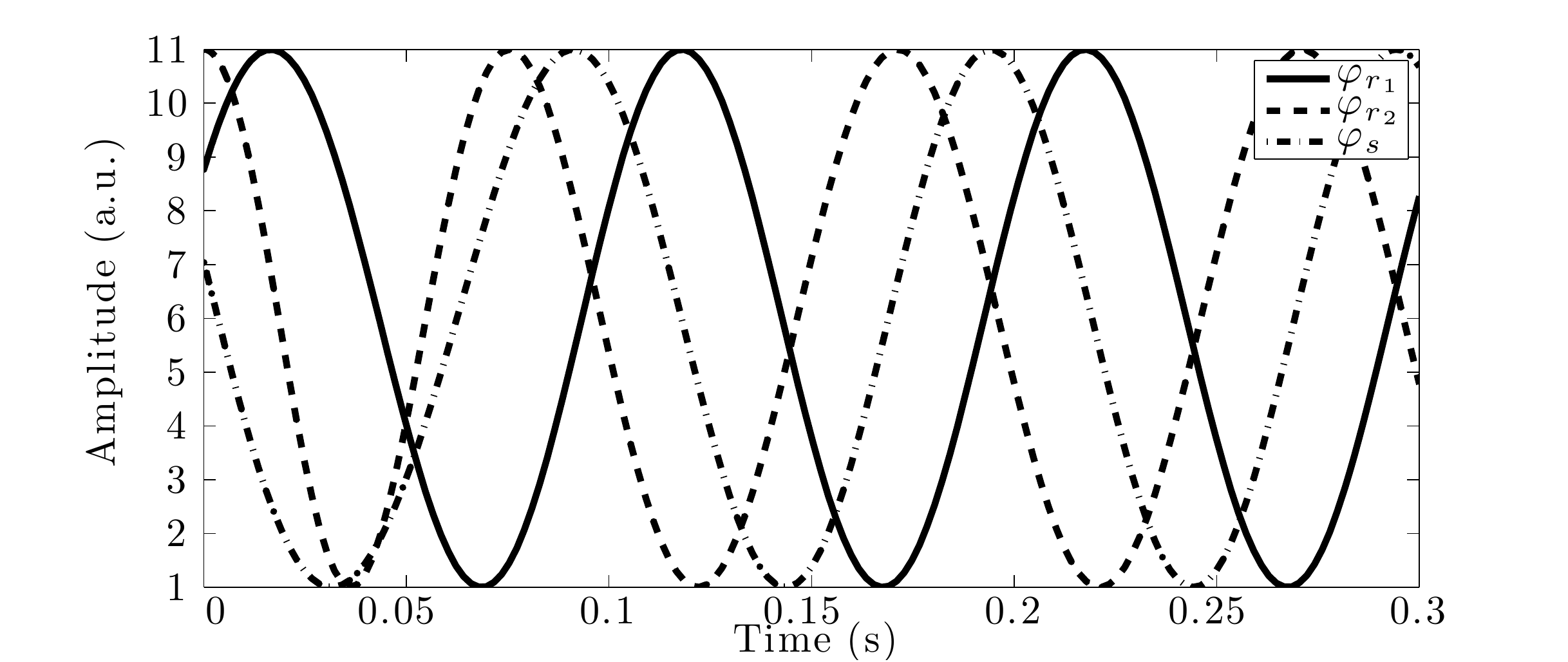}
\par\end{centering}

\begin{centering}
(b)\includegraphics[width=7cm]{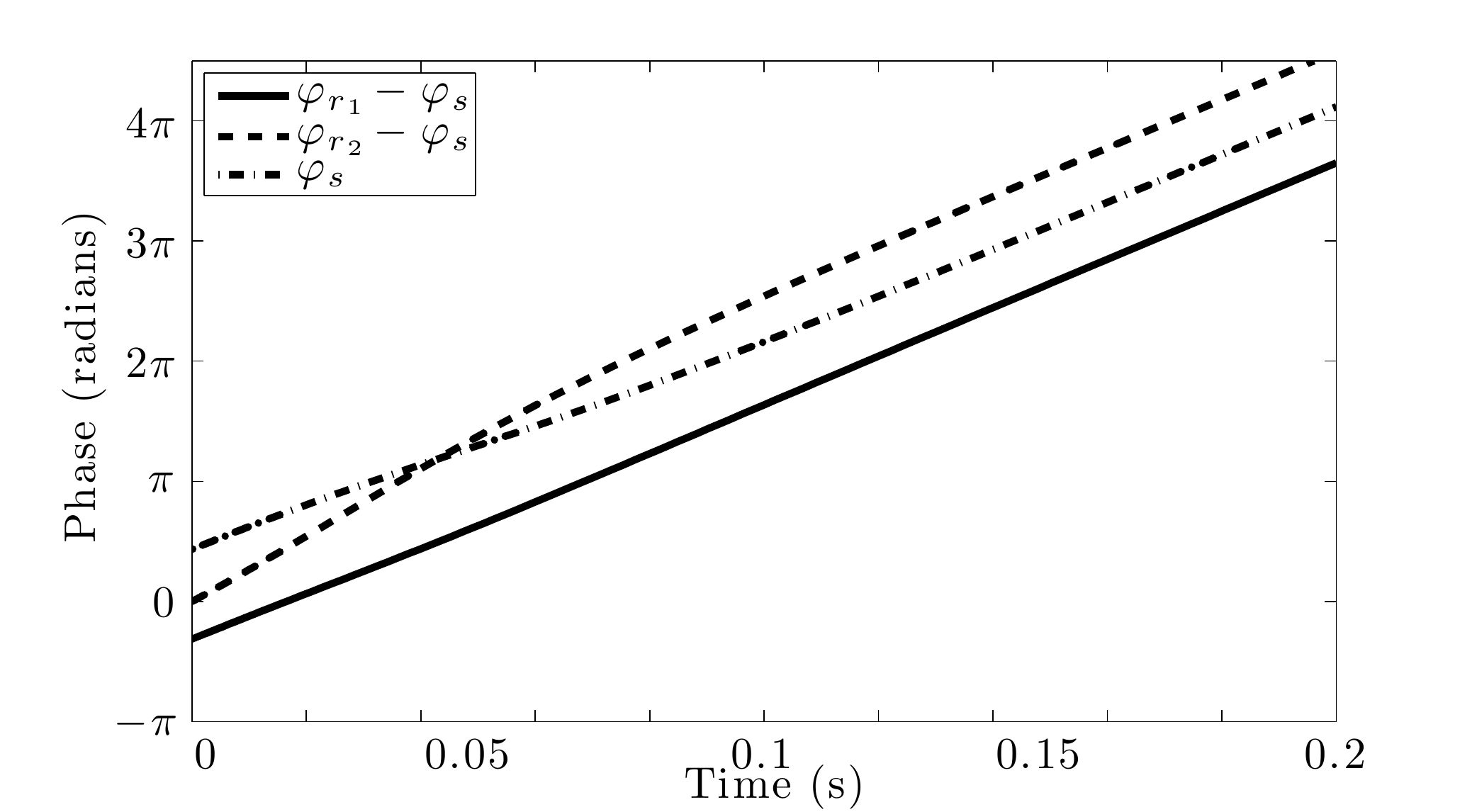}
\par\end{centering}

\begin{centering}
(c)\includegraphics[width=7cm]{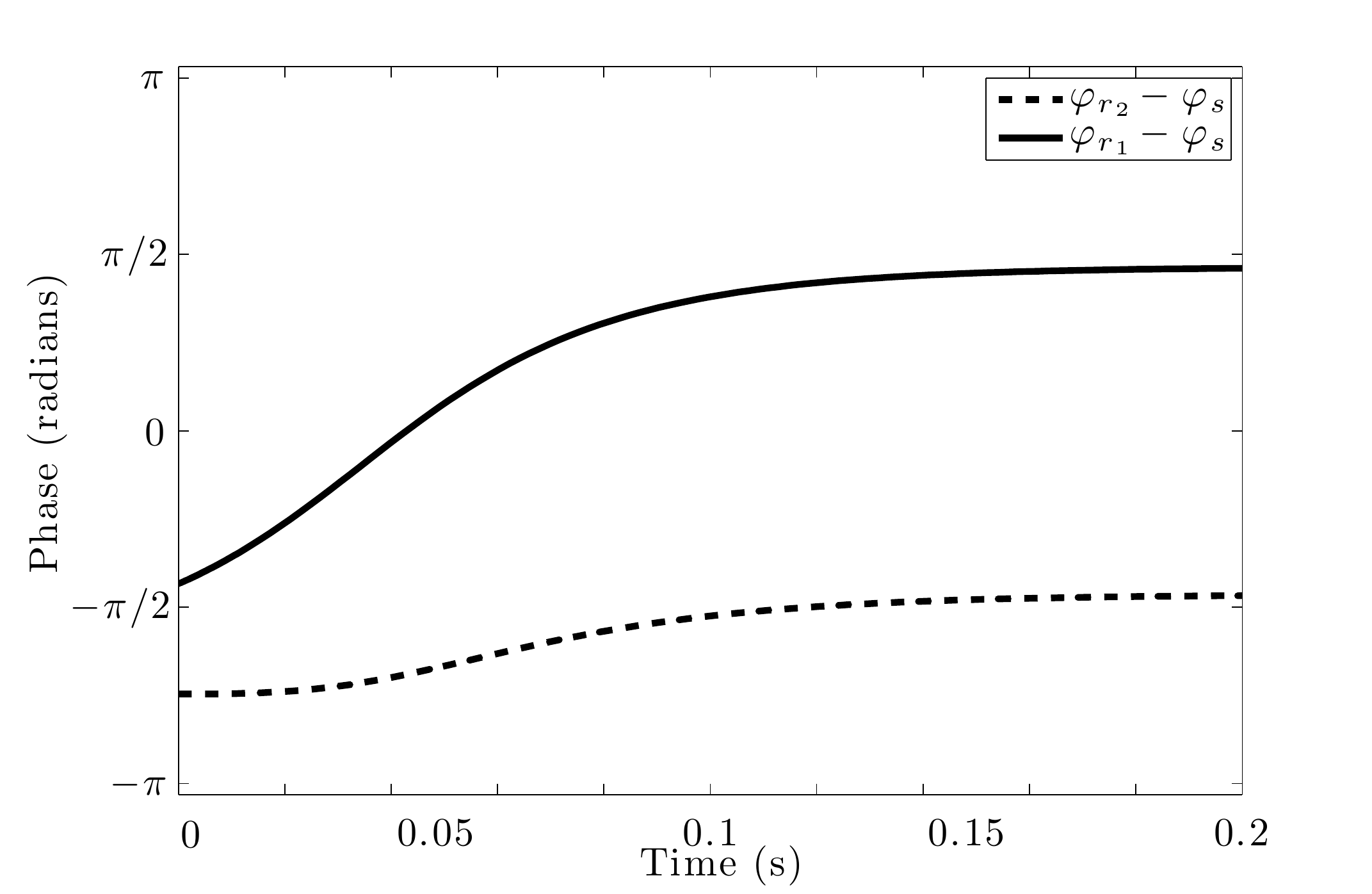}
\par\end{centering}

\begin{centering}
(d)\includegraphics[width=7cm]{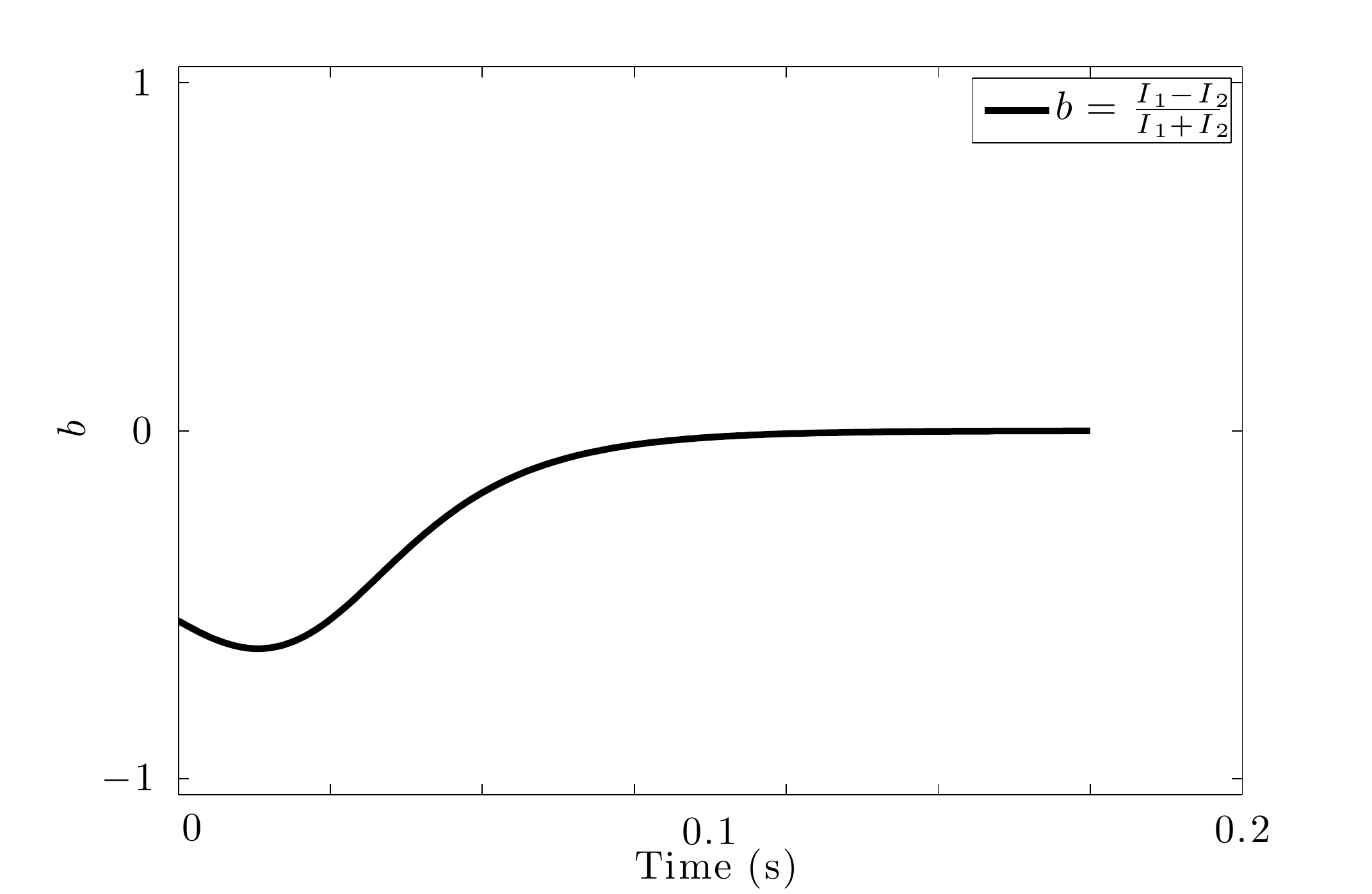}
\par\end{centering}

\caption{\label{fig:Time-evolution-after}Time evolution of the same oscillators
as Figure \ref{fig:Osc-Evol-Response}, but now with the new couplings
generated after the effective reinforcement shown in Figure \ref{fig:Reinforcement-Dynamics}.
Graphs (a) and (b) show the three oscillators, (c) the phase differences,
and (d) the response. We see that the response goes to a value close
to zero, consistent with the reinforcement of zero. }

\end{figure}
 shows the selection of a response for the sampling of the same set
of oscillators after the effective reinforcement.

\section{Oscillator Model for SR-theory\label{sec:Oscillator-Model-SR}}

Now that we have presented the oscillator model from a conceptual
point of view, let us look at the mathematical details of the model.
As mentioned in (iii) and (v) of Section \ref{sub:General-Comments-on-SR},
our goal in this paper is to model the stochastic processes described
by the random variables $\mathbf{X}_{n}$ and $\mathbf{Z}_{n}$ in
terms of oscillators. So, let us start with the oscillator computation
of a response, corresponding to $\mathbf{X}_{n}$. When we think about
neural oscillators, we visualize groups of self-sustaining neurons
that have some coherence and periodicity in their firing patterns.
Here, we assume one of those neural oscillators corresponds to the
sampled stimulus $s$. We will describe below how two neural oscillators
$r_{1}$ and $r_{2}$ can model the computation of randomly selecting
response from a continuum of possible responses in accordance with
a given probability distribution. A way to approach this is to consider
the three harmonic oscillators, $s,$ $r_{1}$, and $r_{2}$, which
for simplicity of computation we chose as having the same natural
frequency $\omega_{0}$ and the same amplitude. We write 
\begin{eqnarray}
s(t) & = & A\cos\left(\omega_{0}t\right)=A\cos\left(\varphi_{s}(t)\right),\label{eq:oscillation-s}\\
r_{1}(t) & = & A\cos\left(\omega_{0}t+\delta\phi_{1}\right)=A\cos\left(\varphi_{r_{1}}(t)\right),\label{eq:oscillation-1}\\
r_{2}(t) & = & A\cos\left(\omega_{0}t+\delta\phi_{2}\right)=A\cos\left(\varphi_{r_{2}}(t)\right),\label{eq:oscillation-2}
\end{eqnarray}
where $s(t)$, $r_{1}(t)$, and $r_{2}(t)$ represent harmonic oscillations,
$\varphi_{s}(t)$, $\varphi_{r_{1}}(t)$, and $\varphi_{r_{2}}(t)$
their phases, $\delta\phi_{1}$ and $\delta\phi_{2}$ are constants,
and $A$ their amplitude. Notice that, since all oscillators have
the same amplitude, their dynamics is completely described by their
phases. Since neural oscillators have a wave-like behavior \citep{Nunez2006a},
their dynamics satisfy the principle of superposition, thus making
oscillators prone to interference effects. As such, the mean intensity,
as usually defined for oscillators, give us a measure of the excitation
carried by the oscillations. The mean intensity $I_{1}$ is the superposition
of $s(t)$ and $r_{1}(t)$, or 
\begin{eqnarray*}
I_{1} & = & \left\langle \left(s(t)+r_{1}(t)\right)^{2}\right\rangle _{t}\\
 & = & \left\langle s(t)^{2}\right\rangle _{t}+\left\langle r_{1}(t)^{2}\right\rangle _{t}+\left\langle 2s(t)r_{1}(t)\right\rangle _{t},
\end{eqnarray*}
where $\left\langle \right\rangle _{t}$ is the time average. A quick
computation gives 
\[
I_{1}=A^{2}\left(1+\cos\left(\delta\phi_{1}\right)\right),
\]
and, similarly for $I_{2}$, 
\[
I_{2}=A^{2}\left(1+\cos\left(\delta\phi_{2}\right)\right).
\]
Therefore, the intensity depends on the phase difference between the
response-computation oscillators and the stimulus oscillator. 

Let us call $I_{1}^{L}$ and $I_{2}^{L}$ the intensities after learning.
The maximum intensity of $I_{1}^{L}$ and $I_{2}^{L}$ is $2A^{2}$,
whereas their minimum intensity is zero. Thus, the maximum difference
between these intensities happens when their relative phases are such
that one of them has a maximum and the other a minimum. For example,
if we choose $\delta\phi_{1}=0$ and $\delta\phi_{2}=\pi$, then $I_{1,max}^{L}=A^{2}\left(1+\cos\left(0\right)\right)=2A^{2}$
and $I_{2,min}^{L}=A^{2}\left(1+\cos\left(\pi\right)\right)=0$. Alternatively,
if we choose $\delta\phi_{1}=\pi$ and $\delta\phi_{2}=0$, $I_{1,min}^{L}=A^{2}\left(1+\cos\left(\pi\right)\right)=0$
and $I_{2,max}^{L}=A^{2}\left(1+\cos\left(0\right)\right)=2A^{2}.$
However, this maximum contrast should only happen when the oscillator
learned a clear response, say the one associated with oscillator $r_{1}(t)$.
When we have in-between responses, we should expect less contrast,
with the minimum happening when the response lies on the mid-point
of the continuum between the responses associated to $r_{1}(t)$ and
$r_{2}(t)$. This happens if we impose
\begin{equation}
\delta\phi_{1}=\delta\phi_{2}+\pi\equiv\delta\phi,\label{eq:ideal-phase-diff}
\end{equation}
which results in 
\begin{equation}
I_{1}^{L}=A^{2}\left(1+\cos\left(\delta\phi\right)\right),\label{eq:phase-1}
\end{equation}
 and
\begin{equation}
I_{2}^{L}=A^{2}\left(1-\cos\left(\delta\phi\right)\right).\label{eq:phase-2}
\end{equation}
From equations (\ref{eq:phase-1}) and (\ref{eq:phase-2}), let $b\in[-1,1]$
be the normalized difference in intensities between $r_{1}$ and $r_{2}$,
i.e. 
\begin{eqnarray}
b & \equiv & \frac{I_{1}^{L}-I_{2}^{L}}{I_{1}^{L}+I_{2}^{L}}=\cos\left(\delta\varphi\right),\label{eq:angle-reinforcement-b}
\end{eqnarray}
 $0\leq\delta\varphi\leq\pi$. So, in principle we can use arbitrary
phase differences between oscillators to code for a continuum of responses
between $-1$ and $1$ (more precisely, because $\delta\varphi$ is
a phase, the interval is in the unit circle $\mathbb{T}$, and not
in a compact interval in $\mathbb{R}$). For arbitrary intervals $(\zeta_{1},\zeta_{2})$,
all that is required is a re-scaling of $b$.

We now turn to the mathematical description of these qualitative ideas.
As we saw, we are assuming that the dynamics is encoded by the phases
in equations (\ref{eq:oscillation-s})--(\ref{eq:oscillation-2}).
We assume that stimulus and response-computation neural oscillators
have natural frequencies $\omega_{0}$, such that their phases are
$\varphi(t)=\omega_{0}t+\mbox{constant}$, when they are not interacting
with other oscillators. The assumption of the same frequency for stimulus
and response-computation oscillators is not necessary, as oscillators
with different natural frequency can entrain \citep{Kuramoto1984},
but it simplifies our analysis, as we focus on phase locking. In real
biological systems, we should expect different neural oscillators
to have different frequencies. We also assume that they are (initially
weakly) coupled to each other with symmetric couplings. The assumption
of coupling between the two response-computation oscillators $r_{1}$
and $r_{2}$ is a detailed feature that has no direct correspondence
in the SR model. We are also not requiring the couplings between oscillators
to be weak after learning, as learning should usually lead to the
strengthening of couplings. This should be contrasted with the usual
requirements of weak couplings in the Kuramoto model \citep{Kuramoto1984}. 

At the beginning of a trial, a stimulus is sampled, and its corresponding
oscillator, $s_{j}$, along with $r_{1}$ and $r_{2}$, are activated.
The sampling of a stimulus is a stochastic process represented in
SR theory by the random variable $\mathbf{S}_{n}$, but we do not
attempt to model it in detail from neural oscillators. Instead, we
just assume that the activation of the oscillators happens in a way
consistent with SR theory; a more detailed model of activation that
includes such sampling would be desirable, but goes beyond the scope
of this paper. Once the stimulus and response-computation oscillators
are activated, the phase of each oscillator resets according to a
normal distribution centered on zero, i. e., $\overline{\varphi}=0$,
with standard deviation $\sigma_{\varphi}$, which we here assume
is the same for all stimulus and response-computation oscillators.
(We choose $\overline{\varphi}=0$ without loss of generality, since
only phase differences are physically meaningful. A Gaussian is used
to represent biological variability. A possible mechanism for this
phase reset can be found in \citet{Wang1995}.) Let $t_{s,n}$ be
the time at which the stimulus oscillator is activated on trial $n$,
and let $\Delta t_{r}$ be the average amount of time it takes for
a response to be selected by oscillators $r_{1}$ and $r_{2}$. We
use $\Delta t_{r}$ as a parameter that is estimated from the experiments,
but we believe that more detailed future models should be able to
predict the value of $\Delta t_{r}$ from the dynamics. Independently
of $n$, the probability density for the phase at time $t_{s,n}$
is given by 
\begin{equation}
f\left(\varphi_{i}\right)=\frac{1}{\sigma_{\varphi}\sqrt{2\pi}}\exp\left(-\frac{\varphi_{i}}{2\sigma_{\varphi}^{2}}\right),\label{eq:phase-density}
\end{equation}
 where $i=s_{j},r_{1},r_{2}$. After the stimulus is sampled, the
active oscillators evolve for the time interval $\Delta t_{r}$ according
to the following set of differential equations, known as the Kuramoto
equations \citep{HoppensteadtIzhikevich1996a,HoppensteadtIzhikevich1996b,Kuramoto1984}.
\begin{equation}
\frac{d\varphi_{i}}{dt}=\omega_{i}-\sum_{i\neq j}k_{y}\sin\left(\varphi_{i}-\varphi_{j}+\delta_{ij}\right),\label{eq:Kuramoto-phase-differences}
\end{equation}
 where $k_{y}$ is the coupling constant between oscillators $i$
and $j$, and $\delta_{ij}$ is an anti-symmetric matrix representing
the phase differences we wish to code, and $i$ and $j$ can be either
$s$, $r_{1}$, or $r_{2}$. 

For $N$ stimulus oscillators $s_{j}$, $j=1,\ldots,N$, we may now
rewrite (\ref{eq:Kuramoto-phase-differences}) for the special case
of the three oscillator equations for $s_{j}$, $r_{1}$, and $r_{2}$.
We introduce in these equation notation for excitatory $\left(k_{j}^{E}\right)$
and inhibitory $\left(k_{y}^{I}\right)$ couplings. These are the
$4N$ excitatory and inhibitory coupling strengths between oscillators
(a more detailed explanation of how (\ref{eq:Kuramoto-3-1-inhib-assym})--(\ref{eq:Kuramoto-3-3-inhib-assym})
are obtained from (\ref{eq:Kuramoto-phase-differences}), as well
as the physical interpretation of the coefficients $k_{s_{1},r_{1}}^{I},\ldots,k_{s_{N},r_{2}}^{E},k_{r_{1},r_{2}}^{E}$
can be found in subsection \ref{sub:Physical-Interpretation-of}).
\begin{eqnarray}
\frac{d\varphi_{s_{j}}}{dt} & = & \omega_{0}-k_{s_{j},r_{1}}^{E}\sin\left(\varphi_{s_{j}}-\varphi_{r_{1}}\right)\nonumber \\
 &  & -k_{s_{j},r_{2}}^{E}\sin\left(\varphi_{s_{j}}-\varphi_{r_{2}}\right)\nonumber \\
 &  & -k_{s_{j},r_{1}}^{I}\cos\left(\varphi_{s_{j}}-\varphi_{r_{1}}\right)\nonumber \\
 &  & -k_{s_{j},r_{2}}^{I}\cos\left(\varphi_{s_{j}}-\varphi_{r_{2}}\right),\label{eq:Kuramoto-3-1-inhib-assym}\\
\frac{d\varphi_{r_{1}}}{dt} & = & \omega_{0}-k_{r_{1},s_{j}}^{E}\sin\left(\varphi_{r_{1}}-\varphi_{s_{j}}\right)\nonumber \\
 &  & -k_{r_{1},r_{2}}^{E}\sin\left(\varphi_{r_{1}}-\varphi_{r_{2}}\right)\nonumber \\
 &  & -k_{r_{1},s_{j}}^{I}\cos\left(\varphi_{r_{1}}-\varphi_{s_{j}}\right)\nonumber \\
 &  & -k_{r_{1},r_{2}}^{I}\cos\left(\varphi_{r_{1}}-\varphi_{r_{2}}\right),\label{eq:Kuramoto-3-2-inhib-assym}\\
\frac{d\varphi_{r_{2}}}{dt} & = & \omega_{0}-k_{r_{2},s_{j}}^{E}\sin\left(\varphi_{r_{2}}-\varphi_{s_{j}}\right)\nonumber \\
 &  & -k_{r_{2},r_{1}}^{E}\sin\left(\varphi_{r_{2}}-\varphi_{r_{1}}\right)\nonumber \\
 &  & -k_{r_{2},s_{j}}^{I}\cos\left(\varphi_{r_{2}}-\varphi_{s_{j}}\right)\nonumber \\
 &  & -k_{r_{2},r_{1}}^{I}\cos\left(\varphi_{r_{2}}-\varphi_{r_{1}}\right),\label{eq:Kuramoto-3-3-inhib-assym}
\end{eqnarray}
 where $\varphi_{s_{j}}$, $\varphi_{r_{1}}$, and $\varphi_{r_{2}}$
are their phases, $\omega_{0}$ their natural frequency. Equations
(\ref{eq:Kuramoto-3-1-inhib-assym})--(\ref{eq:Kuramoto-3-3-inhib-assym})
usually contain the amplitudes of the oscillators as a coupling factor.
For example, instead of just $k_{s_{j},r_{1}}^{E}$ in (\ref{eq:Kuramoto-3-1-inhib-assym}),
the standard form of Kuramoto's equation would have a $A_{s_{j}}A_{r_{1}}k_{s_{j},r_{1}}^{E}$
term multiplying $\sin(\varphi_{s_{j}}-\varphi_{r_{1}})$ \citep{Kuramoto1984}.
For simplicity, we omit this term, as it can be absorbed by $k_{s_{j},r_{1}}^{E}$
if the amplitudes are unchanged. Before any conditioning, the values
for the coupling strengths are chosen following a normal distribution
$g(k)$ with mean $\overline{k}$ and standard deviation $\sigma_{k}.$
It is important to note that reinforcement will change the couplings
while the reinforcement oscillator is acting upon the stimulus and
response oscillators, according to the set of differential equations
presented later in this section. The solutions to (\ref{eq:Kuramoto-3-1-inhib-assym})--(\ref{eq:Kuramoto-3-3-inhib-assym})
and the initial conditions randomly distributed according to $f(\varphi_{i})$
give us the phases at time $t_{r,n}=t_{s,n}+\Delta t_{r}$. (Making
$\Delta t_{r}$ a random variable rather than a constant is a realistic
generalization of the constant value we have used in computations.)
The coupling strengths between oscillators determine their phase locking
and how fast it happens.

\subsection{Oscillator Dynamics of Learning from Reinforcement}

The dynamics of couplings during reinforcement, corresponding to changes
in the conditioning $\mathbf{Z}_{s,n}$. A reinforcement is a strong
external learning event that drives all active oscillators to synchronize
with frequency $\omega_{e}$ to the reinforcement oscillator, while
phase locking to it. We choose $\omega_{e}\neq\omega_{0}$ to keep
its role explicit in our computations. In (\ref{eq:Kuramoto-3-1-inhib-assym})--(\ref{eq:Kuramoto-3-3-inhib-assym})
there is no reinforcement, and as we noted earlier, prior to any reinforcement,
the couplings $k_{s_{1},r_{1}}^{I},\ldots,k_{s_{N},r_{2}}^{E}$ are
normally distributed with mean $\overline{k}$ and standard deviation
$\sigma_{k}.$ To develop equations for conditioning, we assume that
when reinforcement is effective, the reinforcement oscillator deterministically
interferes with the evolution of the other oscillators. This is done
by assuming that the reinforcement event forces the reinforced response-computation
and stimulus oscillators to synchronize with the same phase difference
of $\delta\varphi$, while the two response-computation oscillators
are kept synchronized with a phase difference of $\pi$. Let the reinforcement
oscillator be activated on trial $n$ at time $t_{e,n}$, $t_{r,n+1}>t_{e,n}>t_{r,n}$,
during an interval $\Delta t_{e}$. Let $K_{0}$ be the coupling strength
between the reinforcement oscillator and the stimulus and response-computation
oscillators. In order to match the probabilistic SR axiom governing
the effectiveness of reinforcement, we assume, as something beyond
Kuramoto's equations, that there is a normal probability distribution
governing the coupling strength $K_{0}$ between the reinforcement
and the other active oscillators. It has mean $\overline{K}_{0}$
and standard deviation $\sigma_{K_{0}}$. Its density function is:
\begin{equation}
f\left(K_{0}\right)=\frac{1}{\sigma_{K_{0}}\sqrt{2\pi}}\exp\left\{ -\frac{1}{2\sigma_{K_{0}}^{2}}\left(K_{0}-\overline{K}_{0}\right)^{2}\right\} .\label{eq:K0-density}
\end{equation}
As already remarked, a reinforcement is a disruptive event. When it
is effective, all active oscillators phase-reset at $t_{e,n}$, and
during reinforcement the phases of the active oscillators evolve according
to the following set of differential equations.
\begin{eqnarray}
\frac{d\varphi_{s_{j}}}{dt} & = & \omega_{0}-k_{s_{j},r_{1}}^{E}\sin\left(\varphi_{s_{j}}-\varphi_{r_{1}}\right)\nonumber \\
 &  & -k_{s_{j},r_{2}}^{E}\sin\left(\varphi_{s_{j}}-\varphi_{r_{2}}\right)\nonumber \\
 &  & -k_{s_{j},r_{1}}^{I}\cos\left(\varphi_{s_{j}}-\varphi_{r_{1}}\right)\nonumber \\
 &  & -k_{s_{j},r_{2}}^{I}\cos\left(\varphi_{s_{j}}-\varphi_{r_{2}}\right)\nonumber \\
 &  & -K_{0}\sin\left(\varphi_{s_{j}}-\omega_{e}t\right),\label{eq:learningphaseS-inhib-excite-first}\\
\frac{d\varphi_{r_{1}}}{dt} & = & \omega_{0}-k_{r_{1},s_{j}}^{E}\sin\left(\varphi_{r_{1}}-\varphi_{s_{j}}\right)\nonumber \\
 &  & -k_{r_{1},r_{2}}^{E}\sin\left(\varphi_{r_{1}}-\varphi_{r_{2}}\right)\nonumber \\
 &  & -k_{r_{1},s_{j}}^{I}\cos\left(\varphi_{r_{1}}-\varphi_{s_{j}}\right)\nonumber \\
 &  & -k_{r_{1},r_{2}}^{I}\cos\left(\varphi_{r_{1}}-\varphi_{r_{2}}\right)\nonumber \\
 &  & -K_{0}\sin\left(\varphi_{r_{1}}-\omega_{e}t-\delta\varphi\right),\\
\frac{d\varphi_{r_{2}}}{dt} & = & \omega_{0}-k_{r_{2},s_{j}}^{E}\sin\left(\varphi_{r_{2}}-\varphi_{s_{j}}\right)\nonumber \\
 &  & -k_{r_{2},r_{1}}^{E}\sin\left(\varphi_{r_{2}}-\varphi_{r_{1}}\right)\nonumber \\
 &  & -k_{r_{2},s_{j}}^{I}\cos\left(\varphi_{r_{2}}-\varphi_{s_{j}}\right)\nonumber \\
 &  & -k_{r_{2},r_{1}}^{I}\cos\left(\varphi_{r_{2}}-\varphi_{r_{1}}\right)\nonumber \\
 &  & -K_{0}\sin\left(\varphi_{r_{2}}-\omega_{e}t-\delta\varphi+\pi\right).\label{eq:learningphaseS-inhib-exite-last}
\end{eqnarray}
The excitatory couplings are reinforced if the oscillators are in
phase with each other, according to the following equations. 
\begin{eqnarray}
\frac{dk_{s_{j},r_{1}}^{E}}{dt} & = & \epsilon\left(K_{0}\right)\left[\alpha\cos\left(\varphi_{s_{j}}-\varphi_{r_{1}}\right)-k_{s_{j},r_{1}}^{E}\right],\label{eq:learning-asym-excitatory-1}\\
\frac{dk_{s_{j},r_{2}}^{E}}{dt} & = & \epsilon\left(K_{0}\right)\left[\alpha\cos\left(\varphi_{s_{j}}-\varphi_{r_{2}}\right)-k_{s_{j},r_{2}}^{E}\right],\\
\frac{dk_{r_{1},r_{2}}^{E}}{dt} & = & \epsilon\left(K_{0}\right)\left[\alpha\cos\left(\varphi_{r_{1}}-\varphi_{r_{2}}\right)-k_{r_{1},r_{2}}^{E}\right],\\
\frac{dk_{r_{1},s_{j}}^{E}}{dt} & = & \epsilon\left(K_{0}\right)\left[\alpha\cos\left(\varphi_{s_{j}}-\varphi_{r_{1}}\right)-k_{r_{1},s_{j}}^{E}\right],\\
\frac{dk_{r_{2},s_{j}}^{E}}{dt} & = & \epsilon\left(K_{0}\right)\left[\alpha\cos\left(\varphi_{s_{j}}-\varphi_{r_{2}}\right)-k_{r_{2},s_{j}}^{E}\right],\\
\frac{dk_{r_{2},r_{1}}^{E}}{dt} & = & \epsilon\left(K_{0}\right)\left[\alpha\cos\left(\varphi_{r_{1}}-\varphi_{r_{2}}\right)-k_{r_{2},r_{1}}^{E}\right].
\end{eqnarray}
 Similarly, for inhibitory connections, if two oscillators are perfectly
off sync, then we have a reinforcement of the inhibitory connections.
\begin{eqnarray}
\frac{dk_{s_{j},r_{1}}^{I}}{dt} & = & \epsilon\left(K_{0}\right)\left[\alpha\sin\left(\varphi_{s_{j}}-\varphi_{r_{1}}\right)-k_{s_{j},r_{1}}^{I}\right],\\
\frac{dk_{s_{j},r_{2}}^{I}}{dt} & = & \epsilon\left(K_{0}\right)\left[\alpha\sin\left(\varphi_{s_{j}}-\varphi_{r_{2}}\right)-k_{s_{j},r_{2}}^{I}\right],\\
\frac{dk_{r_{1},r_{2}}^{I}}{dt} & = & \epsilon\left(K_{0}\right)\left[\alpha\sin\left(\varphi_{r_{1}}-\varphi_{r_{2}}\right)-k_{r_{1},r_{2}}^{I}\right],\\
\frac{dk_{r_{1},s_{j}}^{I}}{dt} & = & \epsilon\left(K_{0}\right)\left[\alpha\sin\left(\varphi_{r_{1}}-\varphi_{s_{j}}\right)-k_{r_{1},s_{j}}^{I}\right],\\
\frac{dk_{r_{2},s_{j}}^{I}}{dt} & = & \epsilon\left(K_{0}\right)\left[\alpha\sin\left(\varphi_{r_{2}}-\varphi_{s_{j}}\right)-k_{r_{2},s_{j}}^{I}\right],\\
\frac{dk_{r_{2},r_{1}}^{I}}{dt} & = & \epsilon\left(K_{0}\right)\left[\alpha\sin\left(\varphi_{r_{2}}-\varphi_{r_{1}}\right)-k_{r_{2},r_{1}}^{I}\right].\label{eq:learning-asym-inhibitory-last}
\end{eqnarray}
In the above equations, 
\begin{equation}
\epsilon\left(K_{0}\right)=\left\{ \begin{array}{c}
0\mbox{ if }K_{0}<K'\\
\epsilon_{0}\mbox{ otherwise},
\end{array}\right.\label{eq:epsilon}
\end{equation}
where $\epsilon_{0}\ll\omega_{0}$, $\alpha$ and $K_{0}$ are constant
during $\Delta t_{e}$ \citep{HoppensteadtIzhikevich1996a,HoppensteadtIzhikevich1996b},
and $K'$ is a threshold constant throughout all trials. The function
$\epsilon(K_{0})$ represents nonlinear effects in the brain. These
effects could be replaced by the use of a sigmoid function $\epsilon_{0}(1+\exp\{-\gamma(K_{0}-K')\})^{-1}$
\citep{EeckmanFreeman1991} in (\ref{eq:learning-asym-excitatory-1})--(\ref{eq:learning-asym-inhibitory-last}),
but we believe that our current setup makes the probabilistic features
clearer. In both cases, we can think of $K'$ as a threshold below
which the reinforcement oscillator has no (or very little) effect
on the stimulus and response-computation oscillators.

Before we proceed, let us analyze the asymptotic behavior of these
equations. From (\ref{eq:learning-asym-excitatory-1})--(\ref{eq:learning-asym-inhibitory-last})
and with the assumption that $K_{0}$ is very large, we have, once
we drop the terms that are small compared to $K_{0}$, 
\begin{eqnarray}
\frac{d\varphi_{s_{j}}}{dt} & \approx & \omega_{o}-K_{0}\sin\left(\varphi_{s_{j}}-\omega_{e}t\right),\label{eq:43}\\
\frac{d\varphi_{r_{1}}}{dt} & \approx & \omega_{o}-K_{0}\sin\left(\varphi_{r_{1}}-\omega_{e}t-\delta\varphi\right),\label{eq:44}\\
\frac{d\varphi_{r_{2}}}{dt} & \approx & \omega_{o}-K_{0}\sin\left(\varphi_{r_{2}}-\omega_{e}t-\delta\varphi+\pi\right).\label{eq:45}
\end{eqnarray}
It is straightforward to show that the solutions for (\ref{eq:43})--(\ref{eq:45})
converge, for $t>2/\sqrt{K_{0}^{2}-\left(\omega_{e}-\omega_{0}\right)^{2}}$
and $K_{0}^{2}\gg(\omega_{0}-\omega_{e})^{2}$, to $\varphi_{s_{j}}=\varphi_{r_{1}}=\omega_{e}t-\pi$
and $\varphi_{r_{2}}=\omega_{e}t$ if $\omega_{e}\neq\omega_{0}$.
So, the effect of the new terms added to Kuramoto's equations is to
force a specific phase synchronization between $s_{j}$, $r_{1}$,
and $r_{2}$, and the activated reinforcement oscillator. This leads
to the fixed points $\varphi_{s}=\omega_{e}t,$ $\varphi_{r_{1}}=\omega_{e}t+\delta\varphi,$
$\varphi_{r_{2}}=\omega_{e}t+\delta\varphi-\pi$. With this reinforcement,
the excitatory couplings go to the following asymptotic values $k_{s,r_{1}}^{E}=k_{r_{1},s}^{E}=-k_{s,r_{2}}^{E}=-k_{r_{2},s}^{E}=\alpha\cos\left(\delta\varphi\right)$,
$k_{r_{1},r_{2}}^{E}=k_{r_{2},r_{1}}^{E}=-\alpha$, and the inhibitory
ones go to $k_{s,r_{1}}^{I}=-k_{r_{1},s}^{I}=-k_{s,r_{2}}^{I}=k_{r_{2},s}^{I}=\alpha\sin\left(\delta\varphi\right)$,
and $k_{r_{1},r_{2}}^{I}=k_{r_{2},r_{1}}^{I}=0$. We show in \ref{App:Properties-Oscillator-Model}
that these couplings lead to the desired stable fixed points corresponding
to the phase differences $\delta\varphi$.

We now turn our focus to equations (\ref{eq:learning-asym-excitatory-1})--(\ref{eq:learning-asym-inhibitory-last}).
For simplicity, we will consider the case where $\delta\varphi=0$
is reinforced. For $K_{0}>K'$, $\varphi_{s_{j}}$ and $\varphi_{r_{1}}$
evolve, approximately, to $\omega_{e}t+\pi$, and $\varphi_{r_{2}}$
to $\omega_{e}t$. Thus, some couplings in (\ref{eq:learning-asym-excitatory-1})--(\ref{eq:learning-asym-inhibitory-last})
tend to a solution of the form $\alpha+c_{1}\exp(-\epsilon_{0}t)$,
whereas others tend to $-\alpha+c_{2}\exp(-\epsilon_{0}t)$, with
$c_{1}$ and $c_{2}$ being integration constants. For a finite time
$t>1/\epsilon_{0}$, depending on the couplings, the values satisfy
the stability requirements shown above.

The phase differences between stimulus and response oscillators are
determined by which reinforcement is driving the changes to the couplings
during learning. From (\ref{eq:learning-asym-excitatory-1})--(\ref{eq:learning-asym-inhibitory-last}),
reinforcement may be effective only if $\Delta t_{e}>\epsilon_{0}^{-1}$
(see \citet{SeligerEtAl2002} and \ref{eq:coupling-asymp-delta-excite-6-A}),
setting a lower bound for $\epsilon_{0}$, as $\Delta t_{e}$ is fixed
by the experiment. For values of $\Delta t_{e}>\epsilon_{0}^{-1}$,
the behavioral probability parameter $\theta$ of effective reinforcement
is, from (\ref{eq:K0-density}) and (\ref{eq:epsilon}), reflected
in the equation:
\begin{equation}
\theta=\int_{K'}^{\infty}f\left(K_{0}\right)\, dK_{0}.\label{eq:prob-eff-reinf}
\end{equation}
 This relationship comes from the fact that if $K_{0}<K'$, there
is no effective learning from reinforcement, since there are no changes
to the couplings due to (\ref{eq:learning-asym-excitatory-1})--(\ref{eq:learning-asym-inhibitory-last}),
and (\ref{eq:Kuramoto-3-1-inhib-assym})--(\ref{eq:Kuramoto-3-3-inhib-assym})
describe the oscillators' behavior (more details can be found in \ref{App:Properties-Oscillator-Model}).
Intuitively $K'$ is the effectiveness parameter. The larger it is,
relative to $\overline{K}_{0}$, the smaller the probability the random
value $K_{0}$ of the coupling will be effective in changing the values
of the couplings through (\ref{eq:learning-asym-excitatory-1})--(\ref{eq:learning-asym-inhibitory-last}).

Equations (\ref{eq:learning-asym-excitatory-1})--(\ref{eq:learning-asym-inhibitory-last})
are similar to the evolution equations for neural networks, derived
from some reasonable assumptions in \citet{HoppensteadtIzhikevich1996a,HoppensteadtIzhikevich1996b}.
The general idea of oscillator learning similar to (\ref{eq:learning-asym-excitatory-1})--(\ref{eq:learning-asym-inhibitory-last})
was proposed in \citet{SeligerEtAl2002} and \citet{Nishii1998} as
possible learning mechanisms.

Thus, a modification of Kuramoto's equations, where we include asymmetric
couplings and inhibitory connections, permit the coding of any desired
phase differences. Learning becomes more complex, as several new equations
are necessary to accommodate the lack of symmetries. However, the
underlying ideas are the same, i.e., that neural learning happens
in a Hebb-like fashion, via the strengthening of inhibitory and excitatory
oscillator couplings during reinforcement, which approximates an SR
learning curve.

In summary, the coded phase differences may be used to model a continuum
of responses within SR theory in the following way. At the beginning
of a trial, the oscillators are reset with a small fluctuation, as
in the one-stimulus model, according to the distribution (\ref{eq:phase-density}).
Then, the system evolves according to (\ref{eq:Kuramoto-3-1-inhib-assym})--(\ref{eq:Kuramoto-3-3-inhib-assym})
if no reinforcement is present, and according to (\ref{eq:learningphaseS-inhib-excite-first})--(\ref{eq:learning-asym-inhibitory-last})
if reinforcement is present. The coupling constants and the conditioning
of stimuli are not reset at the beginning of each trial. Because of
the finite amount of time for a response, the probabilistic characteristics
of the initial conditions lead to the smearing of the phase differences
after a certain time, with an effect similar to that of the smearing
distribution in the SR model for a continuum of responses (\citet{Suppes1961test,SuppesEtAl1964}).
We emphasize that the smearing distribution is not introduced as an
extra feature of the oscillator model, but comes naturally from the
stochastic properties of it.

\paragraph{Oscillator Computations for Two-Response Models}

An important case for the SR model is when participants select from
a set of two possible responses, a situation much studied experimentally.
We can, intuitively, think of this as a particular case where the
reinforcement selects from a continuum of responses two well-localized
regions, and one response would be any point selected in one region
and the other response any point selected on the other region. In
our oscillator model, this can be accomplished by reinforcing, say,
$\delta\varphi=0$, thus leading to responses on a continuum that
would be localized in a region close to $b=1$, and small differences
between the actual answer and $1$ would be considered irrelevant.
More accurately, if a stimulus oscillator phase-locks in phase to
a response-computation oscillator, this represents in SR theory that
the stimulus is conditioned to a response, but we do not require precise
phase locking between the stimulus and response-oscillators. The oscillator
computing the response at trial $n$ is the oscillator whose actual
phase difference at time $t_{r,n}$ is the smaller with respect to
$s_{j}$, i.e., the smaller $c_{i}$, 
\begin{equation}
c_{i}=\left|\frac{1}{2\pi}\left|\varphi_{r_{i}}-\varphi_{s_{j}}\right|-\left\lfloor \frac{1}{2\pi}\left|\varphi_{r_{i}}-\varphi_{s_{j}}\right|\right\rfloor -\frac{1}{2}\right|,\label{eq:phasemod}
\end{equation}
where $\left\lfloor x\right\rfloor $ is the largest integer not greater
than $x$, and $i=1,2$.

\subsection{Physical Interpretation of Dynamical Equations\label{sub:Physical-Interpretation-of}}

A problem with equation (\ref{eq:Kuramoto-phase-differences}) is
that it does not have a clear physical or neural interpretation. How
are the arbitrary phase differences $\delta_{ij}$ to be interpreted?
Certainly, they cannot be interpreted by the excitatory firing of
neurons, as they would lead to $\delta_{ij}$ being either $0$ or
$\pi$, as shown in \ref{App:Properties-Oscillator-Model}. Furthermore,
since the $k_{ij}$ represent oscillator couplings, how are the phase
differences stored? To answer these questions, let us rewrite (\ref{eq:Kuramoto-phase-differences})
as\textbf{ 
\begin{eqnarray}
\frac{d\varphi_{i}}{dt} & = & \omega_{i}-\sum_{j}k_{ij}\cos\left(\delta_{ij}\right)\sin\left(\varphi_{i}-\varphi_{j}\right)\nonumber \\
 &  & -\sum_{j}k_{ij}\sin\left(\delta_{ij}\right)\cos\left(\varphi_{i}-\varphi_{j}\right).\label{eq:kura-phase-rewritten}
\end{eqnarray}
}Since the terms involving the phase differences $\delta_{ij}$ are
constant, we can write (\ref{eq:kura-phase-rewritten}) as\textbf{
\begin{equation}
\frac{d\varphi_{i}}{dt}=\omega_{i}-\sum_{j}\left[k_{ij}^{E}\sin\left(\varphi_{i}-\varphi_{j}\right)+k_{ij}^{I}\cos\left(\varphi_{i}-\varphi_{j}\right)\right],\label{eq:kuramoto-equations-inhibition-excitation}
\end{equation}
}where $k_{ij}^{E}=k_{ij}\cos\left(\delta_{ij}\right)$ and $k_{ij}^{I}=k_{ij}\sin\left(\delta_{ij}\right)$.
Equation (\ref{eq:Kuramoto-phase-differences}) now has an immediate
physical interpretation. Since $k_{ij}^{E}$ makes oscillators $\varphi_{i}$
and $\varphi_{j}$ approach each other, as before, we can think of
them as corresponding to excitatory couplings between neurons or sets
of neurons. When a neuron $\varphi_{j}$ fires shortly before it is
time for $\varphi_{i}$ to fire, this makes it more probable $\varphi_{i}$
will fire earlier, thus bringing its rate closer to $\varphi_{j}$.
On the other hand, for $k_{ij}^{I}$ the effect is the opposite: if
a neuron in $\varphi_{j}$ fires, $k_{ij}^{I}$ makes the neurons
in $\varphi_{i}$ fire further away from it. Therefore, $k_{ij}^{I}$
cannot represent an excitatory connection, but must represent inhibitory
ones between $\varphi_{i}$ and $\varphi_{j}$. In other words, we
may give physical meaning to learning phase differences in equations
(\ref{eq:Kuramoto-phase-differences}) by rewriting them to include
excitatory and inhibitory terms. 

We should stress that inhibition and excitation have different meanings
in different disciplines, and some comments must be made to avoid
confusion. First, here we are using inhibition and excitation in a
way analogous to how they are used in neural networks, in the sense
that inhibitions decrease the probability of a neuron firing shortly
after the inhibitory event, whereas excitation increases this probability.
Similarly, the inhibitory couplings in equations (\ref{eq:kuramoto-equations-inhibition-excitation})
push oscillators apart, whereas the excitatory couplings bring them
together. However, though inhibitory and excitatory synaptic couplings
provide a reasonable model for the couplings between neural oscillators,
we are not claiming to have derived equations (\ref{eq:kuramoto-equations-inhibition-excitation})
from such fundamental considerations. In fact, other mechanisms, such
as asymmetries of the extracellular matrix or weak electric-field
effects, could also yield similar results. The second point is to
distinguish our use of inhibition from that in behavioral experiments.
In animal experiments, inhibition is usually the suppression of a
behavior by negative conditioning \citep{dickinson1980contemporary}.
As such, this concept has no direct analogue in the standard mathematical
form of SR theory presented in Section \ref{sec:Stimulus-Response-Theory}.
But we emphasize that our use of inhibition in equations (\ref{eq:kuramoto-equations-inhibition-excitation})
is not the behavioral one, but instead closer to the use of inhibition
in synaptic couplings. 

Before we proceed, another issue needs to be addressed. In a preliminary
analysis of the fixed points, we see three distinct possibilities:
$\phi_{1j}=\phi_{2j}=0$, (ii) $\phi_{1j}=\pi,\,\phi_{2j}=0$, and
(iii) $\phi_{1j}=0,\,\phi_{2j}=\pi$. A detailed investigation of
the different fixed points and their stability is given in \ref{App:Properties-Oscillator-Model}.
For typical biological parameters and the physical ones we have added,
the rate of convergence is such that the solutions are well approximated
by the asymptotic fixed points by the time a response is made. But
given that we have three different fixed points, how can $r_{1}$
be the conditioned response computation oscillator (fixed point iii)
if $r_{2}$ can also phase-lock in phase (fixed point ii)? To answer
this question, we examine the stability of these fixed points, i.e.,
whether small changes in the initial conditions for the phases lead
to the same phase differences. In \ref{App:Properties-Oscillator-Model}
we linearize for small fluctuations around each fixed point. As shown
there (Equation \ref{eq:general-condition}), a sufficient condition
for fixed point $(0,\pi)$ to be stable is

\textbf{
\[
k_{s_{j},r_{2}}k_{r_{1},r_{2}}>k_{s_{j},r_{1}}k_{s_{j},r_{2}}+k_{s_{j},r_{1}}k_{r_{1},r_{2}},
\]
}

\noindent and the other fixed points are unstable (see Fig. \ref{fig:Field-plot}
below); similar results can be computed for the other fixed points.
It is important to notice that, although the above result shows stability,
it does not imply synchronization within some fixed finite time. Because
neural oscillators need to operate in rather small amounts of time,
numerical simulations are necessary to establish synchronization in
an appropriately short amount of time \textbf{ }(though some arguments
are given in \ref{subap:Time-of-convergence} with respect to the
time of convergence and the values of the coupling constants). 
\begin{figure}[h]
\begin{centering}
\includegraphics[clip,width=21pc]{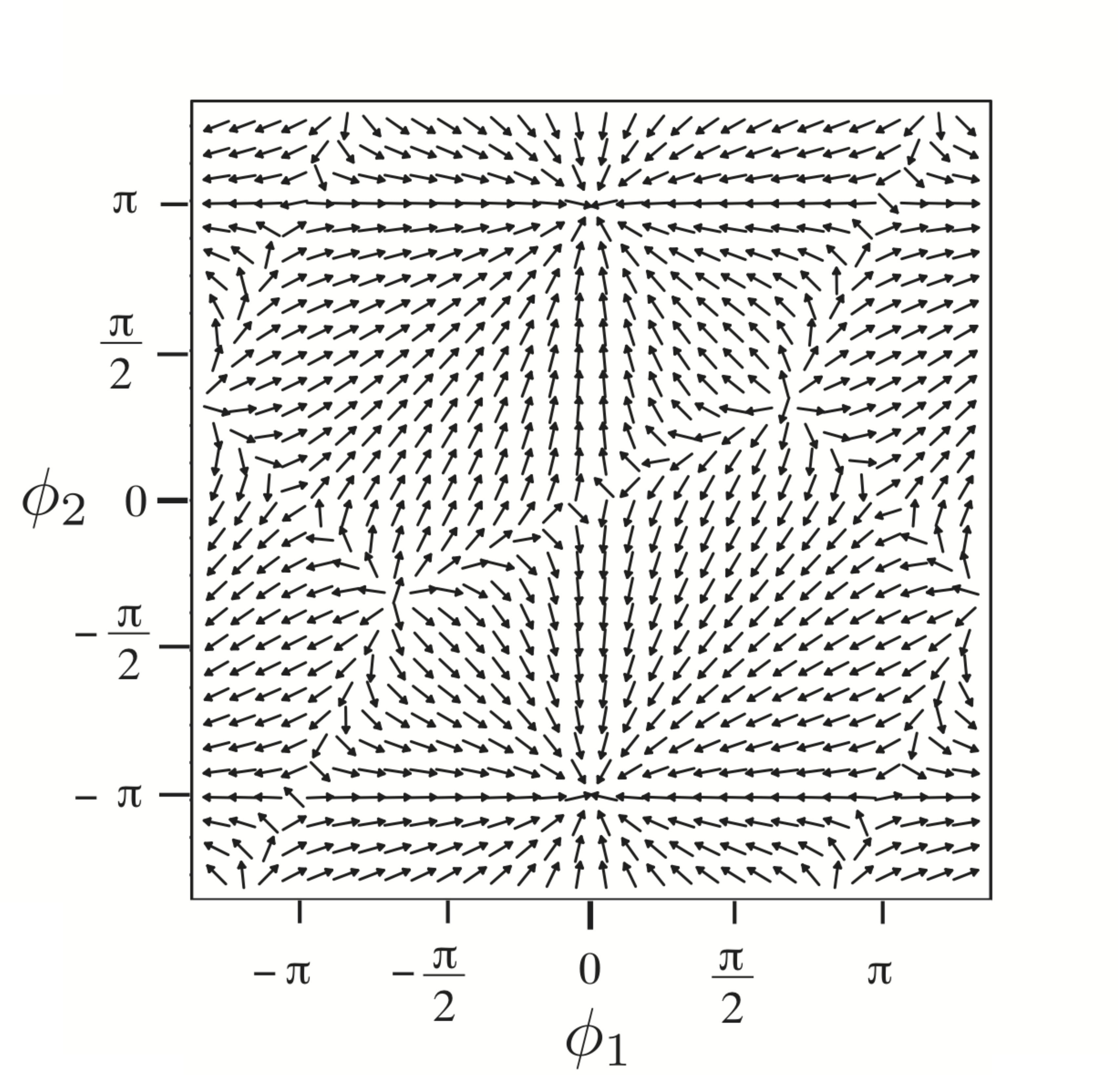} 
\par\end{centering}

\caption{\label{fig:Field-plot}Field plot, generated using Maple 10, for couplings
$k_{s_{j},r_{1}}=k_{r_{1},r_{2}}=-k$ and $k_{s_{j},r_{2}}=k$, $k>0$.
Field lines show that the fixed point at $(0,\pi)$ is stable, whereas
$(0,0)$, $(\pi,0)$, $(\pi,\pi)$, $(\pi/3,2\pi/3)$, and $(2\pi/3,\pi/3)$
are unstable. Thus, for a randomly selected initial condition, in
a finite amount of time $s_{j}$ approaches the phase of $r_{1}$
and departs from $r_{2}$.}
\end{figure}
 If the couplings are zero, the model still picks one of the oscillators:
the one with the minimum value for $c_{i}$ in (\ref{eq:phasemod}).
But in this case, because of the small fluctuations in the initial
conditions, the probability for each response is $1/2$, as expected,
corresponding, obviously, to the SR parameter $p=1/2$ in the one-stimulus
model.

\subsection{Parameter values}

We turn now to the parameters used in the oscillator models. In the
above equations, we have as parameters $N$, $\omega_{0}$, $\omega_{e}$,
$\alpha$, $\Delta t_{r}$, $\Delta t_{e}$, $\overline{k}$, $\sigma_{k}$,
$\overline{\varphi}$, $\sigma_{\varphi}$, $\epsilon_{0}$, $\overline{K}_{0}$,
$\sigma_{K_{0}}$, and $K'$. This large number of parameters stands
in sharp contrast to the small number needed for SR theory, which
is abstract and therefore much simpler. In our view, any other detailed
physical model of SR theory will face a similar problem. Experimental
evidence constrains the range of values for $\omega_{0}$, $\omega_{e}$,
$\Delta t_{r}$, and $\Delta t_{e}$, and throughout this paper we
choose $\omega_{0}$ and $\omega_{e}$ to be at the order of $10\mbox{ Hz}$,
$\Delta t_{r}=200\mbox{ ms}$, and $\Delta t_{e}=400\mbox{ ms}$.
These natural frequencies were chosen because: (i) they are in the
lower range of frequencies measured for neural oscillators \citep{FreemanBarrie1994,FriedrichEtAl2004,KazantsevEtAl2004,MurthyFetz1992,SuppesBrain2000,Tallon-BaudryEtAl2001},
and the lower the frequency, the longer the time it takes for two
oscillators to synchronize, which imposes a lower bound on the time
taken to make a response, (ii) data from \citet{SuppesLuHan1997,SuppesBrain1998,SuppesBrain1999,SuppesBrain1999b,SuppesBrain2000}
suggest that most frequencies used for the brain representation of
language are close to 10 Hz, and (iii) choosing a small range of frequencies
simplifies the oscillator behavior.%
\footnote{However, we should mention that none of the above references involve
reinforcement, though we are working with the assumption that reinforcement
events are also represented by frequencies that are within the range
suggested by evidence. %
} We use $\Delta t_{r}=200$ ms, which is consistent with the part
it plays in the behavioral response times in many psychological experiments;
for extended analysis of the latter see \citet{Luce1986}. The value
for $\Delta t_{e}=400\mbox{ ms}$ is also consistent with psychological
experiments.

Before any conditioning, we should not expect the coupling strengths
between oscillators to favor one response over the other, so we set
$\overline{k}=0$ Hz. The synchronization of oscillators at the beginning
of each trial implies that their phases are almost the same, so for
convenience we use $\overline{\varphi}=0$. The standard deviations
$\sigma_{k}$ and $\sigma_{\varphi}$ are not directly measured, and
we use for our simulations the reasonable values of $\sigma_{k}=10^{-3}\mbox{ Hz}$
and $\sigma_{\varphi}=\pi/4\mbox{ Hz}$. Those values were chosen
because $\sigma_{k}$ should be very small, since it is related to
neuronal connections before any conditioning, and $\sigma_{\varphi}$
should allow for a measurable phase reset at the presentation of a
stimulus, so $\sigma_{\varphi}<1$.

The values of $\alpha$, $\epsilon_{0}$, $\overline{K}_{0}$, and
$\sigma_{K_{0}}$ are based on theoretical considerations. As we discussed
in the previous paragraph, $\overline{K}_{0}$ is the mean strength
for the coupling of the reinforcement oscillator, and $\sigma_{K_{0}}$
is the standard deviation of its distribution. We assume in our model
that this strength is constant during a trial, but that it varies
from trial to trial according to the given probability distribution.
The parameter $\alpha$ is related to the maximum strength allowed
for the coupling constants $k_{s_{j},r_{1}}$, $k_{s_{j},r_{2}}$,
and $k_{r_{1},r_{2}}$. Because of the stability conditions for the
fixed points, shown above, any value of $\alpha$ leads to phase locking,
given enough time. The magnitude of $\alpha$ is monotonically related
to how fast the system phase locks. It needs to be at least of the
order of $1/\Delta t_{r}$ if the oscillators are to phase lock within
the behavioral response latency. Another important parameter is $\epsilon_{0}$,
and we can see its role by fixing the phases and letting the system
evolve. In this situation, the couplings converge exponentially to
$\alpha$ with a characteristic time $\epsilon_{0}^{-1}$, so $\epsilon_{0}$
determines how fast the couplings change. The time of response, $\Delta t_{r}$,
given experimentally, sets a minimum value of $5\mbox{ Hz}$ for $\alpha$,
since $\alpha>\Delta t_{r}^{-1}=5\mbox{ Hz}$ is needed to have phase
locking from Kuramoto-type equations. The time of reinforcement, $\Delta t_{e}$,
and natural frequency $\omega_{0}$ are related to $\epsilon_{0}$
by $\epsilon_{0}>\Delta t_{e}^{-1}=2.5\mbox{ Hz }$ and $\epsilon_{0}\ll\omega_{0}$.
Finally, to have an effective reinforcement, $K_{0}$ must satisfy
$K_{0}\gg\omega_{0}$. These theoretical considerations are consistent
with the values $\alpha=10\mbox{ Hz}$, $\epsilon_{0}=3\mbox{ Hz}$,
$\overline{K}_{0}=4,000\mbox{ Hz}$, and $\sigma_{K_{0}}=1,000\mbox{ Hz}$.
We note one distinction about the roles of the three probability distributions
introduced. Samples are drawn of phases $\varphi$ and reinforcement
oscillator coupling strength $K_{0}$ on each trial, but oscillator
couplings $k_{s_{1},r_{1}}^{E},\ldots,k_{s_{N},r_{2}}^{I},k_{r_{1},r_{2}}^{I}$
are sampled only once at the beginning of a simulated experiment and
then evolve according to (\ref{eq:learning-asym-excitatory-1})--(\ref{eq:learning-asym-inhibitory-last}).

Table 
\begin{table}[h]
\caption{\label{tab:Values-of-parameters}Fixed values of parameters used to
fit the oscillator models to SR experiments.}

\centering{}%
\begin{tabular}{|c|c||c|c|}
\hline 
Parameter  & Value  & Parameter  & Value\tabularnewline
\hline 
\hline 
$\alpha$  & 10 Hz  & $\sigma_{\varphi}$  & $\pi/4$\tabularnewline
\hline 
$\omega_{0}$  & 10 Hz  & $\overline{k}$  & $0$ Hz\tabularnewline
\hline 
$\omega_{e}$  & 12 Hz  & $\sigma_{k}$  & $10^{-3}$ Hz\tabularnewline
\hline 
$\Delta t_{r}$  & 200 ms  & $\sigma_{K_{0}}$  & 1,000 Hz\tabularnewline
\hline 
$\Delta t_{e}$  & 400 ms  & $\overline{K}_{0}$  & 4,000 Hz\tabularnewline
\hline 
$\overline{\varphi}$  & $0$  & $\epsilon_{0}$  & 3 Hz\tabularnewline
\hline 
\end{tabular}
\end{table}
\ref{tab:Values-of-parameters} summarizes the fixed parameter values
used in our simulations, independent of considering the experimental
design or data of a given SR experiment. This leaves us with two remaining
free parameters to estimate from SR experimental data: the number
of stimulus oscillators $N$ and the nonlinear cutoff parameter $K'$.
These two parameters have a straightforward relation to the SR ones.
$N$ corresponds to the number of stimuli, and $K'$ is monotonically
decreasing with the effectiveness of the reinforcement probability
$\theta$, as shown in (\ref{eq:prob-eff-reinf}).

\begin{figure}[h]
\begin{centering}
\includegraphics[width=7cm]{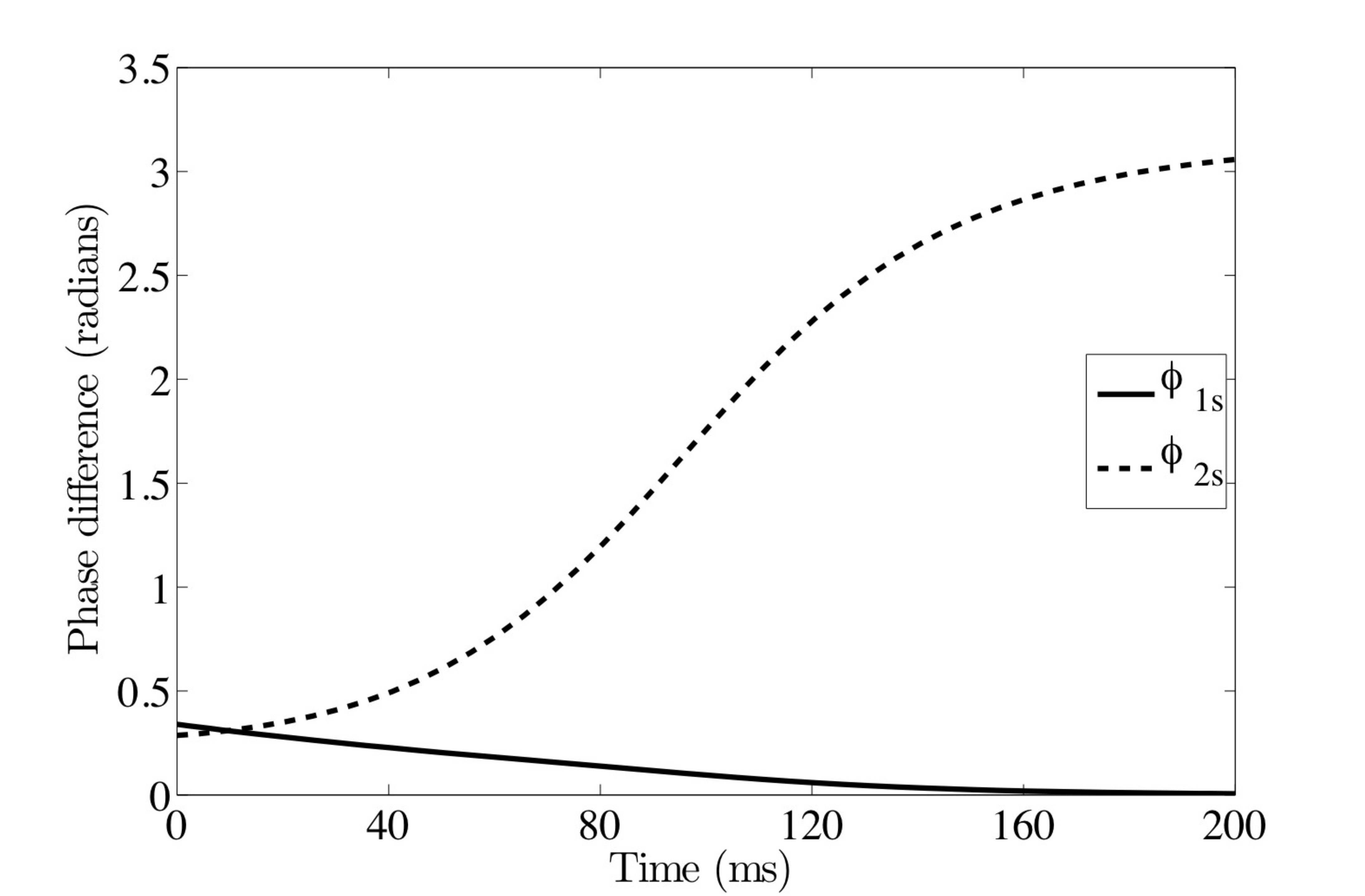} 
\par\end{centering}

\caption{\label{fig:synch-oscillators}MATLAB computation of the phase differences
between three coupled oscillators evolving according to Kuramoto's
equations with couplings $k_{s_{j},r_{1}}=\omega_{0}=-k_{s_{j},r_{2}}=k_{r_{1},r_{2}}$
and $\omega_{0}=20\pi\,\mbox{s}^{-1}$ (10 Hz). The solid line represents
$\phi_{1s_{j}}=\varphi_{r_{1}}-\varphi_{s_{j}}$, and the dashed $\phi_{2s_{j}}=\varphi_{r_{2}}-\varphi_{s}$. }
\end{figure}

Fig. \ref{fig:synch-oscillators} exemplifies the phase-difference
behavior of three oscillators satisfying the Kuramoto equations. 
 In the figure, as shown, all oscillators start with similar phases,
but at 200 ms both $s_{j}$ and $r_{1}$ lock with the same phase,
whereas $r_{2}$ locks with phase difference $\pi$. Details on the
theoretical relations between parameters can be found in \ref{App:Properties-Oscillator-Model}.

\section{\textit{\emph{Comparison with Experiments\label{sec:Comparison-with-Experiment}}}}

Although the oscillator models produce a mean learning curve that
fits quite well the one predicted by SR theory, it is well known that
stochastic models with underlying assumptions that are qualitatively
quite different may predict the same mean learning curves. The asymptotic
conditional probabilities, on the other hand, often have very different
values, depending on the number $N$ of sampled stimuli assumed.\textbf{
}For instance, the observable conditional probability density \textbf{$j(x_{n}|y_{n-1}y_{n-2})$}
for the one-stimulus model is different from the conditional density
for the $N$-stimulus models, $\left(N>1\right)$ as is shown later
in this section.

In the following, we compare some behavioral experimental data to
such theoretical quantities and to the oscillator-simulation data.
We first compare empirical data from an experiment with a continuum
of responses to the oscillator-simulation data. Second, we show that
in a probability matching experiment \cite[Chapter 10]{SuppesAtkinson1960}
the experimental and oscillator-simulated data are quite similar and
fit rather well to the SR theoretical predications. Finally, we examine
the paired-associate learning experiment of \citet{Bower1961}, and
model it in the same way. Here we focus on the model predictions of
stationarity and independence of responses prior to the last error.
Again, the original experimental data, and the oscillator-simulation
data exhibit these two properties to about the same reasonably satisfactory
degree.

\subsection{Continuum of Responses}

As shown in Section \ref{sec:Stimulus-Response-Theory}, to fit the
continuum-of-responses SR model to data, we need to estimate $N$,
the number of stimuli, $\theta$, the probability that a reinforcement
is effective, and $K_{s}(x|z)$, the response smearing distribution
of each stimulus $s$ over the set of possible responses. As we saw
in Section 3, in the oscillator model $\theta$ comes from the dynamics
of learning (see equation (\ref{eq:prob-eff-reinf}) and corresponding
discussion), whereas $N$ is external to the model. The smearing distribution
in the oscillator model comes from the stochastic nature of the initial
conditions coupled with the nonlinear dynamics of Kuramoto's equations. 

Let us now focus on the experiment, described in \citet{SuppesEtAl1964}.
In it, a screen had a 5-foot diameter circle. A noncontingent reinforcement
was given on each trial by a dot of red light on the circumference
of the 5-foot circle, with the position determined by a bimodal distribution
$f(y)$, given by 
\begin{equation}
f(y)=\left\{ \begin{array}{cc}
\frac{2}{\pi^{2}}y, & 0\leq x\leq\frac{\pi}{2}\\
\frac{2}{\pi^{2}}\left(\pi-y\right), & \frac{\pi}{2}<x\leq\pi\\
\frac{2}{\pi^{2}}\left(y-\pi\right), & \pi<x\leq\frac{3\pi}{2}\\
\frac{2}{\pi^{2}}\left(2\pi-y\right), & \frac{3\pi}{2}<x\leq2\pi.
\end{array}\right.\label{eq:bimodaldensityreinforcement}
\end{equation}
 At the center of this circle, a knob connected to an invisible shaft
allowed the red dot to be projected on the screen at any point on
the circle. Participants were instructed to use this knob to predict
on each trial the position of the next reinforcement light. The participants
were 4 male and 26 female Stanford undergraduates. For this experiment,
\citet{SuppesEtAl1964} discuss many points related to the predictions
of SR theory, including goodness of fit. Here we focus mainly on the
relationship between our oscillator-simulated data and the SR predicted
conditional probability densities.

\begin{figure}
\begin{centering}
\includegraphics[width=21pc]{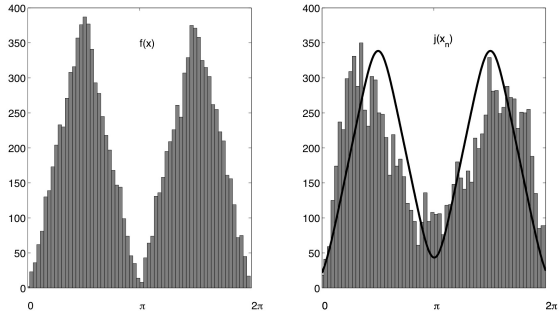} 
\par\end{centering}

\caption{\label{fig:Histogram-density}On the left histogram of the reinforcement
angles used in the oscillator simulation, and on the right both the
SR theoretical density of the one-stimulus model and the histogram
of the oscillator-simulated responses.}
\end{figure}

\begin{figure}
\begin{centering}
\includegraphics[width=250pt]{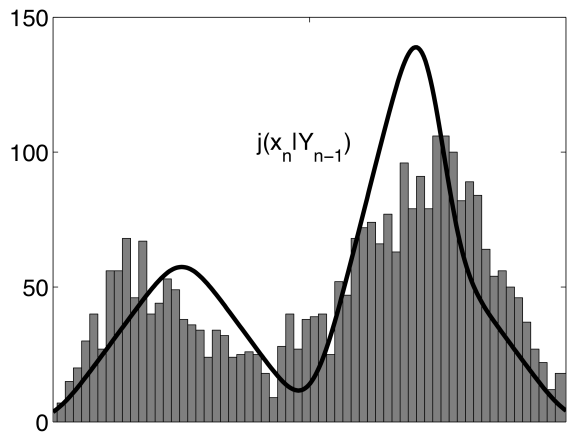} 
\par\end{centering}

\caption{\label{fig:Histogram-density-Condition}Histogram for the oscillator-simulated
response $x_{n}$ on trial $n$ conditioned to a reinforcement on
trial $n-1$ in the interval $(\pi,3\pi/2)$. The black line shows
the fitted SR theoretical predictions of the one-stimulus model. }
\end{figure}

\begin{figure}
\begin{centering}
\includegraphics[width=21pc]{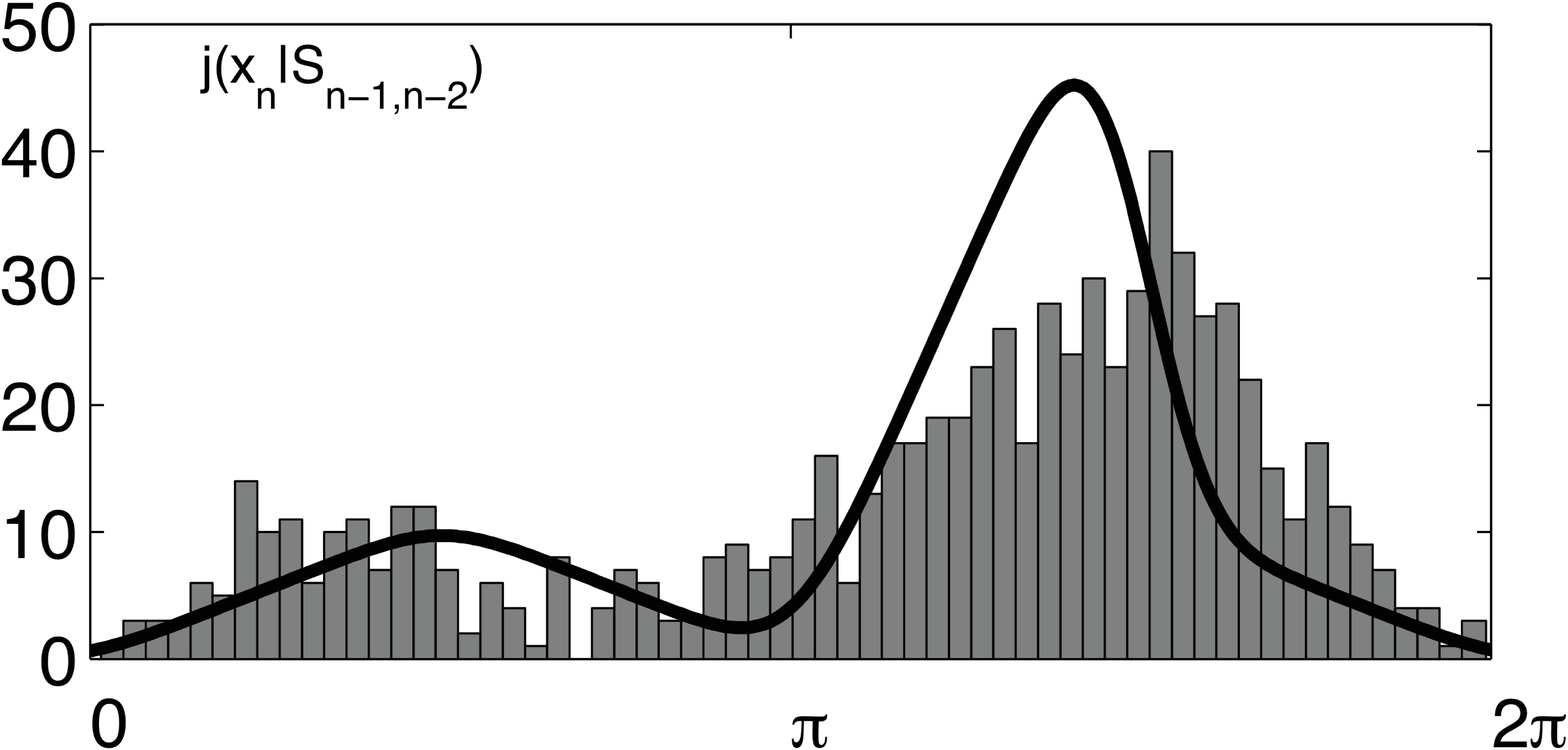}
\par\end{centering}

\begin{centering}
\includegraphics[width=21pc]{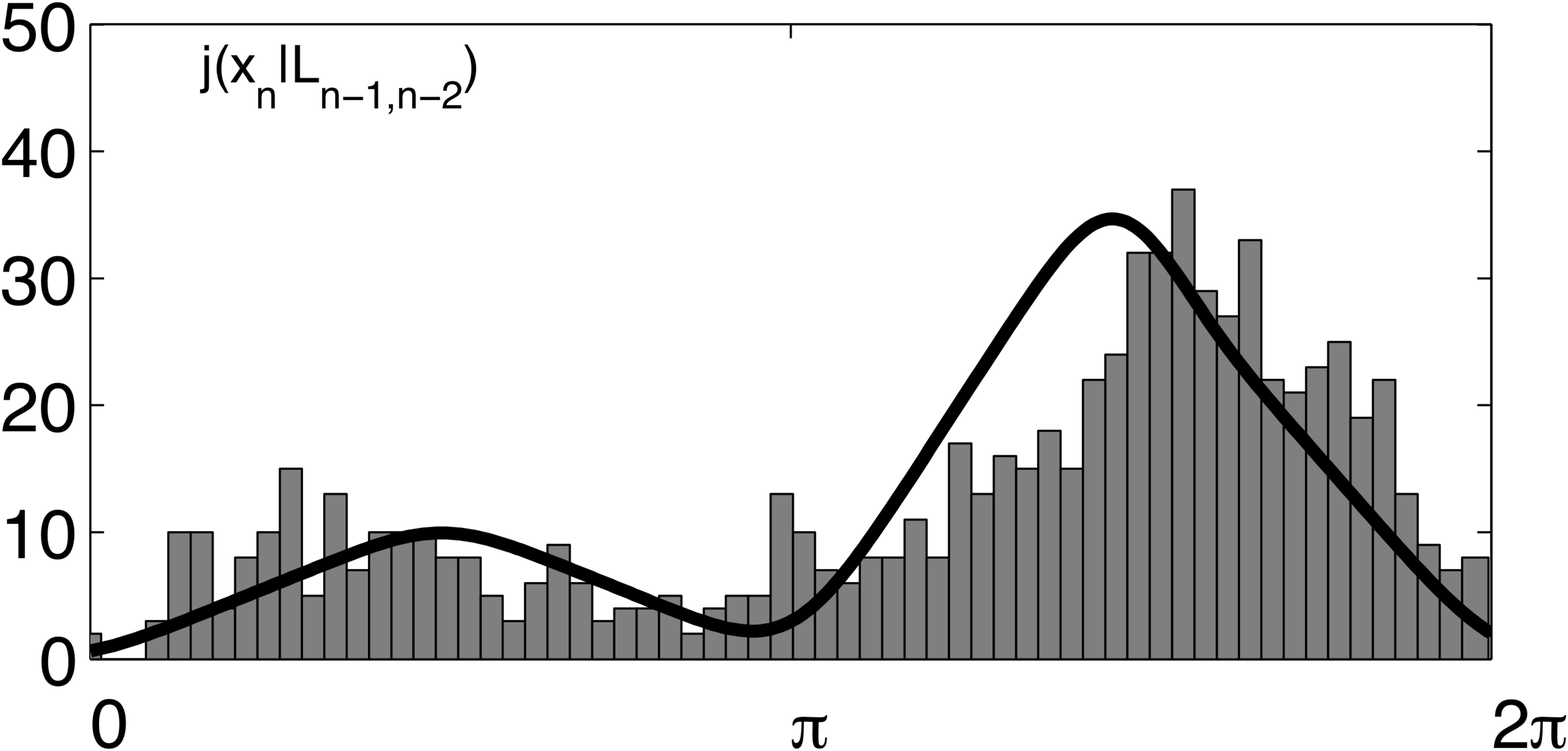} 
\par\end{centering}

\begin{centering}
\includegraphics[width=21pc]{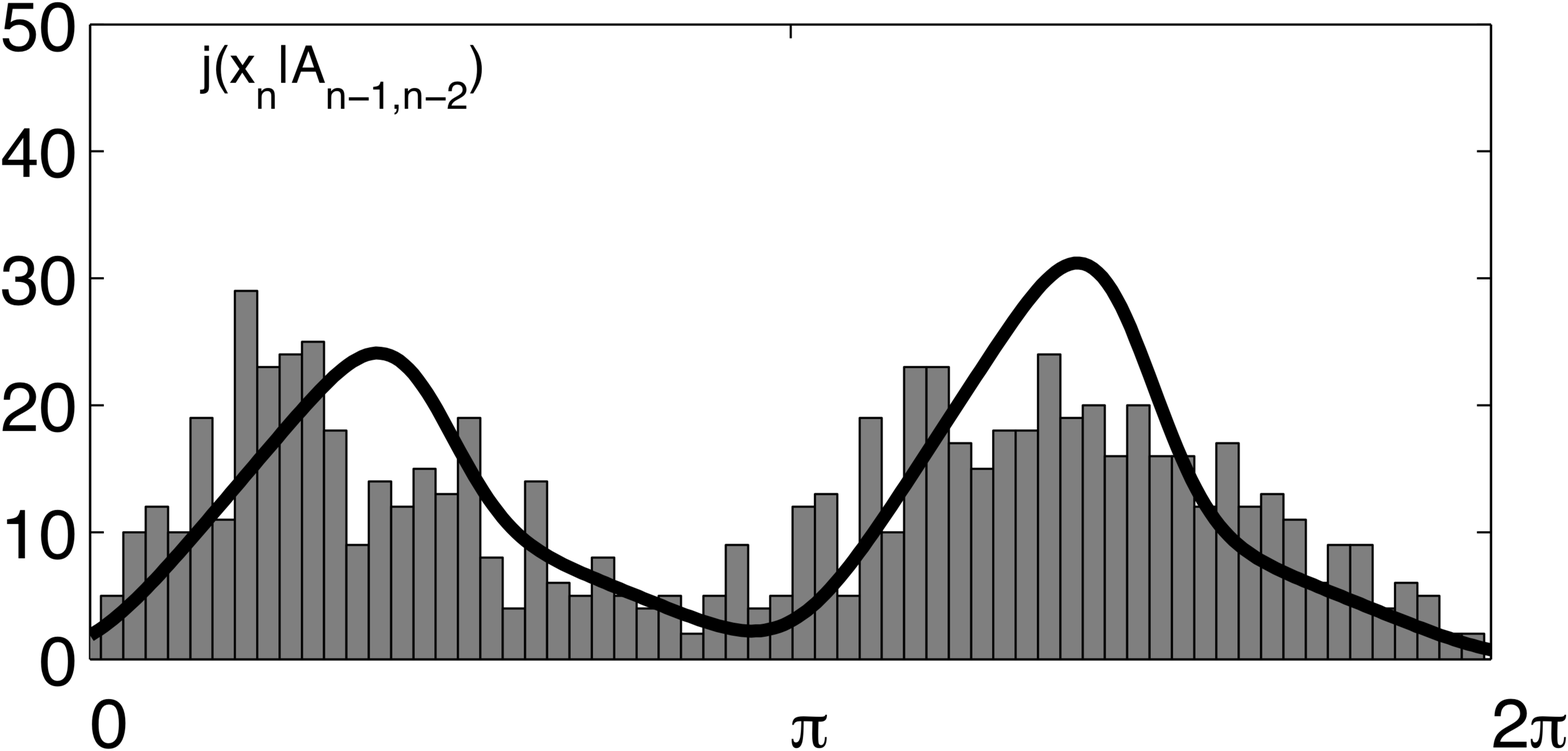} 
\par\end{centering}

\begin{centering}
\includegraphics[width=21pc]{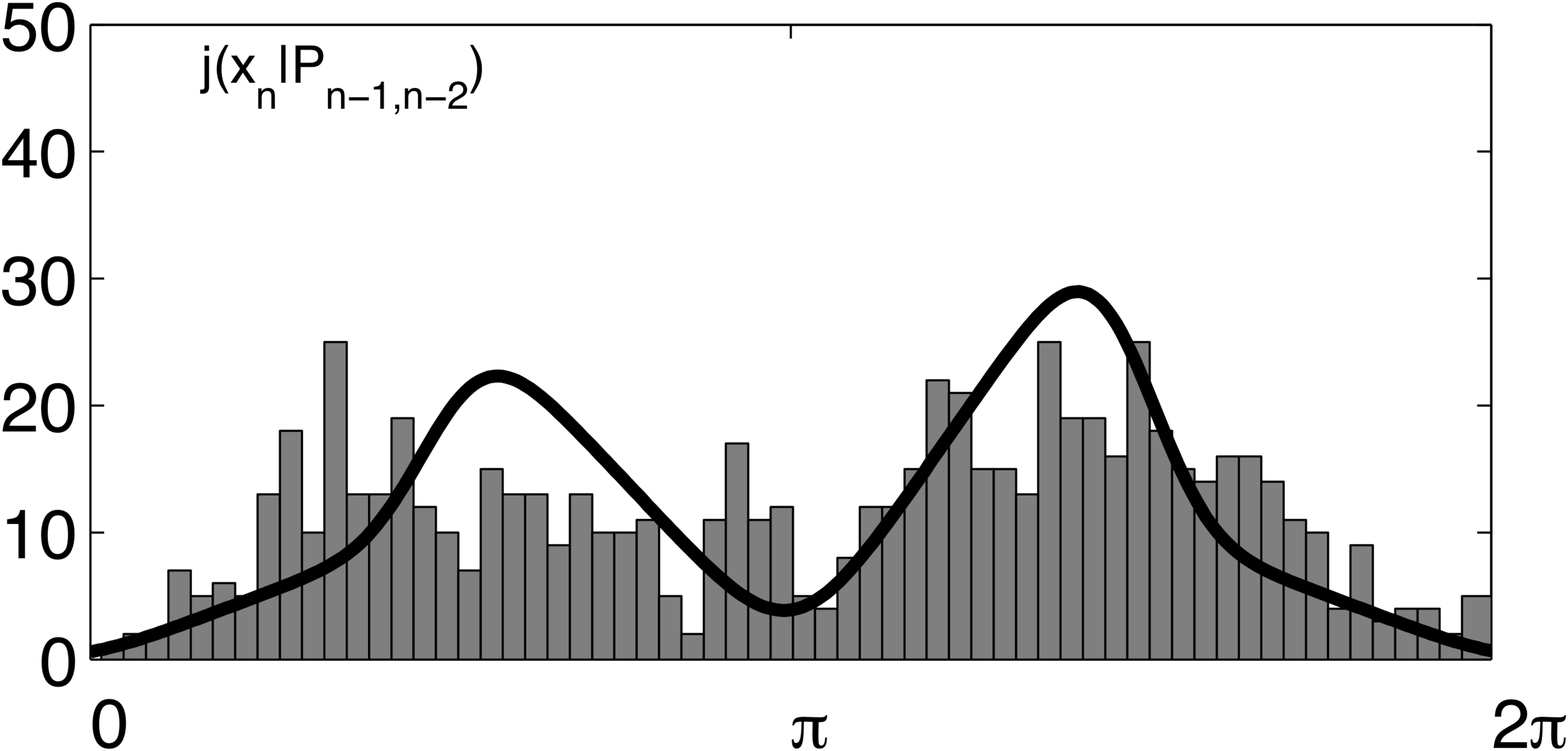} 
\par\end{centering}

\caption{\label{fig:Second-order-conditional-probability}Histograms of oscillators-simulated
responses conditioned to reinforcement on the two previous trial.
All graphs correspond to simulated responses with reinforcement on
trial $n-1$ being in the interval $(\pi,3\pi/2)$. $S$ corresponds
to reinforcements on trial $n-2$ occurring in the same interval.
$L$ corresponds to reinforcement on $n-2$ occurring on the interval
$(3\pi/2,2\pi)$, $A$ on the interval $(0,\pi/2),$ and $P$ on $(\pi/2,\pi)$.
The solid lines show the SR theoretical one-stimulus model predictions. }
\end{figure}

\begin{figure}
\begin{centering}
\includegraphics[width=250pt]{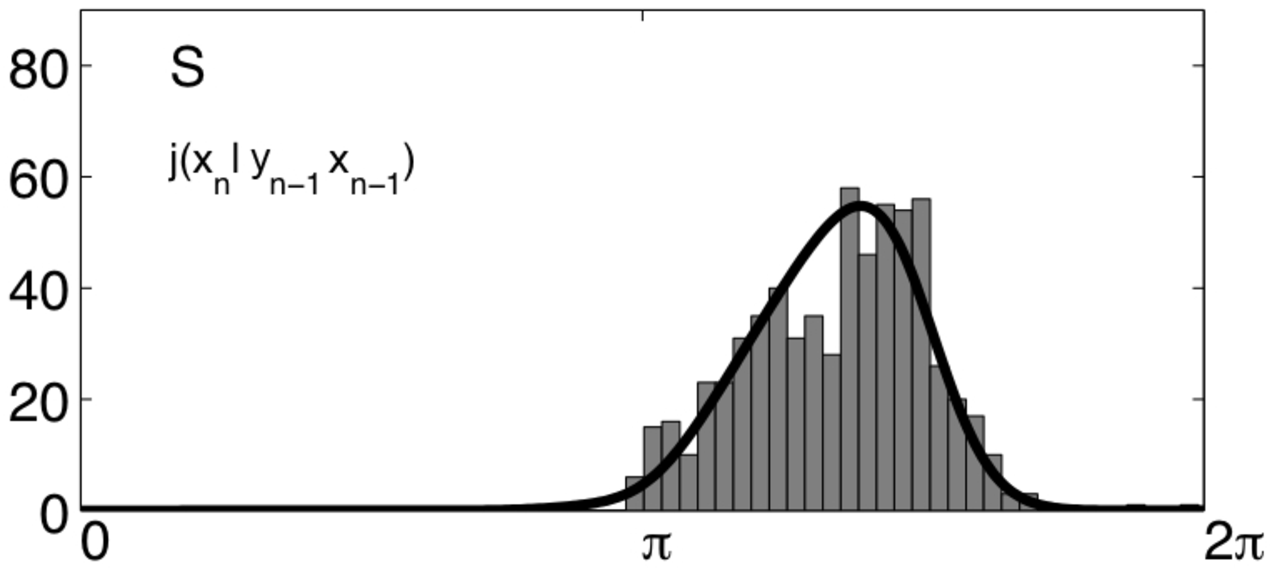}
\par\end{centering}

\begin{centering}
\includegraphics[width=250pt]{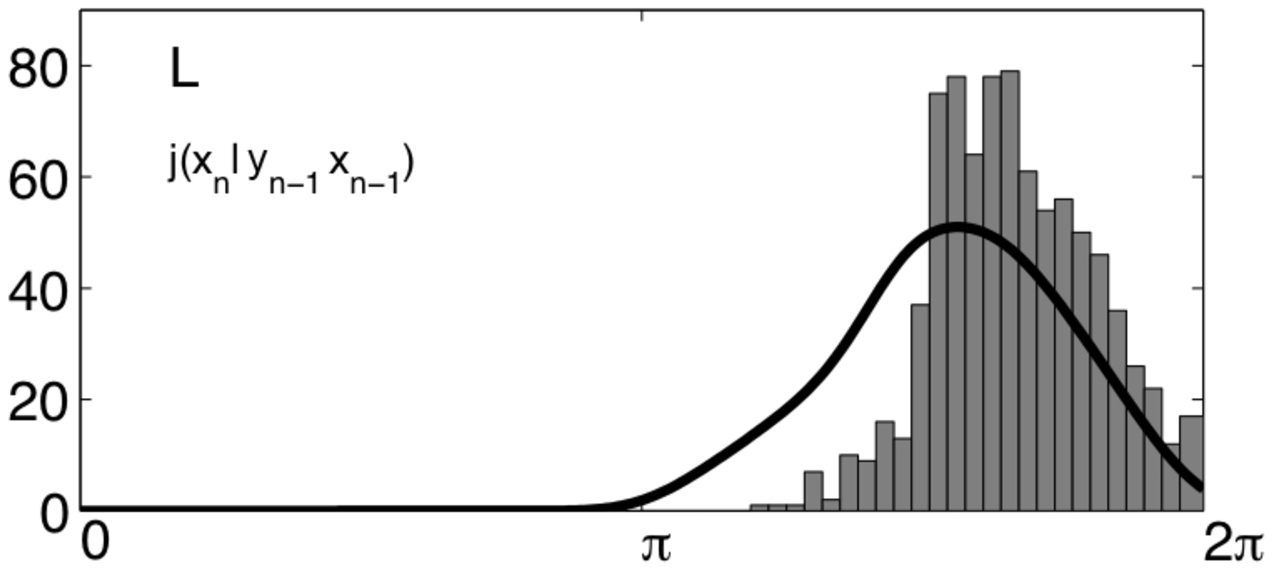}
\par\end{centering}

\begin{centering}
\includegraphics[width=250pt]{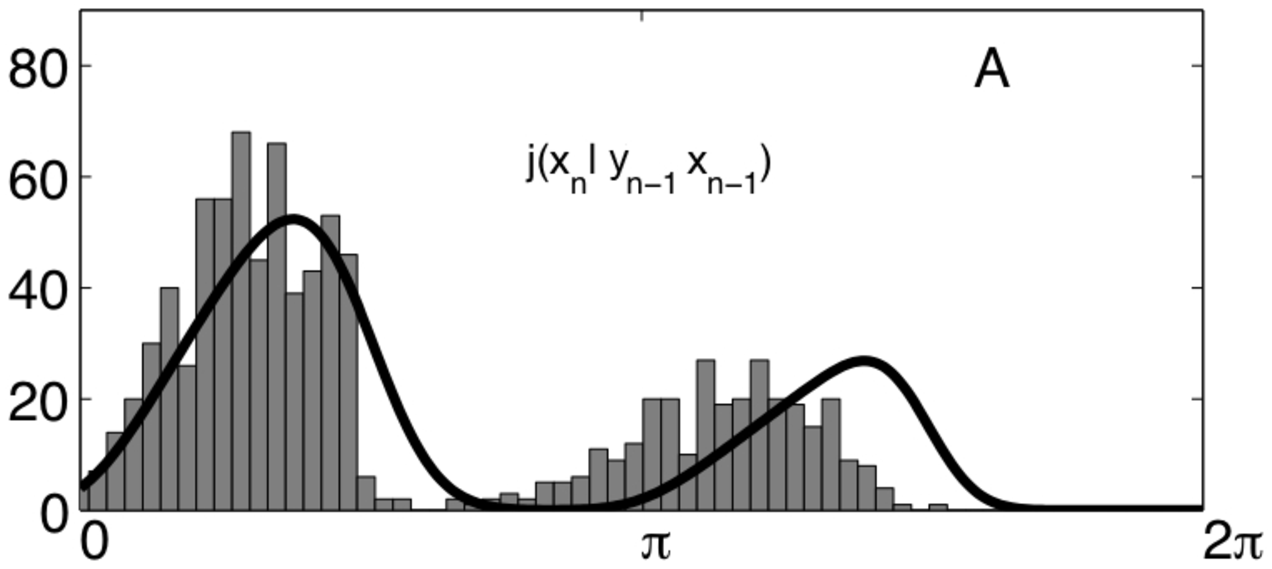}
\par\end{centering}

\begin{centering}
\includegraphics[width=250pt]{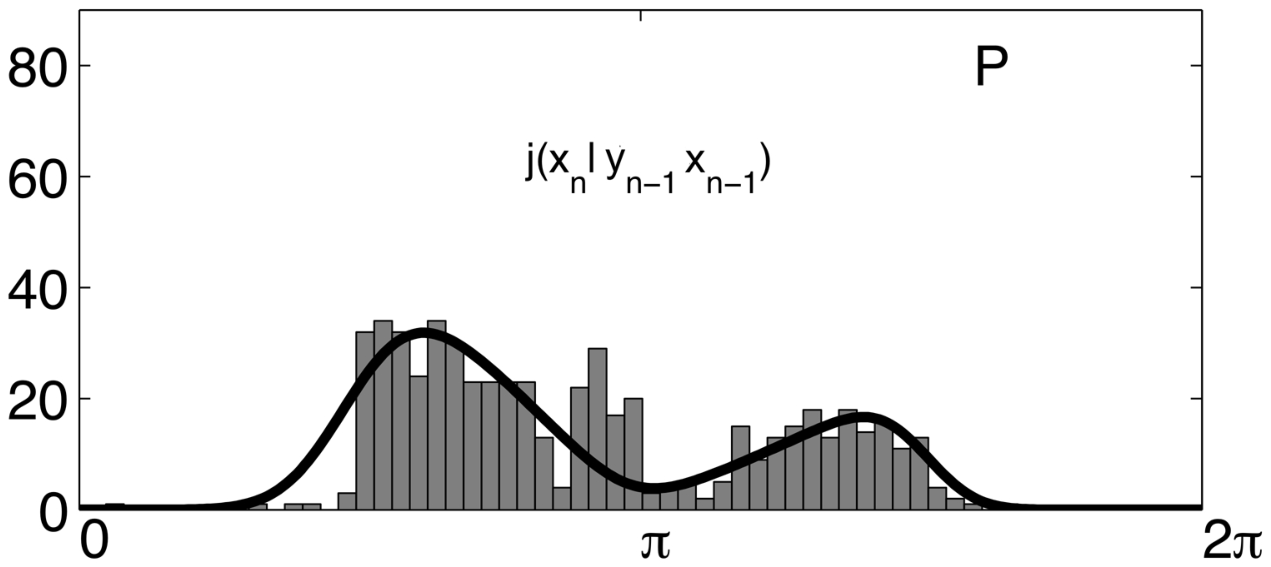}
\par\end{centering}

\caption{\label{fig:Simulated-response-histogramYX}Histograms of oscillators-simulated
responses conditioned to reinforcements and simulated responses on
the previous trial. All graphs correspond to simulated responses with
reinforcement on trial $n-1$ being in the interval $(\pi,3\pi/2)$.
$S$ corresponds to responses on trial $n-1$ occurring in the same
interval. $L$ corresponds to responses on $n-1$ occurring in the
interval $(3\pi/2,2\pi)$, $A$ in the interval $(0,\pi/2),$ and
$P$ in the interval $(\pi/2,\pi)$. The solid lines are the predictions
of the one-stimulus SR model.}
\end{figure}

Our main results are shown in Figures \ref{fig:Histogram-density}--\ref{fig:Simulated-response-histogramYX}.
The histograms were computed using the last 400 trials of the 600
trial simulated data for each of the 30 participants, for a total
of 12,000 sample points, and $b$ was reparametrized to correspond
to the interval $(0,2\pi)$. The parameters used are the same as those
described at the end of Section \ref{sec:Oscillator-Model-SR}, with
the exception of $K_{0}^{'}=4468$, chosen by using equation (\ref{eq:prob-eff-reinf})
such that the learning effectiveness of the oscillator model would
coincide with the observed value of $\theta=0.32$ in \citet{SuppesEtAl1964}.
Each oscillator's natural frequency was randomly selected according
to a Gaussian distribution centered on $10\mbox{ Hz}$, with variance
$1\mbox{ Hz}$. Figure 
\ref{fig:Histogram-density} shows the effects of the smearing distributions
on the oscillator-simulated responses. Figure 
\ref{fig:Histogram-density-Condition} shows the predicted conditional
response density $j(x_{n}|y_{n-1})$ if reinforcement on the previous
trial ($n-1$) happened in the interval $(\pi,3\pi/2)$. As predicted
by SR theory, the oscillator-simulated data also exhibit an asymmetric
histogram matching the predicted conditional density. We computed
from the simulated data the histograms of the conditional distributions
$j(x_{n}|y_{n-1}y_{n-2})$, shown in Figure 
\ref{fig:Second-order-conditional-probability}. Finally, if we use
our simulated data to generate histograms corresponding to the conditional
densities $j(x_{n}|y_{n-1}x_{n-1})$, we obtain a similar result to
that of the one-stimulus model (see Figure 
\ref{fig:Simulated-response-histogramYX}). We emphasize that, as
in the $N$-stimulus oscillator model used for the paired-associate
learning experiment, the continuum-of-response oscillator model can
be modified to yield data similar to the statistical predictions of
the SR model, if we add oscillators to represent additional stimuli.

Up to now we showed how the oscillator-simulated data compare to the
experiments in a tangential way, by fitting them to the predictions
of the SR theory, which we know fit the behavioral data rather well.
Now we examine how our model performed when we compare directly oscillator-simulated
and the behavioral empirical data. For the continuum of responses,
we focus on the data presented in Figures 2 and 5 of \citet{SuppesEtAl1964},
corresponding to our Figures \ref{fig:Histogram-density} and \ref{fig:Histogram-density-Condition}.
As the original data are not available anymore, to make a comparison
to our simulations we normalized the simulated-data histograms and
redrew them with the same number of bins as used in the corresponding
figures of \citet{SuppesEtAl1964}. We then performed a two-sample
Kolmogorov-Smirnov test to compare the empirical behavioral data with
the oscillator-simulated data, the null hypothesis being that the
two histograms originated from the same distribution \citep{keeping1995introduction}.
For the simulated response histogram of Figure \ref{fig:Histogram-density},
the Kolmogorov-Smirnov test yields a $p$-value of $0.53$. So, we
cannot reject the null hypothesis, that the simulated data come from
the same distribution as the corresponding behavioral data. For the
asymmetric conditional probability of Figure \ref{fig:Histogram-density-Condition},
the Kolmogorov-Smirnov test yields a $p$-value of $0.36$, once again
suggesting that the oscillator simulated data are statistically indistinguishable
from the empirical data.

\subsection{Experiment on probability matching}

The second experiment that we modeled with oscillators is a probability-matching
one, described in detail in \citet[Chapter 10]{SuppesAtkinson1960}.
The experiment consisted of 30 participants, who had to choose between
two possible behavioral responses, $R_{1}$ or $R_{2}$. A top light
on a panel would turn on to indicate to the participant the beginning
of a trial. Reinforcement was indicated by a light going on over the
correct response key.%
\footnote{The widely used notation, $R_{i}$ for response $i$ and $E_{j}$
for reinforcement $j$, for experiments with just two responses is
followed in describing the second and third experiments.%
} The correct response was noncontingent, with probability $\beta=0.6$
for reinforcing response $R_{1}$ and $0.4$ for $R_{2}$. The experiment
consisted of 240 trials for each participant.

To compare the oscillator-simulated data with the results for the
probability-matching experiment, we first computed, using the behavioral
data for the last 100 trials, the log pseudomaximum likelihood function
\citep[p. 206]{SuppesAtkinson1960} 
\[
L\left(\theta,N\right)=\sum_{i,j,k=1}^{2}n_{ij,k}\log P_{\infty}\left(R_{k,n+1}|E_{j,n}R_{i,n}\right),
\]
where $n_{ij,k}$ is the observed number of transitions from $R_{i}$
and $E_{j}$ on trial $n$ to $R_{k}$ on trial $n+1$, and the SR
theoretical conditional probabilities \textbf{$P_{\infty}\left(R_{k,n+1}|E_{j,n}R_{i,n}\right)$}
are: 

\begin{eqnarray}
P_{\infty}(R_{1,n+1}|E_{1,n}R_{1,n}) & = & \beta+\frac{1-\beta}{N},\label{eq:cond-1-1}\\
P_{\infty}(R_{1,n+1}|E_{1,n}R_{2,n}) & = & \beta\left(1-\frac{1}{N}\right)+\frac{\theta}{N},\label{eq:cond-2-1}\\
P_{\infty}(R_{1,n+1}|E_{2,n}R_{1,n}) & = & \beta\left(1-\frac{1}{N}\right)+\frac{1-\theta}{N},\label{eq:cond-3-1}\\
P_{\infty}(R_{1,n+1}|E_{2,n}R_{2,n}) & = & \beta\left(1-\frac{1}{N}\right).\label{eq:cond-4-1}
\end{eqnarray}
The log likelihood function $L\left(\theta,N\right)$, with $\beta=0.6$
as in the experiment, had maxima at $\hat{\theta}=0.600$ for $N=3$,
$\hat{\theta}=0.631$ for $N=4$, and $\hat{\theta}=0.611$ for $\hat{N}=3.35$,
where $\hat{N}$ is the log pseudomaximum likelihood estimate of $N$
(for more details, see \citet[Chapter 10]{SuppesAtkinson1960}).\textbf{
}We ran MATLAB oscillator simulations with the same parameters. We
computed 240 trials and 30 sets of oscillators, one for each participant,
for the three-, and four-stimulus oscillator models. Since the behavioral
learning parameter $\theta$ relates to $K'$, we used $K'=65\mbox{ s}^{-1}$
and $K'=56\,\mbox{s}^{-1}$ for the three-, and four-stimulus models,
respectively, corresponding to $\hat{\theta}=0.600$, and $\hat{\theta}=0.631$.
Table \ref{tab:prob-match} compares the experimentally observed conditional
relative frequencies with the predicted asymptotic values for the
$N=$3 or 4 SR models, and the corresponding oscillator simulations.
In Table \ref{tab:prob-match}, $P\left(R_{1,n+1}|E_{1,n}R_{1,n}\right)$
is the asymptotic conditional probability of response $R_{1}$ on
trial $n+1$ if on trial $n$ the response was $R_{1}$ and the reinforcement
was $E_{1}$, and similarly for the other three expressions, for each
of the conditions in Table 2, observed frequency, two oscillator simulations
and two behavioral predictions.

\begin{table}[h]
\caption{\label{tab:prob-match}Comparison between experimental results and
the corresponding oscillator models. Theoretical results for the $N$-stimulus
model are shown for comparison purposes. The first data row shows
experimental values based on the last 100 trials of the 30-subject
group. The other rows show the SR theory asymptotic probabilities
and simulations for the oscillator models averaged over the last 100
trials. The columns show the asymptotic conditional probabilities.
For example, column $R_{1}|E_{1}R_{1}$ shows the asymptotic probability
$P(R{}_{1}|E_{1}R_{1})$ of response $R_{1}$ on trial $n$ given
that on trial $n-1$ the response was $R_{1}$ and the reinforcement
was $E_{1}$. }

\centering{}%
\begin{tabular}{|c|c|c|c|c|}
\hline 
 & $R_{1}|E_{1}R_{1}$ & $R_{1}|E_{1}R_{2}$ & $R_{1}|E_{2}R_{1}$ & $R_{1}|E_{2}R_{2}$\tabularnewline
\hline 
\hline 
Observed & .715  & .602 & .535  & .413 \tabularnewline
\hline 
Oscillator & \multicolumn{1}{c}{} & \multicolumn{1}{c}{} & \multicolumn{1}{c}{} & \tabularnewline
\hline 
\begin{tabular}{c}
\textbf{3-stimulus}\tabularnewline
\textbf{simulation}\tabularnewline
\end{tabular} & .705  & .574 & .522  & .381 \tabularnewline
\hline 
\begin{tabular}{c}
\textbf{4-stimulus}\tabularnewline
\textbf{simulation}\tabularnewline
\end{tabular} & .760 & .610 & .538 & .426\tabularnewline
\hline 
SR-theory & \multicolumn{1}{c}{} & \multicolumn{1}{c}{} & \multicolumn{1}{c}{} & \tabularnewline
\hline 
\begin{tabular}{c}
\textbf{3-stimulus}\tabularnewline
\textbf{prediction}\tabularnewline
\end{tabular} & .733  & .600  & .533  & .400 \tabularnewline
\hline 
\begin{tabular}{c}
\textbf{4-stimulus}\tabularnewline
\textbf{prediction}\tabularnewline
\end{tabular} & .700 & .608 & .542  & .450 \tabularnewline
\hline 
\end{tabular}
\end{table}
In Table \ref{tab:prob-match} we compare the oscillator simulations
to the behavioral data and the best fitting asymptotic SR models.
Each oscillator model was computed using the same estimated $\theta$
value as the corresponding SR model. The fit of the SR model to the
empirical data is quite good, as is the fit for the 4-stimulus oscillator
model.

\subsection{Experiment on paired-associate learning\label{sub:Experiment-on-paired-associate}}

In this experiment, described in detail in \citet{Bower1961}, 29
participants learned a list of ten stimulus items to a criterion of
two consecutive errorless cycles. The order of stimulus presentation
was randomized for each trial. The visual stimuli were different pairs
of consonant letters; the numerical responses were the numbers 1 or
2, each number being assigned as correct to a randomly selected five
stimuli for each participant. A participant was informed of the correct
answer following each response. The one-stimulus model fitted the
behavioral data the best.

To model with oscillators, we used 29 independent sets of oscillators,
each set with ten-stimulus, two-response, and two reinforcements,
to reproduce Bower's setup.\textbf{ }For our oscillator simulations
we set $N=1$ and $K'=94\mbox{ s}^{-1}$, $\overline{K}_{0}=90\mbox{ Hz}$,
and $\sigma_{K_{0}}=10\mbox{ Hz }$. $K'$ was chosen based on (\ref{eq:prob-eff-reinf})
and on Bower's statistical estimation of the probability of effective
reinforcement for the one-stimulus model, which was $\theta=0.344$.

Bower's behavioral data were also tested in \citet{SuppesGinsberg1963}
for the statistical properties of stationarity and independence prior
to the last response error, a prediction of the one-stimulus SR model.
We compare in Table \ref{tab:Comparison-Bower} the behavioral results
with those for our oscillator simulation.

\textbf{Stationarity.} The first test is that of stationarity, i.e.,
whether the probability of a response oscillator phase-locking before
the last error is constant. To perform this test, we restricted our
oscillator-simulated data set to the $M$ responses that happened
\emph{before} the last behavioral error of a given participant. We
then divided this set into two, with the first $M/2$ trials being
early trials and the remaining $M/2$ late trials. Let $n_{1}$ be
the number of correct responses in the first half and $n_{2}$ the
number of correct responses in the second half. If the probability
is stationary, then both $n_{1}/(M/2)$ and $n_{2}/(M/2)$ should
be approximately $(n_{1}+n_{2})/M$. We used a standard $\chi^{2}$
test for this null hypothesis of stationarity for the oscillator-simulated
responses.

\textbf{Independence.} Restricting ourselves again to trials prior
to the last behavioral error, let $n_{ij}$ be the number of transitions
from state $i$ to state $j$, where $i$ and $j$ can take values
0 (correct response) or 1 (incorrect). We use these numbers to estimate
the transition probabilities $p_{ij}$. The null hypothesis of independence
is that $p_{00}=p_{10}$ and $p_{01}=p_{11}$. 
\begin{table}[h]
\caption{\label{tab:Comparison-Bower}Comparison between the paired-associate
experimental data and the simulated one-stimulus oscillator data.
Notice that $N$ is different for the experiment and the simulation,
as $N$ is the number of responses prior to the last error, which
varies for each run of the simulations. }

\begin{centering}
\begin{tabular}{|c|>{\centering}p{8.6pc}|}
\hline 
 & Stationarity\tabularnewline
\hline 
Experiment  & %
\begin{tabular}{c}
\multicolumn{1}{c|}{$\chi^{2}=.97$, $N=549$,}\tabularnewline
$df=6$, $p>.95$\tabularnewline
\end{tabular}\tabularnewline
\hline 
Oscillator simulation & %
\begin{tabular}{c}
$\chi^{2}=2.69$, $N=748$, \tabularnewline
\multicolumn{1}{c}{$df=6$, $p>.8$}\tabularnewline
\end{tabular}\tabularnewline
\hline 
\end{tabular}
\par\end{centering}

\vspace{4pt}

\centering{}%
\begin{tabular}{|c|>{\centering}p{8.6pc}|}
\hline 
 & Independence\tabularnewline
\hline 
Experiment & %
\begin{tabular}{c}
$\chi^{2}=.97$, $N=549$,\tabularnewline
\multicolumn{1}{c}{$df=6$, $p>.95$}\tabularnewline
\end{tabular}\tabularnewline
\hline 
Oscillator simulation & %
\begin{tabular}{c}
$\chi^{2}=0.1$, $N=458$, \tabularnewline
$df=6$, $p>.95$\tabularnewline
\end{tabular}\tabularnewline
\hline 
\end{tabular}
\end{table}
 The sample paths used to compute the oscillator results in Table
\ref{tab:Comparison-Bower} were obtained by running simulations in
MATLAB 7.1.0.183 (R14), SP 3 with the parameter values given earlier.
The simulations consisted of randomly selecting, at each trial, the
initial phase according to (\ref{eq:phase-density}) and then computing
phase evolution during $\Delta t_{r}$ by numerically solving (\ref{eq:Kuramoto-3-1-inhib-assym})--(\ref{eq:Kuramoto-3-3-inhib-assym})
using MATLAB's built-in fourth-fifth order Runge-Kutta method. After
$\Delta t_{r}$ the phase differences were computed and a response
oscillator selected. At the beginning of reinforcement, new initial
conditions and a random value for $K_{0}$ were drawn according to
(\ref{eq:phase-density}) and (\ref{eq:K0-density}), respectively.
If $K_{0}>K'$, then the oscillators' couplings changed according
to (\ref{eq:learningphaseS-inhib-excite-first})--(\ref{eq:learning-asym-inhibitory-last}),
and these equations were numerically solved for the time interval
$\Delta t_{e}$ also using fourth-fifth order Runge-Kutta, and the
new values for the couplings were used on the next trial. 

The results in Table \ref{tab:Comparison-Bower} show that we cannot
sharply differentiate the behavioral and the oscillator-simulation
data in testing the null hypotheses of stationarity and independence.

\section{Conclusions and Final Remarks\label{sec:Conclusions-and-Final}}

We modeled stimulus-response theory using neural oscillators based
on reasonable assumptions from neurophysiology. The effects of the
interaction between the neural oscillators were described by phase
differences obeying Kuramoto's equations. Learning was modeled by
changes in phases and couplings, driven by a reinforcement oscillator,
with phase locking corresponding to behavioral conditioning. We compared
the oscillator-simulated data to the data from three behavioral experiments,
as well as to the corresponding behavioral-model predictions derived
from SR theory. Our simulation results support the claim that neural
oscillators may be used to model stimulus-response theory, as well
as behavioral experiments testing the theory.

From (\ref{eq:prob-eff-reinf}), behavioral and oscillator-simulated
data relate the SR conditioning parameter $\theta$ to $K'$, $\sigma_{K_{0}}$,
and $\overline{K}_{0}$. Despite the large number of parameters, our
numerical oscillator simulations, all done with the same set of parameters,
except $N$ and $K'$, show that the statistical fit reported in Table
\ref{tab:Comparison-Bower} and conditional probabilities in Table
\ref{tab:prob-match} do not vary much even when these parameters
are changed significantly. The most important constraint is this.
Once all parameters are fixed then the stimulus-response conditioning
parameter $\theta$ is a nonlinear monotonically decreasing function
of $K'$ given by (\ref{eq:prob-eff-reinf}). As our oscillator-simulations
are refined and extended to more experiments, further constraints
should narrow the parameter space.

As is sometimes remarked, stimulus-response theory abstracts from
many necessary processes that must have a physical realization. From
a general psychological viewpoint, independent of the details of physical
realization, undoubtedly the most obvious missing process is the perceptual
recognition of the sampling of the same, or nearly the same stimulus,
and of the reinforcement stimulus patterns different on repeated trials.
We should be able to expand the present setup to include oscillators
that recognize, by similarity or congruence computations, when a sampled
stimulus type is one that has been previously sampled. Prior work
on pattern recognition will be useful here. Our own earlier experimental
work on computational models that recognize auditory or visual word,
sentence or other brain images, such as those of red triangles or
blue circles, provides substantial evidence that a single oscillator
with one given natural frequency will be inadequate. Multiple oscillators
with different natural frequencies will be required for such similarity
recognition. For examples of such computational models applied to
EEG-recorded brain data, see \citet{SuppesLuHan1997,SuppesBrain1998,SuppesBrain1999b,SuppesBrain1999}
for Fourier methods, \citet{deBarrosEtAl2006} for a Laplacian model,
\citet{Suppes2009Partial} for congruence between brain and perceptual
feature of language, and \citet{WongEtAl2006} for perceptron models
with regularization and independent component analysis. 

The oscillator computations developed in this paper provide a schematic
model of psychological processes successfully described behaviorally
using stimulus-response theory. The coupled neural phase oscillators
used offer a physical mechanism to explain schematically how conditioning
works in the brain. But many important physical details of such processes
remain to be clarified. The oscillator-simulations do suggest many
physically and biologically relevant measurements that are not part
of stimulus-response theory. For example, although we used the simplifying
assumption that a response occurs at a time $\Delta t_{r}$ after
the onset of the stimulus, a more detailed model should derive a response-time
distribution variance, with mean and variance at least being derived
from biological assumptions about various physical processes, such
as firing thresholds being reached due to oscillator synchronization.
Another feature that could be further explored is the interference
between oscillators. As we saw in Section \ref{sec:Oscillator-Model-SR},
neural oscillators may interfere in ways similar to two electrical
oscillators or two wave sources. This interference may lead to predictions
distinct from those of classical SR theory, when modeling complex
processes. Such predictions could, perhaps, be closer to behavioral
models suggested by very different approaches (see \citep{Busemeyer2006,BruzaBusemeyerGabora2009}
for possible alternatives). 

Finally, we emphasize that neural oscillators may produce, in principle,
measurable electrophysiological signals. Prior to having such measurements
available as psychology and system neuroscience draw closer together,
it is desirable that specific, even if only schematic, physical mechanisms
be proposed to provide conceptually plausible physical realizations
of fundamental psychological processes. The most important idea tested
here is that the main physical mechanism in stimulus-response learning
is phase-locking, a concept widely used in physics, but not in psychology,
or, even as yet, not much in system neuroscience. 

\appendix
\numberwithin{equation}{section}

\section{Properties of the Oscillator Model\label{App:Properties-Oscillator-Model}}

In this appendix we derive many important properties of the oscillator
model. In Section \ref{subap:Fixed-points}, we show that the convergence
values for the reinforced couplings lead to some fixed points for
equations (\ref{eq:Kuramoto-3-1-inhib-assym})--(\ref{eq:Kuramoto-3-3-inhib-assym})
that correspond to the desired phase differences. Section \ref{subap:Stability-Fixed-Points}
examines the different fixed points, and verifies that the couplings
converge to couplings such that the only stable fixed points are those
corresponding to the reinforced phase difference $\delta\varphi$.
Section \ref{subap:General-sufficient-criteria} extends the results
of Section \ref{subap:Stability-Fixed-Points} to the case of different
coupling strengths, and establishes a more general condition for the
stability of the fixed point corresponding to $\delta\varphi$ based
on the relative signs of the couplings. This result is relevant, as
the learning equations do not guarantee the exact values of couplings,
but they can guarantee, when learning is effective, that couplings
do satisfy the conditions presented in Section \ref{subap:General-sufficient-criteria}.
In Section \ref{subap:Time-of-convergence} we analyze the relationship
between coupling strength and the convergence time of equations (\ref{eq:Kuramoto-3-1-inhib-assym})--(\ref{eq:Kuramoto-3-3-inhib-assym})
toward a fixed point. Section \ref{subap:Synchronization} investigates
the behavior of the oscillators during reinforcement. Finally, in
Section \ref{subap:Effectiveness-Reinforcement} we show how some
of the model parameters relate to the effectiveness of reinforcement
$\theta$, deriving equation (\ref{eq:prob-eff-reinf}).

\subsection{Fixed points for coding phase relations\label{subap:Fixed-points}}

Here we derive the asymptotic couplings for the reinforcement angles
coding the phase relation $\delta\varphi$ detailed in Section \ref{sec:Oscillator-Model-SR},
and then show that such couplings lead to a set of differential equations
with fixed points that are stable around solutions corresponding to
the desired phase differences (\ref{eq:ideal-phase-diff}). First,
let us start with the reinforcement schedule set such that $\varphi_{s}=\omega_{e}t,$
$\varphi_{r_{1}}=\omega_{e}t+\delta\varphi,$ and $\varphi_{r_{2}}=\omega_{e}t+\delta\varphi-\pi$.
From (\ref{eq:learningphaseS-inhib-excite-first})--(\ref{eq:learningphaseS-inhib-exite-last})
we obtain the following fixed points for excitatory connections, which
work as an attractor for the asymptotic behavior (for more details,
see \ref{subap:Synchronization}). 
\begin{eqnarray}
k_{s,r_{1}}^{E} & = & \alpha\cos\left(\delta\varphi\right),\label{eq:coupling-asymp-delta-excite-1-A}\\
k_{s,r_{2}}^{E} & = & -\alpha\cos\left(\delta\varphi\right),\\
k_{r_{1},r_{2}}^{E} & = & -\alpha,\\
k_{r_{1},s}^{E} & = & \alpha\cos\left(\delta\varphi\right),\\
k_{r_{2},s}^{E} & = & -\alpha\cos\left(\delta\varphi\right),\\
k_{r_{2},r_{1}}^{E} & = & -\alpha.\label{eq:coupling-asymp-delta-excite-6-A}
\end{eqnarray}
 Similarly, for inhibitory connections we have. 
\begin{eqnarray}
k_{s,r_{1}}^{I} & = & -\alpha\sin\left(\delta\varphi\right),\label{eq:coupling-asymp-delta-inhibit-1-A}\\
k_{s,r_{2}}^{I} & = & \alpha\sin\left(\delta\varphi\right),\\
k_{r_{1},r_{2}}^{I} & = & 0,\\
k_{r_{1},s}^{I} & = & \alpha\sin\left(\delta\varphi\right),\\
k_{r_{2},s}^{I} & = & -\alpha\sin\left(\delta\varphi\right),\\
k_{r_{2},r_{1}}^{I} & = & 0.\label{eq:coupling-asymp-delta-inhibit-6-A}
\end{eqnarray}
Assuming that a sufficient time interval elapsed such that the coupling
coefficients converged to (\ref{eq:coupling-asymp-delta-excite-1-A})--(\ref{eq:coupling-asymp-delta-inhibit-6-A}),
Kuramoto's equations become 
\begin{eqnarray*}
\frac{d\varphi_{s}}{dt} & = & \omega_{0}-\alpha\cos\left(\delta\varphi\right)\sin\left(\varphi_{s}-\varphi_{r_{1}}\right)\\
 &  & +\alpha\cos\left(\delta\varphi\right)\sin\left(\varphi_{s}-\varphi_{r_{2}}\right)\\
 &  & +\alpha\sin\left(\delta\varphi\right)\cos\left(\varphi_{s}-\varphi_{r_{1}}\right)\\
 &  & -\alpha\sin\left(\delta\varphi\right)\cos\left(\varphi_{s}-\varphi_{r_{2}}\right),\\
\frac{d\varphi_{r_{1}}}{dt} & = & \omega_{0}-\alpha\cos\left(\delta\varphi\right)\sin\left(\varphi_{r_{1}}-\varphi_{s}\right)\\
 &  & +\alpha\sin\left(\varphi_{r_{1}}-\varphi_{r_{2}}\right)\\
 &  & -\alpha\sin\left(\delta\varphi\right)\cos\left(\varphi_{r_{1}}-\varphi_{s}\right),\\
\frac{d\varphi_{r_{2}}}{dt} & = & \omega_{0}+\alpha\cos\left(\delta\varphi\right)\sin\left(\varphi_{r_{2}}-\varphi_{s}\right)\\
 &  & +\alpha\sin\left(\varphi_{r_{2}}-\varphi_{r_{1}}\right)\\
 &  & +\alpha\sin\left(\delta\varphi\right)\cos\left(\varphi_{r_{2}}-\varphi_{s}\right).
\end{eqnarray*}
 Regrouping the terms with the same phase oscillators, we have 
\begin{eqnarray*}
\frac{d\varphi_{s}}{dt} & = & \omega_{0}-\alpha\sin\left(\varphi_{s}-\varphi_{r_{1}}+\delta\varphi\right)\\
 &  & +\alpha\sin\left(\varphi_{s}-\varphi_{r_{2}}+\delta\varphi\right),\\
\frac{d\varphi_{r_{1}}}{dt} & = & \omega_{0}-\alpha\sin\left(\varphi_{r_{1}}-\varphi_{s}-\delta\varphi\right)\\
 &  & +\alpha\sin\left(\varphi_{r_{1}}-\varphi_{r_{2}}\right),\\
\frac{d\varphi_{r_{2}}}{dt} & = & \omega_{0}+\alpha\sin\left(\varphi_{r_{2}}-\varphi_{s}-\delta\varphi\right)\\
 &  & +\alpha\sin\left(\varphi_{r_{2}}-\varphi_{r_{1}}\right),
\end{eqnarray*}
 or
\begin{eqnarray}
\frac{d\varphi_{s}}{dt} & = & \omega_{0}-\alpha\sin\left(\varphi_{s}-\varphi_{r_{1}}+\delta\varphi\right)\nonumber \\
 &  & -\alpha\sin\left(\varphi_{s}-\varphi_{r_{2}}+\delta\varphi-\pi\right),\label{eq:kura-couple-phase-1-A}\\
\frac{d\varphi_{r_{1}}}{dt} & = & \omega_{0}-\alpha\sin\left(\varphi_{r_{1}}-\varphi_{s}-\delta\varphi\right)\nonumber \\
 &  & -\alpha\sin\left(\varphi_{r_{1}}-\varphi_{r_{2}}-\pi\right),\label{eq:kura-couple-phase-2-A}\\
\frac{d\varphi_{r_{2}}}{dt} & = & \omega_{0}-\alpha\sin\left(\varphi_{r_{2}}-\varphi_{s}-\delta\varphi+\pi\right)\nonumber \\
 &  & -\alpha\sin\left(\varphi_{r_{2}}-\varphi_{r_{1}}+\pi\right).\label{eq:kura-couple-phase-3-A}
\end{eqnarray}
It is straightforward to verify that equations (\ref{eq:kura-couple-phase-1-A})--(\ref{eq:kura-couple-phase-3-A})
have the desired phase relations as fixed points. In the next section,
we will show that not only are those fixed points stable, but that
they are the only stable points modulo a $2\pi$ transformation (which
in our case is not relevant).

\subsection{Stability of Fixed Points\label{subap:Stability-Fixed-Points}}

In \ref{subap:Fixed-points}, we showed that a specific choice of
couplings in the Kuramoto equations leads to a specific phase relation
between oscillators. Here we derive the stability conditions. We assume
that the couplings are such as required by phase differences $\varphi_{s}=\varphi_{r_{1}}-\delta\varphi$
and $\varphi_{r_{1}}=\varphi_{r_{2}}+\pi$, as coded in couplings
(\ref{eq:coupling-asymp-delta-excite-1-A})--(\ref{eq:coupling-asymp-delta-inhibit-6-A}).
To simplify our approach, we start with equations (\ref{eq:kura-couple-phase-1-A})--(\ref{eq:kura-couple-phase-3-A})
and make the following change of variables:
\begin{align*}
\phi_{s} & =\varphi_{s}+\delta\varphi,\\
\phi_{1} & =\varphi_{r_{1}},\\
\phi_{2} & =\varphi_{r_{2}}+\pi.
\end{align*}
Equations (\ref{eq:kura-couple-phase-1-A})--(\ref{eq:kura-couple-phase-3-A})
then become
\[
\]
\begin{eqnarray}
\frac{d\phi_{s}}{dt} & = & \omega_{0}-\alpha\sin\left(\phi_{s}-\phi_{1}\right)\nonumber \\
 &  & -\alpha\sin\left(\phi_{s}-\phi_{2}\right),\label{eq:kura-phase-newvar-S-A}\\
\frac{d\phi_{1}}{dt} & = & \omega_{0}-\alpha\sin\left(\phi_{1}-\phi_{s}\right)\nonumber \\
 &  & -\alpha\sin\left(\phi_{1}-\phi_{2}\right),\label{eq:kura-phase-newvar-1-A}\\
\frac{d\phi_{2}}{dt} & = & \omega_{0}-\alpha\sin\left(\phi_{2}-\phi_{s}\right)\nonumber \\
 &  & -\alpha\sin\left(\phi_{2}-\phi_{1}\right).\label{eq:kura-phase-newvar-2-A}
\end{eqnarray}
This change of variables allow us to re-write the dynamical equations
without any references to the specific phase differences, and therefore
we can obtain general results for any phase differences consistent
with the desired relations for conditioning. Additionally, since we
are mainly interested in phase differences, we make yet another change
of variables, further simplifying our equations, by defining 
\begin{align*}
\phi_{1s} & =\phi_{1}-\phi_{s},\\
\phi_{2s} & =\phi_{2}-\phi_{s}.
\end{align*}
Substituting in (\ref{eq:kura-phase-newvar-S-A})--(\ref{eq:kura-phase-newvar-2-A}),
we reduce the system to the following pair of coupled differential
equations.
\begin{eqnarray}
\frac{d\phi_{1s}}{dt} & = & -2\alpha\sin\left(\phi_{1s}\right)-\alpha\sin\left(\phi_{2s}\right)\label{eq:kuramoto-1-2}\\
 &  & -\alpha\sin\left(\phi_{1s}-\phi_{2s}\right),\nonumber \\
\frac{d\phi_{2s}}{dt} & = & -\alpha\sin\left(\phi_{1s}\right)-2\alpha\sin\left(\phi_{2s}\right)\label{eq:kuramoto-2-2}\\
 &  & +\alpha\sin\left(\phi_{1s}-\phi_{2s}\right).\nonumber 
\end{eqnarray}
Because we are interested in general properties of stability, and
we want to understand what conditions lead to the stable desired fixed
points, it is convenient to not restrict our couplings to the ones
imposed by a specific phase relation. Instead, we rewrite the above
differential equations in terms of couplings that are not necessarily
all the same, i.e., 
\[
\]

\begin{eqnarray}
\frac{d\phi_{1s}}{dt} & = & -2k_{s,r_{1}}\sin\left(\phi_{1s}\right)-k_{s,r_{2}}\sin\left(\phi_{2s}\right)\label{eq:kuramoto-1}\\
 &  & -k_{r_{1},r_{2}}\sin\left(\phi_{1s}-\phi_{2s}\right),\nonumber \\
\frac{d\phi_{2s}}{dt} & = & -k_{s,r_{1}}\sin\left(\phi_{1s}\right)-2k_{s,r_{2}}\sin\left(\phi_{2s}\right)\label{eq:kuramoto-2}\\
 &  & +k_{r_{1},r_{2}}\sin\left(\phi_{1s}-\phi_{2s}\right).\nonumber 
\end{eqnarray}
The only constraint we will make later on in our analysis is that
such couplings need to have the same absolute value, a requirement
consistent with asymptotic behavior of the learning equations (see
\ref{subap:Synchronization}). Right now, the question we want to
answer is the following. Are there stable solutions to (\ref{eq:kuramoto-1})\textbf{
}and\textbf{ }(\ref{eq:kuramoto-2})? If so, how do they depend on
the couplings?

To answer the above questions, we should first compute whether there
are any fixed points for differential equations (\ref{eq:kuramoto-1})\textbf{
}and\textbf{ }(\ref{eq:kuramoto-2}). Since a fixed point is a point
where the system remains if initially placed in it, they are characterized
by having both time derivatives of $\phi_{1j}$ and $\phi_{2j}$ equal
to zero. Thus, the fixed points are given by 
\begin{eqnarray}
-2k_{s,r_{1}}\sin\left(\phi_{1s}\right)-k_{s,r_{2}}\sin\left(\phi_{2s}\right)\label{eq:fixed-1}\\
-k_{r_{1},r_{2}}\sin\left(\phi_{1s}-\phi_{2s}\right) & = & 0,\nonumber \\
-k_{s,r_{1}}\sin\left(\phi_{1s}\right)-2k_{s,r_{2}}\sin\left(\phi_{2s}\right)\label{eq:fixed-2}\\
+k_{r_{1},r_{2}}\sin\left(\phi_{1s}-\phi_{2s}\right) & = & 0.\nonumber 
\end{eqnarray}
 Because of the periodicity of the sine function, there are infinitely
many fixed points. Since we are interested in phases, we do not need
to examine all points, but it suffices to focus on the region determined
by $0\leq\phi_{1s}<2\pi$ and $0\leq\phi_{2s}<2\pi$. The trivial
fixed points, corresponding to each sine function in (\ref{eq:fixed-1})\textbf{
}and\textbf{ }(\ref{eq:fixed-2}) being zero, are $(0,0)$, $(\pi,0)$,
$(0,\pi)$, and $(\pi,\pi)$. The other solutions depend on the values
of the coupling constants. To simplify our analysis, let us consider
the case where all couplings have the same magnitude, i.e. $\left|k_{s,r_{1}}\right|=\left|k_{r_{1},r_{2}}\right|=\left|k_{s,r_{2}}\right|$.
Since the relative coupling strengths are relevant for matters of
stability, let us consider the following scenarios: $k\equiv k_{s,r_{1}}=k_{s,r_{2}}=k_{r_{1},r_{2}}$,
$k\equiv-k_{s,r_{1}}=k_{s,r_{2}}=k_{r_{1},r_{2}}$, $k\equiv k_{s,r_{1}}=-k_{s,r_{2}}=k_{r_{1},r_{2}}$,
and $k\equiv k_{s,r_{1}}=k_{s,r_{2}}=-k_{r_{1},r_{2}}$. Other cases,
like $-k_{s,r_{1}}=-k_{s,r_{2}}=k_{r_{1},r_{2}}$, are contained in
those, since $k$ can be either positive or negative. Let us examine
each case separately.

\subsubsection{$k\equiv k_{s,r_{1}}=k_{s,r_{2}}=k_{r_{1},r_{2}}$\label{subsubap:kkk}}

For this case, in addition to $\{(0,0),(0,\pi),(\pi,0),(\pi,\pi)\}$,
we have $(\frac{2}{3}\pi,-\frac{2}{3}\pi)$ and $(-\frac{2}{3}\pi,\frac{2}{3}\pi)$
as fixed points. Let us start with $\{(0,0),(0,\pi),(\pi,0),(\pi,\pi)\}$.
In order to analyze their local stability, we linearize Kuramoto's
equations (\ref{eq:kuramoto-1})\textbf{ }and (\ref{eq:kuramoto-2})
around $\{(0,0),(0,\pi),(\pi,0),(\pi,\pi)\}$. To linearize the equations,
first we substitute $\phi_{1s}$ and $\phi_{2s}$ by $a+\epsilon_{1}$
and $b+\epsilon_{2}$, where $a$ and $b$ are the coordinates of
the fixed points. Then we make a linear approximation for the sine
function for values close to the fixed point, i.e. for $\epsilon_{1}$
and $\epsilon_{2}$ being small. Below we show the linearized equations
for different fixed points $(a,b)$.

For example, for $(0,\pi)$ we define $\phi_{1s}=\epsilon_{1}$ and
$\phi_{2s}=\pi+\epsilon_{2}$. $ $ Then (\ref{eq:kuramoto-1})\textbf{
}and (\ref{eq:kuramoto-2})\textbf{ }become 
\begin{eqnarray}
\frac{d\epsilon_{1}}{dt} & = & -2k\sin\left(\epsilon_{1}\right)-k\sin\left(\pi+\epsilon_{2}\right)\label{eq:kurafix1}\\
 &  & -k\sin\left(\epsilon_{1}-\epsilon_{2}-\pi\right),\nonumber \\
\frac{d\epsilon_{2}}{dt} & = & -k\sin\left(\epsilon_{1}\right)-2k\sin\left(\pi+\epsilon_{2}\right)\label{eq:kurafix2}\\
 &  & +k\sin\left(\epsilon_{1}-\epsilon_{2}-\pi\right).\nonumber 
\end{eqnarray}
 Because close to the fixed points $\epsilon_{1}\ll1$ and $\epsilon_{2}\ll1$,
we make the approximation that $\sin(\epsilon_{1})\approx\epsilon_{1}$
and $\sin(\epsilon_{2})\approx\epsilon_{2}$ , and (\ref{eq:kurafix1})\textbf{
}and (\ref{eq:kurafix2})\textbf{ }can be written as
\begin{eqnarray}
\frac{d\epsilon_{1}}{dt} & = & -2k\epsilon_{1}+k\epsilon_{2}+k\left(\epsilon_{1}-\epsilon_{2}\right),\label{eq:linear-A}\\
\frac{d\epsilon_{2}}{dt} & = & -k\epsilon_{1}+2k\epsilon_{2}-k\left(\epsilon_{1}-\epsilon_{2}\right),\label{eq:linear-2-A}
\end{eqnarray}
 or 
\begin{eqnarray}
\frac{d\epsilon_{1}}{dt} & = & -k\epsilon_{1},\label{eq:linear-1}\\
\frac{d\epsilon_{2}}{dt} & = & -2k\epsilon_{1}+3k\epsilon_{2}.\label{eq:linear-2}
\end{eqnarray}
The eigenvalues for the linearized (\ref{eq:linear-1})\textbf{ }and
(\ref{eq:linear-2})\textbf{ }are $-k$ and $3k$. A fixed point is
Liapunov stable if its eigenvalues are negative \citep{GuckenheimerHolmes1983}.
Therefore, for $k\equiv k_{s,r_{1}}=k_{s,r_{2}}=k_{r_{1},r_{2}}$,
the fixed point $(0,\pi)$ cannot be stable.

We perform the same computations for the other fixed points. The linearized
equations are 
\begin{eqnarray}
\frac{d\epsilon_{1}}{dt} & = & -3k\epsilon_{1},\label{eq:kkk-me-ne-1}\\
\frac{d\epsilon_{2}}{dt} & = & -3k\epsilon_{2},\label{eq:kkk-me-ne-2}
\end{eqnarray}
 for $(0,0)$, 
\begin{eqnarray}
\frac{d\epsilon_{1}}{dt} & = & -k\epsilon_{1},\\
\frac{d\epsilon_{2}}{dt} & = & -2k\epsilon_{1}+3k\epsilon_{2},
\end{eqnarray}
 once again, for $(0,\pi)$, 
\begin{eqnarray}
\frac{d\epsilon_{1}}{dt} & = & 3k\epsilon_{1}-2k\epsilon_{2},\label{eq:eqdup1}\\
\frac{d\epsilon_{2}}{dt} & = & -k\epsilon_{2},\label{eq:eqdup2}
\end{eqnarray}
 for $(\pi,0)$, and 
\begin{eqnarray}
\frac{d\epsilon_{1}}{dt} & = & k\epsilon_{1}+2k\epsilon_{2},\label{eq:kkk-me-no-1}\\
\frac{d\epsilon_{2}}{dt} & = & 2k\epsilon_{1}+k\epsilon_{2},\label{eq:kkk-mo-no-2}
\end{eqnarray}
 for $(\pi,\pi)$. Only (\ref{eq:kkk-me-ne-1}) and (\ref{eq:kkk-me-ne-2})
have negative eigenvalues, when $k>0$, thus corresponding to stable
fixed points.

Now let us examine fixed points $(\frac{2}{3}\pi,-\frac{2}{3}\pi)$
and $(-\frac{2}{3}\pi,\frac{2}{3}\pi).$ Substituting $\phi_{1s}=\pm\frac{2}{3}\pi+\epsilon_{1}\left(t\right)$
and $\phi_{2s}=\mp\frac{2}{3}\pi+\epsilon_{2}\left(t\right)$ in (\ref{eq:kuramoto-1})\textbf{
}and\textbf{ }(\ref{eq:kuramoto-2}), we have 
\begin{eqnarray}
\frac{1}{k}\frac{d\epsilon_{1}}{dt} & = & -2\sin\left(\pm\frac{2}{3}\pi+\epsilon_{1}\right)-\sin\left(\mp\frac{2}{3}\pi+\epsilon_{2}\right)\nonumber \\
 &  & -\sin\left(\pm\frac{2}{3}\pi+\epsilon_{1}\left(t\right)\pm\frac{2}{3}\pi-\epsilon_{2}\right),\\
\frac{1}{k}\frac{d\epsilon_{2}}{dt} & = & -\sin\left(\pm\frac{2}{3}\pi+\epsilon_{1}\right)-2\sin\left(\mp\frac{2}{3}\pi+\epsilon_{2}\right)\nonumber \\
 &  & +\sin\left(\pm\frac{2}{3}\pi+\epsilon_{1}\left(t\right)\pm\frac{2}{3}\pi-\epsilon_{2}\right).
\end{eqnarray}
 Using the linear approximation for $\epsilon_{1}\ll1$ and $\epsilon_{2}\ll1$
we get 
\begin{eqnarray}
\frac{1}{k}\frac{d\epsilon_{1}}{dt} & = & \mp2\sin\left(\frac{2}{3}\pi\right)-2\cos\left(\frac{2}{3}\pi\right)\epsilon_{1}\nonumber \\
 &  & \pm\sin\left(\frac{2}{3}\pi\right)-\cos\left(\frac{2}{3}\pi\right)\epsilon_{2}\nonumber \\
 &  & \mp\sin\left(\frac{4}{3}\pi\right)\nonumber \\
 &  & -\cos\left(\frac{4}{3}\pi\right)\left(\epsilon_{1}-\epsilon_{2}\right),\\
\frac{1}{k}\frac{d\epsilon_{2}}{dt} & = & \mp\sin\left(\frac{2}{3}\pi\right)-\cos\left(\frac{2}{3}\pi\right)\epsilon_{1}\nonumber \\
 &  & \pm2\sin\left(\frac{2}{3}\pi\right)-2\cos\left(\frac{2}{3}\pi\right)\epsilon_{2}\nonumber \\
 &  & \pm\sin\left(\frac{4}{3}\pi\right)\nonumber \\
 &  & +\cos\left(\frac{4}{3}\pi\right)\left(\epsilon_{1}-\epsilon_{2}\right).
\end{eqnarray}
 or 
\begin{eqnarray}
\frac{1}{k}\frac{d\epsilon_{1}}{dt} & = & \frac{3}{2}\epsilon_{1},\\
\frac{1}{k}\frac{d\epsilon_{2}}{dt} & = & \frac{3}{2}\epsilon_{2}.
\end{eqnarray}
 This point is stable if $k<0.$ Thus, for $k\equiv k_{s,r_{1}}=k_{s,r_{2}}=k_{r_{1},r_{2}}$,
there are three stable fixed points: $(0,0)$, $(\frac{2}{3}\pi,-\frac{2}{3}\pi)$,
and $(-\frac{2}{3}\pi,\frac{2}{3}\pi)$. However, if $k>0$ the only
fixed point that is stable is $(0,0)$, corresponding to the situation
when all oscillators are exactly in phase.

\subsubsection{$k\equiv-k_{s,r_{1}}=k_{s,r_{2}}=k_{r_{1},r_{2}}$\label{subsubap:-kkk}}

Now the fixed points are $(0,0),$ $(0,\pi),$ $(\pi,0)$, $(\pi,\pi)$,
$(\frac{2}{3}\pi,\frac{1}{3}\pi)$, and $(-\frac{2}{3}\pi,-\frac{1}{3}\pi)$.
The linearized equations are 
\begin{eqnarray}
\frac{d\epsilon_{1}}{dt} & = & k\epsilon_{1},\label{eq:-kkk-me-ne-1}\\
\frac{d\epsilon_{2}}{dt} & = & 2k\epsilon_{1}-3k\epsilon_{2},\label{eq:-kkk-me-ne-2}
\end{eqnarray}
 for $(0,0)$, 
\begin{eqnarray}
\frac{d\epsilon_{1}}{dt} & = & 3k\epsilon_{1},\label{eq:-kkk-me-no-1}\\
\frac{d\epsilon_{2}}{dt} & = & 3k\epsilon_{2},\label{eq:-kkk-me-no-2}
\end{eqnarray}
 for $(0,\pi)$, 
\begin{eqnarray}
\frac{d\epsilon_{1}}{dt} & = & -k\epsilon_{1}-2k\epsilon_{2},\label{eq:-kkk-mo-ne-1}\\
\frac{d\epsilon_{2}}{dt} & = & -2k\epsilon_{1}-k\epsilon_{2},\label{eq:-kkk-mo-ne-2}
\end{eqnarray}
 for $(\pi,0)$, and 
\begin{eqnarray}
\frac{d\epsilon_{1}}{dt} & = & -3k\epsilon_{1}+2k\epsilon_{2},\\
\frac{d\epsilon_{2}}{dt} & = & k\epsilon_{2},\label{eq:-kkk-mo-no-2}
\end{eqnarray}
 for $(\pi,\pi)$. Only (\ref{eq:-kkk-me-no-1}) and (\ref{eq:-kkk-me-no-2})
correspond to fixed points that can be stable, if $k<0$, as there
are no other choices of $k$ that would allow the other points to
be stable.

For fixed points $(\frac{2}{3}\pi,\frac{1}{3}\pi)$ and $(-\frac{2}{3}\pi,-\frac{1}{3}\pi)$,
we substitute $\phi_{1s}=\pm\frac{2}{3}\pi+\epsilon_{1}\left(t\right)$
and $\phi_{2s}=\pm\frac{1}{3}\pi+\epsilon_{2}\left(t\right)$ in (\ref{eq:kuramoto-1})\textbf{
}and\textbf{ }(\ref{eq:kuramoto-2}), and we have 
\begin{eqnarray}
\frac{d\epsilon_{1}}{dt} & = & 2k\sin\left(\pm\frac{2}{3}\pi+\epsilon_{1}\right)-k\sin\left(\pm\frac{1}{3}\pi+\epsilon_{2}\right)\nonumber \\
 &  & -k\sin\left(\pm\frac{2}{3}\pi\mp\frac{1}{3}\pi+\epsilon_{1}-\epsilon_{2}\right),\\
\frac{d\epsilon_{2}}{dt} & = & k\sin\left(\pm\frac{2}{3}\pi+\epsilon_{1}\right)-2k\sin\left(\pm\frac{1}{3}\pi+\epsilon_{2}\right)\nonumber \\
 &  & +k\sin\left(\pm\frac{2}{3}\pi\mp\frac{1}{3}\pi+\epsilon_{1}-\epsilon_{2}\right).
\end{eqnarray}
 Using the linear approximation for $\epsilon_{1}\ll1$ and $\epsilon_{2}\ll1$
we get 
\begin{eqnarray}
\frac{1}{k}\frac{d\epsilon_{1}}{dt} & = & -\frac{3}{2}\epsilon_{1},\label{eq:linear-kuramoto-1}\\
\frac{1}{k}\frac{d\epsilon_{2}}{dt} & = & -\frac{3}{2}\epsilon_{2}.\label{eq:linear-kuramoto-w}
\end{eqnarray}
 Points $(\frac{2}{3}\pi,\frac{1}{3}\pi)$ and $(-\frac{2}{3}\pi,-\frac{1}{3}\pi)$
are stable if $k>0.$ Thus, for $k\equiv-k_{s,r_{1}}=k_{s,r_{2}}=k_{r_{1},r_{2}}$,
there are three relevant stable fixed points: $(0,\pi)$, $(\frac{2}{3}\pi,\frac{1}{3}\pi)$
and $(-\frac{2}{3}\pi,-\frac{1}{3}\pi)$. However, if $k<0$ the only
fixed point that is stable is $(0,\pi)$, corresponding to the situation
when the oscillators $s$ and $r_{1}$ are in phase and $s$ and $r_{2}$
are off phase by $\pi$.

\subsubsection{$k\equiv k_{s,r_{1}}=-k_{s,r_{2}}=k_{r_{1},r_{2}}$\label{subsubap:k-kk}}

For the points $\{(0,0),(0,\pi),(\pi,0),(\pi,\pi)\}$ the linearized
form of (\ref{eq:linear-kuramoto-1})\textbf{ }and (\ref{eq:linear-kuramoto-w})
around the stability points are 
\begin{eqnarray}
\frac{d\epsilon_{1}}{dt} & = & -3k\epsilon_{1}+2k\epsilon_{2},\label{eq:k-kk-me-ne-1}\\
\frac{d\epsilon_{2}}{dt} & = & k\epsilon_{2},\label{eq:k-kk-me-ne-2}
\end{eqnarray}
 for $(0,0)$, 
\begin{eqnarray}
\frac{d\epsilon_{1}}{dt} & = & -k\epsilon_{1}-2k\epsilon_{2},\label{eq:k-kk-me-no-1}\\
\frac{d\epsilon_{2}}{dt} & = & -2k\epsilon_{1}-k\epsilon_{2},
\end{eqnarray}
 for $(0,\pi)$, 
\begin{eqnarray}
\frac{d\epsilon_{1}}{dt} & = & 3k\epsilon_{1},\label{eq:k-kk-mo-ne-1}\\
\frac{d\epsilon_{2}}{dt} & = & 3k\epsilon_{2},\label{eq:k-kk-mo-ne-2}
\end{eqnarray}
 for $(\pi,0)$, and 
\begin{eqnarray}
\frac{d\epsilon_{1}}{dt} & = & k\epsilon_{1},\\
\frac{d\epsilon_{2}}{dt} & = & 2k\epsilon_{1}-3k\epsilon_{2},\label{eq:k-kk-mo-no-2}
\end{eqnarray}
 for $(\pi,\pi)$. Only (\ref{eq:k-kk-mo-ne-1}) and (\ref{eq:k-kk-mo-ne-2})
correspond to fixed points that can be stable, if we choose $k<0$.
There are no choices of $k$ that would allow the other points to
be stable.

The other stable points are $(\frac{1}{3}\pi,\frac{2}{3}\pi)$ and
$(-\frac{1}{3}\pi,-\frac{2}{3}\pi).$ The linearized (\ref{eq:kuramoto-1})\textbf{
}and\textbf{ }(\ref{eq:kuramoto-2}) become 
\begin{eqnarray}
\frac{1}{k}\frac{d\epsilon_{1}}{dt} & = & -2\sin\left(\pm\frac{1}{3}\pi+\epsilon_{1}\right)+\sin\left(\pm\frac{2}{3}\pi+\epsilon_{2}\right)\nonumber \\
 &  & -\sin\left(\pm\frac{1}{3}\pi+\epsilon_{1}\mp\frac{2}{3}\pi-\epsilon_{2}\right),\\
\frac{1}{k}\frac{d\epsilon_{2}}{dt} & = & -\sin\left(\pm\frac{1}{3}\pi+\epsilon_{1}\right)+2\sin\left(\pm\frac{2}{3}\pi+\epsilon_{2}\right)\nonumber \\
 &  & +\sin\left(\pm\frac{1}{3}\pi+\epsilon_{1}\mp\frac{2}{3}\pi-\epsilon_{2}\right).
\end{eqnarray}
 Using the linear approximation for $\epsilon_{1}\ll1$ and $\epsilon_{2}\ll1$
we get 
\begin{eqnarray*}
\frac{1}{k}\frac{d\epsilon_{1}}{dt} & = & -\frac{3}{2}\epsilon_{1},\\
\frac{1}{k}\frac{d\epsilon_{2}}{dt} & = & -\frac{3}{2}\epsilon_{2}.
\end{eqnarray*}
 This point is stable if $k>0.$ Thus, for $k\equiv k_{s,r_{1}}=-k_{s,r_{2}}=k_{r_{1},r_{2}}$
there are three relevant stable fixed points: $(\pi,0)$, $(\frac{1}{3}\pi,\frac{2}{3}\pi)$,
and $(-\frac{1}{3}\pi,-\frac{2}{3}\pi)$, but if $k<0$ only the point
$(\pi,0)$ is stable, corresponding to oscillator $r_{2}$ in phase
with oscillator $s$, and $r_{1}$ off phase with $s$.

\subsubsection{$k\equiv k_{s,r_{1}}=k_{s,r_{2}}=-k_{r_{1},r_{2}}$\label{subsubap:kk-k}}

For the points $\{(0,0),(0,\pi),(\pi,0),(\pi,\pi)\}$, linearizing
(\ref{eq:linear-kuramoto-1})\textbf{ }and (\ref{eq:linear-kuramoto-w})
gives us 
\begin{eqnarray}
\frac{d\epsilon_{1}}{dt} & = & -3k\epsilon_{1},\label{eq:kk-k-me-ne-1}\\
\frac{d\epsilon_{2}}{dt} & = & k\epsilon_{2},\label{eq:kk-k-me-ne-2}
\end{eqnarray}
 for $(0,0)$, 
\begin{eqnarray}
\frac{d\epsilon_{1}}{dt} & = & -3k\epsilon_{1}+2k\epsilon_{2},\label{eq:kk-k-me-no-1}\\
\frac{d\epsilon_{2}}{dt} & = & k\epsilon_{2},
\end{eqnarray}
 for $(0,\pi)$, 
\begin{eqnarray}
\frac{d\epsilon_{1}}{dt} & = & k\epsilon_{1},\\
\frac{d\epsilon_{2}}{dt} & = & 2k\epsilon_{1}-3k\epsilon_{2},\label{eq:kk-k-mo-ne-2}
\end{eqnarray}
 for $(\pi,0)$, and 
\begin{eqnarray}
\frac{d\epsilon_{1}}{dt} & = & 3k\epsilon_{1},\label{eq:kk-k-mo-no-1}\\
\frac{d\epsilon_{2}}{dt} & = & 3k\epsilon_{2},\label{eq:kk-k-mo-no-2}
\end{eqnarray}
 for $(\pi,\pi)$. Once again, only (\ref{eq:kk-k-mo-no-1}) and (\ref{eq:kk-k-mo-no-2})
correspond to fixed points that can be stable, if we choose $k<0$.

The other fixed points are $(-\frac{1}{3}\pi,\frac{1}{3}\pi)$ and
$(\frac{1}{3}\pi,-\frac{1}{3}\pi).$ Substituting $\phi_{1s}=\mp\frac{1}{3}\pi+\epsilon_{1}\left(t\right)$
and $\phi_{2s}=\pm\frac{1}{3}\pi+\epsilon_{2}\left(t\right)$ in (\ref{eq:kuramoto-1})\textbf{
}and\textbf{ }(\ref{eq:kuramoto-2}), after a linear approximation
for $\epsilon_{1}\ll1$ and $\epsilon_{2}\ll1$, we obtain

\begin{eqnarray}
\frac{1}{k}\frac{d\epsilon_{1}}{dt} & = & -\frac{3}{2}\epsilon_{1},\\
\frac{1}{k}\frac{d\epsilon_{2}}{dt} & = & -\frac{3}{2}\epsilon_{2}.
\end{eqnarray}
 This point is stable if $k>0.$ For $k\equiv k_{s,r_{1}}=k_{s,r_{2}}=-k_{r_{1},r_{2}}$there
are three relevant stable fixed points: $(\pi,\pi)$, $(\frac{2}{3}\pi,-\frac{2}{3}\pi)$,
and $(-\frac{2}{3}\pi,\frac{2}{3}\pi)$.%
\footnote{In fact we have, as discussed above, an infinite number of stable
fixed points. But since we are only interested in phase differences,
and the fixed points other than $(\frac{2}{3}\pi,-\frac{2}{3}\pi)$
and $(-\frac{2}{3}\pi,\frac{2}{3}\pi)$ are periodic, with periodicity
$2\pi$, we omit them from our discussion. %
} However, if $k<0$ the only fixed point that is stable is $(\pi,\pi)$,
corresponding to the situation when both oscillators $r_{1}$ and
$r_{2}$ are off phase with respect to oscillator $s$ by $\pi$.

The results of Sections \ref{subsubap:kkk}--\ref{subsubap:kk-k}
are summarized in Table \ref{tab:stability}. 
\begin{table}
\begin{centering}
\begin{tabular}{|c|c|c|}
\hline 
 & $k_{s,r_{1}}>0$  & $k_{s,r_{1}}<0$\tabularnewline
\hline 
\hline 
$k_{s,r_{1}}=k_{s,r_{2}}=k_{r_{1},r_{2}}$  & $(0,0)$  & %
\begin{tabular}{c}
$(\frac{2}{3}\pi,-\frac{2}{3}\pi)$\tabularnewline
$(-\frac{2}{3}\pi,\frac{2}{3}\pi)$\tabularnewline
\end{tabular}\tabularnewline
\hline 
$-k_{s,r_{1}}=k_{s,r_{2}}=k_{r_{1},r_{2}}$  & %
\begin{tabular}{c}
$(\frac{2}{3}\pi,\frac{1}{3}\pi)$\tabularnewline
$(-\frac{2}{3}\pi,-\frac{1}{3}\pi)$ \tabularnewline
\end{tabular} & $(0,\pi)$\tabularnewline
\hline 
$k_{s,r_{1}}=-k_{s,r_{2}}=k_{r_{1},r_{2}}$  & %
\begin{tabular}{c}
$(\frac{1}{3}\pi,\frac{2}{3}\pi)$\tabularnewline
$(-\frac{1}{3}\pi,-\frac{2}{3}\pi)$\tabularnewline
\end{tabular}  & $(\pi,0)$\tabularnewline
\hline 
$k_{s,r_{1}}=k_{s,r_{2}}=-k_{r_{1},r_{2}}$  & %
\begin{tabular}{c}
$(-\frac{1}{3}\pi,\frac{1}{3}\pi)$\tabularnewline
$(\frac{1}{3}\pi,-\frac{1}{3}\pi)$\tabularnewline
\end{tabular} & $(\pi,\pi)$\tabularnewline
\hline 
\end{tabular}
\par\end{centering}

\caption{\label{tab:stability}Stable fixed points for $\left|k_{s,r_{1}}\right|=\left|k_{s,r_{2}}\right|=\left|k_{r_{1},r_{2}}\right|$.
Since the absolute values of the coupling strengths are all the same,
which fixed points are stable is determined by the relative signs
of the couplings. }
\end{table}
 If $k_{s,r_{1}}<0$, the relative signs of the other coupling constants
determine unique stability points, corresponding to oscillator $r_{1}$
in phase with oscillator $s$ and $r_{2}$ off phase by $\pi$ (row
2), $r_{2}$ in phase and $r_{1}$ off phase by $\pi$ (row 3), and
both $r_{1}$ and $r_{2}$ off phase by $\pi$ (row 4). The only possibility
for all oscillators $r_{1}$, $r_{2}$, and $s$ to be in phase is
when all couplings are positive.

\subsection{General sufficient criteria\label{subap:General-sufficient-criteria}}

In the section above, we showed how the fixed points behaved for couplings
with the same strength but different signs. But, since the learning
equations (\ref{eq:learningphaseS-inhib-excite-first})-(\ref{eq:learning-asym-inhibitory-last})
and the initial conditions cannot guarantee couplings with the same
strength, we need to show that a fixed point is stable under even
if the couplings have different strengths. Here, we show a more general
condition for stability. Let us take, for instance, the point $(0,\pi)$.
Close to this point, Equations (\ref{eq:kuramoto-1})-(\ref{eq:kuramoto-2}),
after linearization, become
\begin{eqnarray}
\frac{d\epsilon_{1}}{dt} & = & -2k_{s,r_{1}}\epsilon_{1}+k_{s,r_{2}}\epsilon_{2}\nonumber \\
 &  & +k_{r_{1},r_{2}}\left(\epsilon_{1}-\epsilon_{2}\right),\label{eq:linear-A-1-1}\\
\frac{d\epsilon_{2}}{dt} & = & -k_{s,r_{1}}\epsilon_{1}+2k_{s,r_{2}}\epsilon_{2}\nonumber \\
 &  & -k_{r_{1},r_{2}}\left(\epsilon_{1}-\epsilon_{2}\right).\label{eq:linear-2-A-1-1}
\end{eqnarray}
Equations (\ref{eq:linear-A-1-1}) and (\ref{eq:linear-2-A-1-1})
have eigenvalues 
\begin{eqnarray}
\lambda_{1} & = & -k_{s,r_{1}}+k_{r_{1},r_{2}}+k_{s,r_{2}}-\sqrt{D},\label{eq:eigenvalue1}
\end{eqnarray}
and 
\begin{eqnarray}
\lambda_{2} & = & -k_{s,r_{1}}+k_{r_{1},r_{2}}+k_{s,r_{2}}+\sqrt{D},\label{eq:eigenvalue2}
\end{eqnarray}
where 
\begin{eqnarray*}
D & = & k_{s,r_{1}}^{2}+k_{s,r_{1}}k_{r_{1},r_{2}}+k_{r_{1},r_{2}}^{2}\\
 &  & +k_{s,r_{1}}k_{s,r_{2}}-k_{s,r_{2}}k_{r_{1},r_{2}}+k_{s,r_{2}}^{2}.
\end{eqnarray*}
The stability point $(0,\pi)$ is stable if the eigenvalues $\lambda_{1}$
and $\lambda_{2}$ are negative. Since $\lambda_{2}<0$ implies $\lambda_{1}<0$,
we focus on Equation (\ref{eq:eigenvalue2}). As long as the couplings
are real, we can show that $D\geq0$. Thus, the requirement that in
(\ref{eq:eigenvalue2}), $\lambda_{2}$ is negative is equivalent
to
\begin{eqnarray*}
\left(k_{s,r_{1}}-k_{s,r_{2}}-k_{r_{1},r_{2}}\right)^{2} & > & k_{s,r_{1}}^{2}+k_{s,r_{1}}k_{r_{1},r_{2}}\\
 &  & +k_{r_{1},r_{2}}^{2}+k_{s,r_{1}}k_{s,r_{2}}\\
 &  & -k_{s,r_{2}}k_{r_{1},r_{2}}+k_{s,r_{2}}^{2},
\end{eqnarray*}
which simplifies to 
\begin{equation}
k_{s,r_{2}}k_{r_{1},r_{2}}>k_{s,r_{1}}k_{r_{1},r_{2}}+k_{s,r_{1}}k_{s,r_{2}}.\label{eq:general-condition}
\end{equation}
Equation (\ref{eq:general-condition}) is a sufficient condition for
the stability of the point $(0,\pi)$. Using the same technique, similar
conditions can be obtained for different fixed points.

\subsection{Time of convergence\label{subap:Time-of-convergence}}

As we showed in Equation (\ref{eq:general-condition}), what determines
whether a point is stable or not are the relations between the different
oscillator couplings. For example, Equation (\ref{eq:general-condition})
could be satisfied for a set of $k$'s that are very small, such as
$k_{s,r_{2}}=.01$, $k_{r_{1},r_{2}}=.01$, and $k_{s,r_{1}}=.0001$.
However, since we are interested in biologically relevant models,
showing that the equations of motion asymptotically converges to a
fixed point for these values is not sufficient. We need to show that
such convergence can happen within a reasonably short amount of time
$\Delta t_{r}$, such that responses associated to the fixed points
are correctly selected, after effective reinforcement. In this section
we will estimate the values of the couplings such that the we should
expect fast convergence. To do this, let us focus on the case of $ $$k_{s,r_{1}}=-k_{s,r_{2}}=-k_{r_{1},r_{2}}\equiv k$,
$k>0$, discussed in \ref{subsubap:-kkk}. In this case, the Kuramoto
equations are
\begin{eqnarray}
\frac{d\phi_{1s}}{dt} & = & -2k\sin\left(\phi_{1s}\right)+k\sin\left(\phi_{2s}\right)\label{eq:kuramoto-1-1}\\
 &  & +k\sin\left(\phi_{1s}-\phi_{2s}\right),\nonumber \\
\frac{d\phi_{2s}}{dt} & = & -k\sin\left(\phi_{1s}\right)+2k\sin\left(\phi_{2s}\right)\label{eq:kuramoto-2-1}\\
 &  & -k\sin\left(\phi_{1s}-\phi_{2s}\right),\nonumber 
\end{eqnarray}
and for $(0,\pi)$ we define $\phi_{1s}=\epsilon_{1}$ and $\phi_{2s}=\pi+\epsilon_{2}$.
$ $ Then (\ref{eq:kuramoto-1-1})\textbf{ }and (\ref{eq:kuramoto-2-1})\textbf{
}become
\begin{eqnarray}
\frac{d\epsilon_{1}}{dt} & = & -3k\epsilon_{1},\label{eq:-kkk-me-no-1-1}\\
\frac{d\epsilon_{2}}{dt} & = & -3k\epsilon_{2}.\label{eq:-kkk-me-no-2-1}
\end{eqnarray}
We emphasize that equations are just an approximation for the system
close to the fixed point $(0,\pi)$. However, because of the sine
terms in Kuramoto's equation, this is a reasonably good approximation,
with an error of the order of $O(\epsilon_{1}^{3},\epsilon_{2}^{3})$.
Thus, for points close to $(0,\pi)$, the solutions to (\ref{eq:kuramoto-1-1})\textbf{
}and (\ref{eq:kuramoto-2-1}) are 
\[
\phi_{1s}\left(t\right)=\phi_{1s}\left(0\right)e^{-3kt},
\]
and 
\[
\phi_{2s}\left(t\right)=\pi+\left[\phi_{2s}\left(0\right)-\pi\right]e^{-3kt},
\]
where $\phi_{1s}\left(0\right)$ and $\phi_{2s}\left(0\right)$ are
the initial conditions. This is an exponential decay converging to
the fixed point, and of course neither of the solutions go, within
a finite amount of time, to $\left(0,\pi\right)$. However, the constant
$\tau=1/3k$ gives us the mean lifetime of the system, thus giving
a measure of how fast the convergence to the fixed point is happening.
Thus, if we expect the mean lifetime to be at least of the order of
$\Delta t_{r}=200\mbox{ ms}$, $k$ needs to be approximately $2\mbox{ Hz}$.
For practical purposes, if we have $k\gg2\mbox{ Hz}$, our system
should converge most of the time within the expected time $\Delta t_{r}$,
a result consistent with our numerical simulations.

\subsection{Synchronization of oscillators during reinforcement\label{subap:Synchronization}}

We are interested in knowing the qualitative behavior of the phases
when a reinforcement occurs. During reinforcement, the oscillators
satisfy equations 
\begin{eqnarray}
\frac{d\varphi_{s}}{dt} & = & +K_{0}\sin\left(\varphi_{S}-\omega_{e}t\right)\nonumber \\
 &  & \omega_{0}-k_{s,r_{1}}\sin\left(\varphi_{s}-\varphi_{r_{1}}\right)\nonumber \\
 &  & -k_{s,r_{2}}\sin\left(\varphi_{s}-\varphi_{r_{2}}\right),\label{eq:learningphaseS-B}\\
\frac{d\varphi_{r_{1}}}{dt} & = & K_{0}\sin\left(\varphi_{r_{1}}-\omega_{e}t-\pi\left(1-\delta_{E_{n},1}\right)\right)\nonumber \\
 &  & +\omega_{0}-k_{s,r_{1}}\sin\left(\varphi_{r_{1}}-\varphi_{s}\right)\nonumber \\
 &  & -k_{r_{1},r_{2}}\sin\left(\varphi_{r_{1}}-\varphi_{r_{2}}\right),\label{eq:learningphase1-B}\\
\frac{d\varphi_{r_{2}}}{dt} & = & K_{0}\sin\left(\varphi_{r_{2}}-\omega_{e}t-\pi\left(1-\delta_{E_{n},2}\right)\right)\nonumber \\
 &  & +\omega_{0}-k_{s,r_{2}}\sin\left(\varphi_{r_{2}}-\varphi_{s}\right)\nonumber \\
 &  & -k_{r_{1},r_{2}}\sin\left(\varphi_{r_{2}}-\varphi_{r_{1}}\right),\label{eq:learningphase2-B}\\
\frac{dk_{s,r_{1}}}{dt} & = & \epsilon\left(K_{0}\right)\left[\alpha\cos\left(\varphi_{s}-\varphi_{r_{1}}\right)-k_{s,r_{1}}\right],\label{eq:learningcouplingS1-B}\\
\frac{dk_{s,r_{2}}}{dt} & = & \epsilon\left(K_{0}\right)\left[\alpha\cos\left(\varphi_{s}-\varphi_{r_{2}}\right)-k_{s,r_{2}}\right],\label{eq:learningcouplingS2-B}\\
\frac{dk_{r_{1},r_{2}}}{dt} & = & \epsilon\left(K_{0}\right)\left[\alpha\cos\left(\varphi_{r_{1}}-\varphi_{r_{2}}\right)-k_{r_{1},r_{2}}\right],\label{eq:learningcoupling12-B}
\end{eqnarray}
where $E_{n}$ is either $1$ or $2$ depending on which finite response
is being reinforced, and $\delta_{i,j}$ is Kronecker's delta, i.e.,
$\delta_{E_{n},2}$ is one if $E_{n}=2$ and zero otherwise. We show
here that the first three equations lead the oscillators $s$, $r_{1}$,
and $r_{2}$ to synchronize and phase lock with reinforcement oscillators
$e_{1}$ and $e_{2}$ if $K_{0}$ is sufficiently large. We start
with the approximation 
\[
\dot{\varphi}_{s}\approx\omega_{0}+K_{0}\sin\left(\varphi_{s}-\omega_{e}t\right),
\]
 for $K_{0}\gg k_{r_{1},r_{2}},k_{s,r_{1}},k_{s,r_{2}}$. An exact
solution to this equation can be found, namely
\[
\varphi_{s}=\omega_{e}t+2\arctan\left(-\frac{\Gamma}{\delta_{e0}}\right),
\]
where $\Gamma=K_{0}-\sqrt{\delta_{e0}^{2}-K_{0}^{2}}\tan\left(\frac{1}{2}\left(t+c_{1}\right)\sqrt{\delta_{e0}^{2}-K_{0}^{2}}\right)$,
$\delta_{e0}=\omega_{0}-\omega_{e}$, $c_{1}$ is an integration constant,
and we assumed $\omega_{e}\neq\omega_{0}$. This equation shows a
term with frequency $\omega_{e}$ plus another term that depends on
$t$ in a more complicated way. Let us assume, as we did in the main
text, that $K_{0}$ is large enough, such that $K_{0}>\left|\omega_{0}-\omega_{e}\right|$.
This implies that the term inside the square root in the above equation
is negative. Let 
\[
K_{1}=i\sqrt{K_{0}^{2}-\delta_{e0}^{2}}=i\kappa,
\]
 and we have
\begin{eqnarray*}
\varphi_{s} & = & \omega_{e}t+2\arctan\left(F\left(t\right)\right),
\end{eqnarray*}
where 
\[
F\left(t\right)=-\frac{K_{0}-i\kappa\tan\left(\frac{1}{2}\left(t+c_{1}\right)i\kappa\right)}{\delta_{e0}}.
\]
 If we take the limit when $t$ goes to infinity for the $\arctan$
term we have 
\[
\lim_{t\rightarrow\infty}\arctan\left(F\left(t\right)\right)=\arctan\left(\frac{K_{0}+\kappa}{\omega_{e}-\omega_{0}}\right),
\]
 or 
\[
\arctan\left(\frac{K_{0}+\sqrt{K_{0}^{2}-\left(\omega_{0}-\omega_{e}\right)^{2}}}{\omega_{e}-\omega_{0}}\right).
\]
 If $K_{0}\gg\left|\omega_{0}-\omega_{e}\right|$, then 
\begin{align*}
\lim_{t\rightarrow\infty}\arctan\left(F\left(t\right)\right) & \approx\arctan\left(\frac{2K_{0}}{\omega_{e}-\omega_{0}}\right)\\
 & \approx\pm\frac{\pi}{2}.
\end{align*}
 So, for large $K_{0}$ and $t$, 
\[
\varphi_{s}\approx\omega_{e}t\pm\pi
\]
 and 
\[
\varphi_{r_{1}}\approx\omega_{e}t\pm\pi.
\]
 We now turn to 
\[
\dot{\varphi}_{r_{2}}=\omega_{0}+K_{0}\sin\left(\varphi_{r_{2}}-\omega_{e}t-\pi\right).
\]
 The solution is 
\[
\varphi_{r_{2}}=\omega_{e}t-2\arctan\left(\frac{\Gamma'}{\delta_{e0}}\right),
\]
 where $\Gamma'=K_{0}+\sqrt{\delta_{e0}^{2}-K_{0}^{2}}\tan\left(\frac{1}{2}\left(t+c_{2}\right)\sqrt{\delta_{e0}^{2}-K_{0}^{2}}\right)$,
and $c_{2}$ is another integration constant. Following the same arguments
as before, we obtain
\begin{align*}
\lim_{t\rightarrow\infty}\arctan\left(F\left(t\right)\right) & \approx\arctan\left(0\right)\\
 & \approx0,
\end{align*}
 or, for large $K_{0}$ and $t$, 
\[
\varphi_{r_{2}}\approx\omega_{e}t.
\]

Of course, the above arguments only tell us the behavior of the equations
when $t$ goes to infinity, but in our models we deal with finite
times. To better understand how fast the solution converges to $\omega_{e}t\pm\pi$
or $\omega_{e}t$, let us rewrite equation
\[
\varphi_{s}=\omega_{e}t+2\arctan\left(-\frac{\Gamma}{\delta_{e0}}\right)
\]
 in terms of dimensionless quantities. Let $\gamma=\left(\omega_{e}-\omega_{0}\right)/K_{0},$
then
\begin{align*}
\varphi_{s} & =\omega_{e}t\\
 & +2\arctan\left(\frac{1}{\gamma}-\frac{\sqrt{\gamma^{2}-1}}{\gamma}\tan\left(\frac{\left(t+c_{1}\right)K_{0}\sqrt{\gamma^{2}-1}}{2}\right)\right).
\end{align*}
Since $\gamma$ is real and $\gamma\ll1$ for large values of $K_{0}$,
we rewrite
\begin{align*}
\varphi_{s} & =\omega_{e}t\\
 & +2\arctan\left(\frac{1+\sqrt{1-\gamma^{2}}\tanh\left(\frac{K_{0}\left(t+c_{1}\right)\sqrt{1-\gamma^{2}}}{2}\right)}{\gamma}\right).
\end{align*}
 We're only interested in the last term, 
\begin{equation}
2\arctan\left(\frac{1+\sqrt{1-\gamma^{2}}\tanh\left(\frac{K_{0}\left(t+c_{1}\right)\sqrt{1-\gamma^{2}}}{2}\right)}{\gamma}\right).\label{eq:A}
\end{equation}
 But for large values of $t$, 
\begin{equation}
\tanh\frac{K_{0}\left(t+c_{1}\right)\sqrt{1-\gamma^{2}}}{2},\label{eq:tanh}
\end{equation}
which is equal to 
\[
\frac{e^{\frac{K_{0}}{2}\left(t+c_{1}\right)\sqrt{1-\gamma^{2}}}-e^{-\frac{K_{0}}{2}\left(t+c_{1}\right)\sqrt{1-\gamma^{2}}}}{e^{\frac{K_{0}}{2}\left(t+c_{1}\right)\sqrt{1-\gamma^{2}}}+e^{-\frac{K_{0}}{2}\left(t+c_{1}\right)\sqrt{1-\gamma^{2}}}},
\]
goes to $1$. In fact, (\ref{eq:tanh}), and therefore (\ref{eq:A}),
shows a characteristic time 
\[
t_{c}=\frac{2}{K_{0}\sqrt{1-\gamma^{2}}}=\frac{2}{\sqrt{K_{0}^{2}-\left(\omega_{e}-\omega_{0}\right)^{2}}}.
\]
 Thus, for $t>t_{c}$ and for large $K_{0}$ (i.e., $K_{0}\gg\left|\omega_{e}-\omega_{0}\right|$),
a good approximation for $\varphi_{s}$ is 
\[
\varphi_{s}\approx\omega_{e}t\pm\pi.
\]
 Similar arguments can be made for $\varphi_{r_{1}}$ and $\varphi_{r_{2}}$.

\subsection{Effectiveness of Reinforcement for the Oscillator Model\label{subap:Effectiveness-Reinforcement}}

Here we show how to compute $K'$ from the behavioral parameter $\theta$
and the other oscillator parameters. First, $K_{0}$ satisfies the
normal distribution density
\begin{equation}
p\left(K_{0}\right)=\frac{1}{\sigma\sqrt{2\pi}}e^{-\frac{1}{2\sigma^{2}}\left(K_{0}-\overline{K}_{0}\right)^{2}},
\end{equation}
 where $\sigma$ is the standard deviation. The sigmoid function 
\begin{equation}
\epsilon\left(K_{0}\right)=\frac{\epsilon_{0}}{1+e^{-\gamma\left(K_{0}-K'\right)}}
\end{equation}
 determines whether the couplings are affected. We also assume that
the time of reinforcement, $\Delta t_{e}$, is large compared to $1/\epsilon$,
as discussed in the text. If $\gamma\gg1$, $\epsilon(K_{0})$ approaches
a Heaviside function $H(K_{0})$ as $\epsilon(K_{0})\approx\epsilon_{0}H(K_{0}-K')$.
Thus, for large values of $\gamma$, learning happens only if $K_{0}>K'$,
and the probability of an oscillator reinforcement being effective
\begin{equation}
\theta=\frac{1}{\sigma\sqrt{2\pi}}\int_{K'}^{\infty}e^{-\frac{1}{2\sigma^{2}}\left(K_{0}-\overline{K}_{0}\right)^{2}}dK_{0},
\end{equation}
 or
\begin{equation}
\theta=\frac{1}{2}\left(1+\mbox{erf }\left(\frac{\sqrt{2}}{2}\frac{\overline{K}_{0}-K'}{\sigma}\right)\right).
\end{equation}
 Since $\theta$ is monotonically decreasing with $K'$, it is possible
to solve the above equation for $K'$. For example, if we set $\theta'=.19$
we obtain that 
\begin{equation}
\overline{K}_{0}-K'=-0.8779\sigma.\label{eq:k0kprimesigma}
\end{equation}
 Choosing $\sigma=10$ and $\overline{K}_{0}=100$, from (\ref{eq:k0kprimesigma})
we obtain $K'=187.79$.

\bibliographystyle{elsart-harv} \bibliographystyle{elsarticle-harv}
\bibliographystyle{elsarticle-harv}
\bibliography{Biophysics,Biophysics_add}

\begin{thebibliography}{62}
\expandafter\ifx\csname natexlab\endcsname\relax\def\natexlab#1{#1}\fi
\expandafter\ifx\csname url\endcsname\relax
  \def\url#1{\texttt{#1}}\fi
\expandafter\ifx\csname urlprefix\endcsname\relax\def\urlprefix{URL }\fi

\bibitem[{Acebron et~al.(2005)Acebron, Bonilla, Vicente, Ritort, and
  Spigler}]{AcebronEtAl2005}
Acebron, J.~A., Bonilla, L.~L., Vicente, C. J.~P., Ritort, F., Spigler, R.,
  2005. The kuramoto model: A simple paradigm for synchronization phenomena.
  Reviews of Modern Physics 77~(1), 137--185.

\bibitem[{Billock and Tsou(2005)}]{billock2005sensory}
Billock, V.~A., Tsou, B.~H., 2005. Sensory recoding via neural synchronization:
  integrating hue and luminance into chromatic brightness and saturation. J.
  Opt. Soc. Am. A 22~(10), 2289--2298.

\bibitem[{Billock and Tsou(2011)}]{billock2011tohonor}
Billock, V.~A., Tsou, B.~H., 2011. To honor fechner and obey stevens:
  Relationships between psychophysical and neural nonlinearities. Psychological
  Bulletin 137~(1), 1--18.

\bibitem[{Bower(1961)}]{Bower1961}
Bower, G., 1961. Application of a model to paired-associate learning.
  Psychometrika 26, 255--280.

\bibitem[{Bower and Beeman(2003)}]{BowerBeeman2003}
Bower, J.~M., Beeman, D., 2003. The Book of Genesis: Exploring Realistic Neural
  Models with the {GE}neral {NE}ural {SI}mulation {S}ystem. Internet Edition.
\newline\urlprefix\url{http://www.genesis-sim.org/GENESIS}

\bibitem[{Bruza et~al.(2009)Bruza, Busemeyer, and
  Gabora}]{BruzaBusemeyerGabora2009}
Bruza, P., Busemeyer, J., Gabora, L., 2009. {Introduction to the special issue
  on quantum cognition}. Journal of Mathematical Psychology 53, 303--305.

\bibitem[{Busemeyer et~al.(2006)Busemeyer, Wang, and Townsend}]{Busemeyer2006}
Busemeyer, J.~R., Wang, Z., Townsend, J.~T., 2006. Quantum dynamics of human
  decision-making. Journal of Mathematical Psychology 50, 220--241.

\bibitem[{de~Barros et~al.(2006)de~Barros, Carvalhaes, de~Mendonca, and
  Suppes}]{deBarrosEtAl2006}
de~Barros, J.~A., Carvalhaes, C.~G., de~Mendonca, J. P. R.~F., Suppes, P.,
  2006. Recognition of words from the eeg laplacian. Revista Brasileira de
  Engenharia Biomedica 21, 45--59.

\bibitem[{de~Barros and Suppes(2009)}]{deBarrosSuppes2009a}
de~Barros, J.~A., Suppes, P., 2009. {Quantum mechanics, interference, and the
  brain}. Journal of Mathematical Psychology 53, 306--313.

\bibitem[{Dickinson(1980)}]{dickinson1980contemporary}
Dickinson, A., 1980. Contemporary animal learning theory. Cambridge Univ Press,
  Cambridge, Great Britain.

\bibitem[{Eckhorn et~al.(1988)Eckhorn, Bauer, Jordan, Brosch, Kruse, Munk, and
  Reitboeck}]{EckhornEtAl1988}
Eckhorn, R., Bauer, R., Jordan, W., Brosch, M., Kruse, W., Munk, M., Reitboeck,
  H., 1988. Coherent oscillations: A mechanism of feature linking in the visual
  cortex? Biological Cybernetics 60~(2), 121--130.

\bibitem[{Eeckman and Freeman(1991)}]{EeckmanFreeman1991}
Eeckman, F.~H., Freeman, W.~J., 1991. Asymmetric sigmoid non-linearity in the
  rat olfactory system. Brain Research 557, 13--21.

\bibitem[{Estes(1950)}]{estes1950towarda2}
Estes, W.~K., 1950. Toward a statistical theory of learning. Psychological
  Review 57~(2), 94--107.
\newline\urlprefix\url{http://psycnet.apa.org/journals/rev/57/2/94/}

\bibitem[{Estes(1959)}]{EstesSuppes1959}
Estes, W.~K., 1959. Component and pattern models with markovian
  interpretations. In: Bush, R.~R., Estes, W.~K. (Eds.), Studies in
  Mathematical Leaning Theory. Stanford University Press, Stanford, CA, pp.
  9--52.

\bibitem[{Freeman(1979)}]{Freeman1979}
Freeman, W.~J., 1979. Nonlinear dynamics of paleocortex manifested in the
  olfactory eeg. Biological Cybernetics 35, 21--37.

\bibitem[{Freeman and Barrie(1994)}]{FreemanBarrie1994}
Freeman, W.~J., Barrie, J.~M., 1994. Temporal Coding in the Brain. Springer,
  New York, Ch. Chaotic oscillations and the genesis of meaning in cerebral
  cortex, pp. 13--37, corrected Internet edition, available at
  http://sulcus.berkeley.edu/wjf/AB.Genesis.of.Meaning.pdf.

\bibitem[{Friedrich et~al.(2004)Friedrich, Habermann, and
  Laurent}]{FriedrichEtAl2004}
Friedrich, R.~W., Habermann, C.~J., Laurent, G., 2004. Multiplexing using
  synchrony in the zebrafish olfactory bulb. Nature neuroscience 7~(8),
  862--871.

\bibitem[{Gerstner and Kistler(2002)}]{GerstnerKistler2002}
Gerstner, W., Kistler, W., 2002. Spiking Neuron Models. Cambridge University
  Press, Cambridge.

\bibitem[{Guckenheimer and Holmes(1983)}]{GuckenheimerHolmes1983}
Guckenheimer, J., Holmes, P., 1983. Nonlinear Oscillations, Dynamical Systems,
  and Bifurcation of Vector Fields. Springer-Verlag, New York.

\bibitem[{Hoppensteadt and
  Izhikevich(1996{\natexlab{a}})}]{HoppensteadtIzhikevich1996a}
Hoppensteadt, F.~C., Izhikevich, E.~M., 1996{\natexlab{a}}. Synaptic
  organizations and dynamical properties of weakly connected neural oscillators
  i. analysis of a canonical model. Biological Cybernetics 75~(2), 117--127.

\bibitem[{Hoppensteadt and
  Izhikevich(1996{\natexlab{b}})}]{HoppensteadtIzhikevich1996b}
Hoppensteadt, F.~C., Izhikevich, E.~M., 1996{\natexlab{b}}. Synaptic
  organizations and dynamical properties of weakly connected neural oscillators
  ii. learning phase information. Biological Cybernetics 75~(2), 129--135.

\bibitem[{Izhikevich(2007)}]{Izhikevich2007}
Izhikevich, E.~M., 2007. Dynamical Systems in Neuroscience: The Geometry of
  Excitability and Bursting. The MIT Press, Cambridge, Massachusetts.

\bibitem[{Kazantsev et~al.(2004)Kazantsev, Nekorkin, Makarenko, and
  Llinas}]{KazantsevEtAl2004}
Kazantsev, V.~B., Nekorkin, V.~I., Makarenko, V.~I., Llinas, R., 2004.
  Self-referential phase reset based on inferior olive oscillator dynamics.
  Proceedings of the National Academy of Sciences of the United States of
  America 101~(52), 18183--18188.

\bibitem[{Keeping(1995)}]{keeping1995introduction}
Keeping, E.~S., 1995. Introduction to statistical inference. Dover Publications
  Inc., Mineola, New York.

\bibitem[{Kuramoto(1984)}]{Kuramoto1984}
Kuramoto, Y., 1984. Chemical Oscillations, Waves, and Turbulence. Dover
  Publications, Inc., Mineola, New York.

\bibitem[{Leznik et~al.(2002)Leznik, Makarenko, and Llinas}]{LeznikEtAl2002}
Leznik, E., Makarenko, V., Llinas, R., 2002. {E}lectrotonically {M}ediated
  {O}scillatory {P}atterns in {N}euronal {E}nsembles: {A}n {I}n {V}itro
  {V}oltage-{D}ependent {D}ye-{I}maging {S}tudy in the {I}nferior {O}live. J.
  Neurosci. 22~(7), 2804--2815.

\bibitem[{Luce(1986)}]{Luce1986}
Luce, R.~D., 1986. Response Times. Oxford University Press, Ney York.

\bibitem[{Lutz et~al.(2002)Lutz, Lachaux, Martinerie, and
  Varela}]{LutzEtAl2002}
Lutz, A., Lachaux, J.-P., Martinerie, J., Varela, F.~J., 2002. Guiding the
  study of brain dynamics by using first-person data: Synchrony patterns
  correlate with ongoing conscious states during a simple visual task.
  Proceedings of the National Academy of Sciences 99~(3), 1586--1591.

\bibitem[{Lytton and Sejnowski(1991)}]{LyttonSejnowski1991}
Lytton, W.~W., Sejnowski, T.~J., 1991. Simulations of cortical pyramidal
  neurons synchronized by inhibitory interneurons. J Neurophysiol 66~(3),
  1059--1079.

\bibitem[{Murthy and Fetz(1992)}]{MurthyFetz1992}
Murthy, V.~N., Fetz, E.~E., 1992. {C}oherent 25- to 35-{H}z {O}scillations in
  the {S}ensorimotor {C}ortex of {A}wake {B}ehaving {M}onkeys. Proceedings of
  the National Academy of Sciences 89~(12), 5670--5674.

\bibitem[{Nishii(1998)}]{Nishii1998}
Nishii, J., 1998. A learning model for oscillatory networks. Neural Networks
  11~(2), 249--257.

\bibitem[{Nunez and Srinivasan(2006)}]{Nunez2006a}
Nunez, P., Srinivasan, R., 2006. Electric Fields of the Brain: The Neurophysics
  of EEG, 2nd Ed. Oxford University Press.

\bibitem[{Park et~al.(2003)Park, Soa, Barreto, Gluckman, and
  Schi}]{ParkEtAl2003}
Park, E.-H., Soa, P., Barreto, E., Gluckman, B.~J., Schi, S.~J., 2003. Electric
  field modulation of synchronization in neuronal networks. Neurocomputing
  52-54, 169--175.

\bibitem[{Rees et~al.(2002)Rees, Kreiman, and Koch}]{ReesEtAl2002}
Rees, G., Kreiman, G., Koch, C., 2002. Neural correlates of consciousness in
  humans. Nat Rev Neurosci 3~(4), 261--270.

\bibitem[{Rodriguez et~al.(1999)Rodriguez, George, Lachaux, Martinerie, and
  Varela}]{RodriguezEtAl1999}
Rodriguez, E., George, N., Lachaux, J.-P., Martinerie, B.~R., Varela, F.~J.,
  1999. Perception's shadow: long-distance synchronization of human brain
  activity. Nature 397, 430--433.

\bibitem[{Seliger et~al.(2002)Seliger, Young, and Tsimring}]{SeligerEtAl2002}
Seliger, P., Young, S.~C., Tsimring, L.~S., 2002. Plasticity and learning in a
  network of coupled phase oscillators. Physical Review E 65, 041906--1--7.

\bibitem[{Sompolinsky et~al.(1990)Sompolinsky, Golomb, and
  Kleinfeld}]{SompolinskyEtAl1990}
Sompolinsky, H., Golomb, D., Kleinfeld, D., 1990. {Global Processing of Visual
  Stimuli in a Neural Network of Coupled Oscillators}. PNAS 87~(18),
  7200--7204.

\bibitem[{Steinmetz et~al.(2000)Steinmetz, Roy, Fitzgerald, Hsiao, Johnson, and
  Niebur}]{SteinmetzEtAl2000}
Steinmetz, P.~N., Roy, A., Fitzgerald, P.~J., Hsiao, S.~S., Johnson, K.~O.,
  Niebur, E., 2000. Attention modulates synchronized neuronal firing in primate
  somatosensory cortex. Nature 404~(6774), 187--190.

\bibitem[{Strogatz(2000)}]{Strogatz2000}
Strogatz, S.~H., 2000. From kuramoto to crawford: exploring the onset of
  synchronization in populations of coupled oscillators. Physica D: Nonlinear
  Phenomena 143~(1-4), 1--20.

\bibitem[{Suppes(1959)}]{suppes1959alinear}
Suppes, P., 1959. A linear learning model for a continuum of responses. In:
  Bush, R.~R., Estes, W.~K. (Eds.), Studies in Mathematical Leaning Theory.
  Stanford University Press, Stanford, {CA}, pp. 400--414.

\bibitem[{Suppes(1960)}]{suppes1960stimulus}
Suppes, P., 1960. Stimulus-sampling theory for a continuum of responses. In:
  K.~Arrow, S.~K., Suppes, P. (Eds.), Mathematical Methods in the Social
  Sciences, 1959; proceedings of the first Stanford Symposium. Stanford
  University Press, Stanford, {CA}, Ch.~23, pp. 348--365.

\bibitem[{Suppes(1969)}]{suppes1969stimulusresponse}
Suppes, P., Oct. 1969. Stimulus-response theory of finite automata. Journal of
  Mathematical Psychology 6~(3), 327--355.
\newline\urlprefix\url{http://www.sciencedirect.com/science/article/pii/0022249669900108}

\bibitem[{Suppes(2002)}]{SuppesRISS2002}
Suppes, P., 2002. Representation and Invariance of Scientific Structures. CSLI
  Publications, Stanford, CA.

\bibitem[{Suppes and Atkinson(1960)}]{SuppesAtkinson1960}
Suppes, P., Atkinson, R.~C., 1960. Markov Learning Models for Multiperson
  Interactions. Stanford University Press, Stanford, CA.

\bibitem[{Suppes and de~Barros(2007)}]{SuppesdeBarros2007}
Suppes, P., de~Barros, J.~A., 2007. Quantum mechanics and the brain. In:
  Quantum Interaction: Papers from the AAAI Spring Symposium. Technical Report
  SS-07-08. AAAI Press, Menlo Park, CA, pp. 75--82.

\bibitem[{Suppes and Frankmann(1961)}]{Suppes1961test}
Suppes, P., Frankmann, R., 1961. {Test of stimulus sampling theory for a
  continuum of responses with unimodal noncontingent determinate
  reinforcement.} Journal of Experimental Psychology 61~(2), 122--132.

\bibitem[{Suppes and Ginsberg(1963)}]{SuppesGinsberg1963}
Suppes, P., Ginsberg, R., 1963. A fundamental property of all-or-none models,
  binomial distribution of responses prior to conditioning, with application to
  concept formation in children. Psychological Review 70, 139--161.

\bibitem[{Suppes and Han(2000)}]{SuppesBrain2000}
Suppes, P., Han, B., 2000. Brain-wave representation of words by superposition
  of a few sine waves. Proceedings of the National Academy of Sciences 97,
  8738--8743.

\bibitem[{Suppes et~al.(1999{\natexlab{a}})Suppes, Han, Epelboim, and
  Lu}]{SuppesBrain1999b}
Suppes, P., Han, B., Epelboim, J., Lu, Z.-L., 1999{\natexlab{a}}. Invariance
  between subjects of brain wave representations of language. Proceedings of
  the National Academy of Sciences 96, 12953--12958.

\bibitem[{Suppes et~al.(1999{\natexlab{b}})Suppes, Han, Epelboim, and
  Lu}]{SuppesBrain1999}
Suppes, P., Han, B., Epelboim, J., Lu, Z.-L., 1999{\natexlab{b}}. Invariance of
  brain-wave representations of simple visual images and their names.
  Proceedings of the National Academy of Sciences 96, 14658--14663.

\bibitem[{Suppes et~al.(1998)Suppes, Han, and Lu}]{SuppesBrain1998}
Suppes, P., Han, B., Lu, Z.-L., 1998. Brain-wave recognition of sentences.
  Proceedings of the National Academy of Sciences 95, 15861--15866.

\bibitem[{Suppes et~al.(1997)Suppes, Lu, and Han}]{SuppesLuHan1997}
Suppes, P., Lu, Z.-L., Han, B., 1997. Brain wave recognition of word.
  Proceedings of the National Academy of Sciences 94, 14965--14969.

\bibitem[{Suppes et~al.(2009)Suppes, Perreau-Guimaraes, and
  Wong}]{Suppes2009Partial}
Suppes, P., Perreau-Guimaraes, M., Wong, D., 2009. Partial orders of similarity
  differences invariant between eeg-recorded brain and perception
  respresentations of language. Neural Computation 21, 3228--3269.

\bibitem[{Suppes et~al.(1964)Suppes, Rouanet, Levine, and
  Frankmann}]{SuppesEtAl1964}
Suppes, P., Rouanet, H., Levine, M., Frankmann, R.~W., 1964. Empirical
  comparison of models for a continuum of responses with noncontingent bimodal
  reinforcement. In: Atkinson, R.~C. (Ed.), Studies in Mathematical Psychology.
  Stanford University Press, Stanford, CA, pp. 358--379.

\bibitem[{Tallon-Baudry et~al.(2001)Tallon-Baudry, Bertrand, and
  Fischer}]{Tallon-BaudryEtAl2001}
Tallon-Baudry, C., Bertrand, O., Fischer, C., 2001. {O}scillatory {S}ynchrony
  between {H}uman {E}xtrastriate {A}reas during {V}isual {S}hort-{T}erm
  {M}emory {M}aintenance. J. Neurosci. 21~(20), 177RC--1--5.

\bibitem[{Trevisan et~al.(2005)Trevisan, Bouzat, Samengo, and
  Mindlin}]{TrevisanEtAl2005}
Trevisan, M.~A., Bouzat, S., Samengo, I., Mindlin, G.~B., 2005. Dynamics of
  learning in coupled oscillators tutored with delayed reinforcements. Physical
  Review E (Statistical, Nonlinear, and Soft Matter Physics) 72~(1),
  011907--1--7.

\bibitem[{Vassilieva et~al.(2011)Vassilieva, Pinto, de~Barros, and
  Suppes}]{vassilieva2011learning}
Vassilieva, E., Pinto, G., de~Barros, J., Suppes, P., 2011. Learning pattern
  recognition through quasi-synchronization of phase oscillators. IEEE
  Transactions on Neural Networks 22~(99), 84--95.

\bibitem[{Wang(1995)}]{Wang1995}
Wang, D., 1995. Emergent synchrony in locally coupled neural oscillators.
  Neural Networks, IEEE Transactions on Neural Networks 6~(4), 941--948.

\bibitem[{Winfree(2002)}]{Winfree2002}
Winfree, A.~T., 2002. {OSCILLATING} {SYSTEMS}: {O}n {E}merging {C}oherence.
  Science 298~(5602), 2336--2337.

\bibitem[{Wong et~al.(2006)Wong, Uy, Guimaraes, Yang, and
  Suppes}]{WongEtAl2006}
Wong, D.~K., Uy, E.~T., Guimaraes, M.~P., Yang, W., Suppes, P., 2006.
  Interpretation of perceptron weights as construct time series for eeg
  classification. NeurocomputingIn press.

\bibitem[{Wright and Liley(1995)}]{WrightLiley1995}
Wright, J., Liley, D., 1995. Simulation of electrocortical waves. Biological
  Cybernetics 72~(4), 347--356.

\bibitem[{Yamanishi et~al.(1980)Yamanishi, Kawato, and
  Suzuki}]{YamanishiEtAl1980}
Yamanishi, J.-i., Kawato, M., Suzuki, R., 1980. Two coupled oscillators as a
  model for the coordinated finger tapping by both hands. Biological
  Cybernetics 37~(4), 219--225.

\end{thebibliography}

\end{document}